%% file: from_arxiv.tex
\documentclass[prx,twocolumn,unsortedaddress,nofootinbib,superscriptaddress]{revtex4-2}
\usepackage{amsthm}
\usepackage{amsfonts}
\usepackage{siunitx}
\usepackage{amsmath}
\usepackage{amssymb}
\usepackage{graphicx}
\usepackage{verbatim}
\usepackage[colorlinks]{hyperref}
\usepackage{tikz}
\usepackage{braket}
\usepackage{xcolor}

\setcitestyle{super,open={},close={}}
\renewcommand{\onlinecite}[1]{\nocite{#1}\citenum{#1}}

\makeatletter
\def\@ssect@ltx#1#2#3#4#5#6[#7]#8{%
  \def\H@svsec{\phantomsection}%
  \@tempskipa #5\relax
  \@ifdim{\@tempskipa>\z@}{%
    \begingroup
      \interlinepenalty \@M
      #6{%
       \@ifundefined{@hangfroms@#1}{\@hang@froms}{\csname @hangfroms@#1\endcsname}%
       {\hskip#3\relax\H@svsec}{#8}%
      }%
      \@@par
    \endgroup
    \@ifundefined{#1smark}{\@gobble}{\csname #1smark\endcsname}{#7}%
  }{%
    \def\@svsechd{%
      #6{%
       \@ifundefined{@runin@tos@#1}{\@runin@tos}{\csname @runin@tos@#1\endcsname}%
       {\hskip#3\relax\H@svsec}{#8}%
      }%
      \@ifundefined{#1smark}{\@gobble}{\csname #1smark\endcsname}{#7}%
      \addcontentsline{toc}{#1}{\protect\numberline{}#8}%
    }%
  }%
  \@xsect{#5}%
}%
\makeatother

\definecolor{linkcolor}{RGB}{0,83,166}
\hypersetup{
  colorlinks = true,
  allcolors = {linkcolor}
}

\begin{document}
\newcommand{\mytitle}{Beyond-classical computation in quantum simulation}
\title{\mytitle}

\newcommand{\affildw}{D-Wave Quantum Inc., Burnaby, British Columbia, Canada}
\newcommand{\affilsfu}{Department of Physics, Simon Fraser University, Burnaby, British Columbia, Canada}
\newcommand{\affilubc}{Department of Physics and Astronomy and Quantum Matter Institute, University of British Columbia, Vancouver, British Columbia, Canada}
\newcommand{\affiljag}{Jagiellonian University, Institute of Theoretical Physics, {\L}ojasiewicza 11, PL-30348 Krak\'ow, Poland}

\author{Andrew D.~King}
\email[]{aking@dwavesys.com}
\affiliation{\affildw}

\author{Alberto Nocera}
\affiliation{\affilubc}
\author{Marek M. Rams}
\affiliation{\affiljag}
\author{Jacek Dziarmaga}
\affiliation{\affiljag}
\author{Roeland Wiersema}
\affiliation{Vector Institute, MaRS  Centre,  Toronto,  Ontario,  M5G  1M1,  Canada}
\affiliation{Department of Physics and Astronomy, University of Waterloo, Ontario, N2L 3G1, Canada} 
\author{William Bernoudy}
\affiliation{\affildw}
\author{Jack Raymond}
\affiliation{\affildw}
\author{Nitin Kaushal}
\affiliation{\affilubc}
\author{Niclas Heinsdorf}
\affiliation{\affilubc}
\affiliation{Max Planck Institute for Solid State Research, Stuttgart, 70569, Germany}
\author{Richard Harris}
\affiliation{\affildw}
\author{Kelly Boothby}
\affiliation{\affildw}
\author{Fabio Altomare}
\affiliation{\affildw}
\author{Mohsen Asad}
\affiliation{\affildw}
\author{Andrew J.~Berkley}
\affiliation{\affildw}
\author{Martin Boschnak}
\affiliation{\affildw}
\author{Kevin Chern}
\affiliation{\affildw}
\author{Holly Christiani}
\affiliation{\affildw}
\author{Samantha Cibere}
\affiliation{\affildw}
\author{Jake Connor}
\affiliation{\affildw}
\author{Martin H. Dehn}
\affiliation{\affildw}
\author{Rahul Deshpande}
\affiliation{\affildw}
\author{Sara Ejtemaee}
\affiliation{\affildw}
\author{Pau Farr{\'{e}}}
\affiliation{\affildw}
\author{Kelsey Hamer}
\affiliation{\affildw}
\author{Emile Hoskinson}
\affiliation{\affildw}
\author{Shuiyuan Huang}
\affiliation{\affildw}
\author{Mark W.~Johnson}
\affiliation{\affildw}
\author{Samuel Kortas}
\affiliation{\affildw}
\author{Eric Ladizinsky}
\affiliation{\affildw}
\author{Trevor Lanting}
\affiliation{\affildw}
\author{Tony Lai}
\affiliation{\affildw}
\author{Ryan Li}
\affiliation{\affildw}
\author{Allison J.R.~MacDonald}
\affiliation{\affildw}
\author{Gaelen Marsden}
\affiliation{\affildw}
\author{Catherine C.~McGeoch}
\affiliation{\affildw}
\author{Reza Molavi}
\affiliation{\affildw}
\author{Travis Oh}
\affiliation{\affildw}
\author{Richard Neufeld}
\affiliation{\affildw}
\author{Mana Norouzpour}
\affiliation{\affildw}
\author{Joel Pasvolsky}
\affiliation{\affildw}
\author{Patrick Poitras}
\affiliation{\affildw}
\author{Gabriel Poulin-Lamarre}
\affiliation{\affildw}
\author{Thomas Prescott}
\affiliation{\affildw}
\author{Mauricio Reis}
\affiliation{\affildw}
\author{Chris Rich}
\affiliation{\affildw}
\author{Mohammad Samani}
\affiliation{\affildw}
\author{Benjamin Sheldan}
\affiliation{\affildw}
\author{Anatoly Smirnov}
\affiliation{\affildw}
\author{Edward Sterpka}
\affiliation{\affildw}
\author{Berta Trullas Clavera}
\affiliation{\affildw}
\author{Nicholas Tsai}
\affiliation{\affildw}
\author{Mark Volkmann}
\affiliation{\affildw}
\author{Alexander Whiticar}
\affiliation{\affildw}
\author{Jed D.~Whittaker}
\affiliation{\affildw}
\author{Warren Wilkinson}
\affiliation{\affildw}
\author{Jason Yao}
\affiliation{\affildw}
\author{T.J. Yi}
\affiliation{\affildw}

\author{Anders W.~Sandvik}
\affiliation{Department of Physics, Boston University, Boston, MA, USA.}

\author{Gonzalo Alvarez}
\affiliation{Computational Sciences and Engineering Division, Oak Ridge National Laboratory, Oak Ridge, TN 37831, USA}

\author{Roger G.~Melko}
\affiliation{Department of Physics and Astronomy, University of Waterloo, Ontario, N2L 3G1, Canada}
\affiliation{Perimeter Institute for Theoretical Physics, Waterloo, Ontario, N2L 2Y5, Canada}
\author{Juan Carrasquilla}
\affiliation{Institute for Theoretical Physics, ETH Z\"urich, 8093, Switzerland}
\affiliation{Vector Institute, MaRS  Centre,  Toronto,  Ontario,  M5G  1M1,  Canada}
\affiliation{Department of Physics and Astronomy, University of Waterloo, Ontario, N2L 3G1, Canada}
\author{Marcel Franz}
\affiliation{\affilubc}

\author{Mohammad H.~Amin}
\email[]{amin@dwavesys.com}
\affiliation{\affildw}
\affiliation{\affilsfu}

\date{\today}
\begin{abstract}
Quantum computers hold the promise of solving certain problems that lie beyond the reach of conventional computers. Establishing this capability, especially for impactful and meaningful problems, remains a central challenge. Here we show that superconducting quantum annealing processors can rapidly generate samples in close agreement with solutions of the Schrödinger equation. We demonstrate area-law scaling of entanglement in the model quench dynamics of two-, three- and infinite-dimensional spin glasses, supporting the observed stretched-exponential scaling of effort for matrix-product-state approaches. We show that several leading approximate methods based on tensor networks and neural networks cannot achieve the same accuracy as the quantum annealer within a reasonable time frame. Thus quantum annealers can answer questions of practical importance that may remain out of reach for classical computation.
\end{abstract}

\maketitle

\def\title#1{\gdef\@title{#1}\gdef\THETITLE{#1}}

\section*{Introduction}

For decades, the theory of quantum computing has promised monumental speedups for select tasks.  Only in the last few years has this come to fruition, with a growing variety of quantum processing units (QPUs) performing computations enormously faster than sophisticated classical approaches---these QPUs include photonic\cite{Zhong2021,Madsen2022}, neutral-atom\cite{Ebadi2022,Shaw2023}, and superconducting\cite{Arute2019,Wu2021,Morvan2023,Kim2023} systems.  As these processors expand in size and precision, their growing capacity to tackle complex computations promises to surpass the reach of any existing classical computer~\cite{Preskill2012}.

So far, nearly all attempts to showcase beyond-classical computation have revolved around random-number generation, be it through boson sampling or random-circuit sampling\cite{10.1038/s41534-017-0018-2,10.1038/s41534-023-00703-x,Morvan2023}.  Classical simulation methods can faithfully replicate the quantum evolution of the circuits, but generally require time and memory resources that scale exponentially with problem size.  However, the presence of noise and errors can substantially diminish the required resources\cite{Aharonov2022}.  Moreover, great strides have been made in classical simulations of noisy quantum systems, curtailing some claims\cite{Pednault2019,Oh2022,Kechedzhi2024,Tindall2023}.  While it is no longer controversial to say that near-term QPUs can solve some problems far faster than the best classical methods, this has not yet been firmly established for problems of practical interest.
Ref.~\onlinecite{Kim2023} marked the first attempt to demonstrate beyond-classical computation in a problem with potential applications across various domains---discretized dynamics of a kicked Ising model---although the claim was quickly brought into question\cite{Tindall2023,beguvsic2023fast,liao2023simulation,anand2023classical,torre2023dissipative}.

\begin{figure}
\includegraphics[scale=.8]{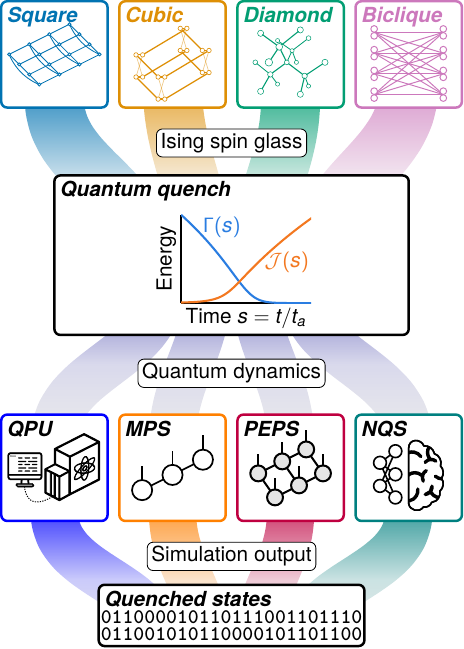}
  \caption{{\bf Sampling post-quench states in Ising spin glasses.} Random inputs are generated in topologies of varying dimension.  For each input, we consider the task of sampling from the distribution of states following a quantum quench, i.e., rapid change of transverse field $\Gamma(t/t_a)$ and longitudinal field $\mathcal J(t/t_a)$ within time $t_a$.  This task is performed using quantum annealing and classical methods based on tensor networks (MPS and PEPS) and neural networks (NQS).}\label{fig:1}
\end{figure}

Here we approach a more general and practical problem in the same category, simulating the continuous-time quantum dynamics of the transverse-field Ising model (TFIM), using superconducting quantum annealing (QA) processors\cite{Kadowaki1998, Harris2018}.  QA has already been employed to simulate quantum phase transitions (QPTs) in Ising-like systems, using both superconducting qubits\cite{Harris2018,Bando2020,King2022,King2023} and Rydberg arrays\cite{Keesling2019,Scholl2021,Ebadi2022}.  Ref.~\onlinecite{King2023} demonstrated quantum critical dynamics in cubic spin glasses on over 5000 qubits, with measured critical exponents closely matching estimates from the expected universality class---compelling  evidence of Schr\"{o}dinger evolution, which is not generally simulable with classical methods at this scale.  Still, it remains important to establish that within the parametric range studied, despite the limited correlation length and finite experimental precision, approximate classical methods cannot match the solution quality of the QPU in a reasonable amount of time.

The aim of this paper is to evaluate the computational resources necessary to sample the state of a TFIM following a quench through a QPT, at a comparable accuracy to the QPU.  Simulating quantum critical dynamics is an important problem with relevance to condensed matter physics~\cite{Zurek1985,Sachdev2011,Calabrese2006,Canovi2014,Polkovnikov2011,delCampo2014,Rossini2021} and optimization~\cite{Brooke1999,Das2008,Ebadi2022,King2023,Bernaschi2023}.  The ability of superconducting QA processors to solve this simulation problem with high accuracy has been established for 1D chains~\cite{King2022} and small spin glasses~\cite{King2023}.
Here we simulate large programmable topologies of varying dimension as illustrated in Fig.~\ref{fig:1}.

To estimate QPU error for small problems, we compare simulations with ground truths computed using well-converged matrix product state (MPS) simulations run on the Summit and Frontier supercomputers at Oak Ridge National Laboratory.  
We then employ state-of-the-art approximate classical methods based on tensor networks and neural networks to estimate the computational cost of approximately simulating the quantum dynamics to match QPU accuracy.  This direct, non-universal verification can only be achieved at small scales, where the ground truth can be computed.  To extrapolate beyond the classically simulable regime, we verify universal quantum critical scaling in QPU results, and estimate classical resource requirements based on area-law scaling that is both expected and observed.

\section*{Quenching a quantum spin glass}

We consider a time-dependent Hamiltonian that interpolates between a driving Hamiltonian $\mathcal H_D$ and a classical Ising problem Hamiltonian $\mathcal H_P$:
\begin{align}
  &\mathcal H(t) = \Gamma(t/t_a)\mathcal H_D + \mathcal J(t/t_a)\mathcal H_P,\\
&\mathcal H_D = -\sum_{i}\sigma_i^x,\ \ \ \ \ \mathcal H_P = \hspace{-1ex}\sum_{i<j} J_{ij}\sigma_i^z\sigma_j^z,\label{eq:ham}
\end{align}
where $\sigma^{x,z}_i$ are Pauli matrices acting on qubit $i$, and $\Gamma$ and $\mathcal J$ are the transverse-field and Ising energy scales\cite{SM}, respectively.
The evolution starts from time $t{=}0$ in a paramagnetic phase with $\Gamma(0) \gg \mathcal J(0)$ and ends at quench time $t{=}t_a$ with $\Gamma(1) \ll \mathcal J(1)$, deep in the spin-glass phase. These two phases are separated by a QPT whose universal behavior is dictated by the topology of the programmed Ising model. 
Nonzero couplings $J_{ij}$ are chosen randomly within various topologies: square lattices, dimerized cubic lattices, diamond lattices, and dimerized biclique graphs\cite{SM}.  
The first three represent lattice structures of real materials\cite{Bertinshaw2019, Brooke1999} and the latter has applications in generative artificial intelligence (AI)\cite{Khoshaman2018}.

We use two QPUs of distinct design: one Advantage system (ADV1) and one prototype Advantage2 system (ADV2).  Although these processors have significantly different annealing schedules, we simulate the same model quench (Fig.~\ref{fig:1}) on both.  Each QPU generates at least 1000 samples per second, with parallelization exploited for small problems\cite{SM}.

\section*{Classical simulation techniques}

We compare QPU results with classical algorithms designed to approximate the solution to the time-dependent Schr\"{o}dinger equation---see Supplementary Material (SM, Ref.~\onlinecite{SM}) for details on these methods and other potential approaches. We use two tensor-network techniques---MPS and projected entangled-pair states (PEPS)~\cite{ 10.1016/j.aop.2010.09.012, Stoudenmire2012, Verstraete2008, Orus2014, Ran2020, Cirac_review_2021}---as well as neural quantum state (NQS)~\cite{carleo2017solving} simulations.  MPS, PEPS, and NQS are currently recognized as the leading approximation frameworks in this context~\cite{Schmitt2022}.

Tensor-network ans\"atze are designed to leverage the area law of entanglement entropy~\cite{Verstraete2006, Eisert2010} to reduce the complexity of state representation. In particular, for a system of $N$ spins in $d$ spatial dimensions, MPS (or, more generally, tree tensor network) methods reduce the complexity from exponential in $N$ to exponential in $N^{(d-1)/d}$. The benefits diminish as $d$ becomes larger, e.g., for biclique problems, where $d$ diverges. The tree-like nature of such network geometries allows for the existence of canonical representations---and easy transformations between them---that underlie the time-dependent variational principle (TDVP)~\cite{Haegeman2016}, a general algorithm to simulate time evolution. The main trade-off between approximation quality and computational cost is controlled by the maximum MPS bond dimension $\chi$. Required memory scales as $O(N\chi^2)$ and time of TDVP simulation scales as $O(N\chi^3)$ (other geometries of tree tensor networks beyond MPS have higher powers of $\chi$ appearing in the corresponding scalings). The access to canonical forms enables efficient and monotonic control over Hilbert-space truncation, making MPS a suitable technique for ground-truth estimation.

While PEPS ans\"atze are better aligned with the geometry of the problem and so can more efficiently encode states with area law beyond one dimension, in general they do not benefit from the existence of canonical forms. 
Inference from PEPS representation, or its exact contraction, is known to be a $\#P$-complete problem in general~\cite{Schuch2007, Haferkamp2020}. 
All this makes it susceptible to truncation error during the time evolution~\cite{SM}. 
In principle, for short-ranged quantum correlations, PEPS can employ local or approximate heuristic contractions to reduce computational cost~\cite{Verstraete2008, Orus2014, Ran2020}. This can make PEPS efficient for sufficiently fast quenches with short-ranged correlations, such as for dynamics of a kicked Ising model on a ``heavy-hex'' lattice~\cite{Tindall2023,beguvsic2023fast,liao2023simulation}. However, we demonstrate that the cost of expanding the neighborhood for the longest experimental annealing times becomes prohibitive compared to MPS.
Moreover, as the size of the tensors is exponential in $d$, this methodology quickly becomes impractical, with extreme restrictions on feasible simulations even in three dimensions~\cite{VlaarCorboz3D}. 

\begin{figure*}
  \includegraphics[scale=1]{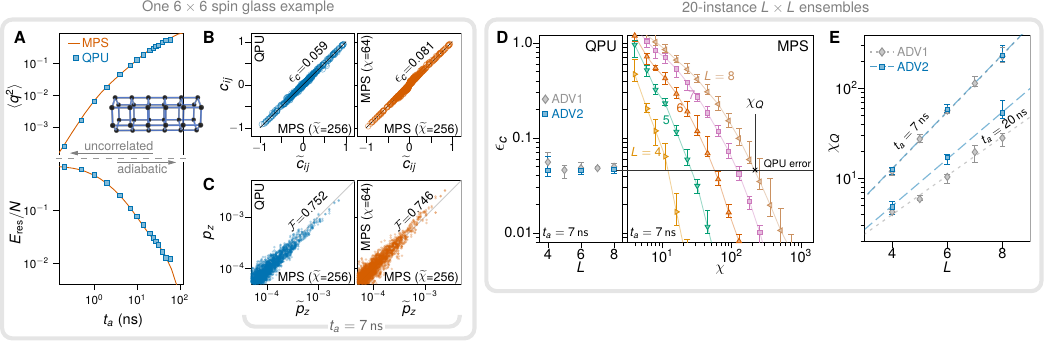}
  \caption{{\bf QPU and MPS on quenched 2D spin glasses.} ({\bf A}--{\bf C}) Data for a typical realization of a $6{\times} 6$ cylindrical spin glass. 
    ({\bf A}) Squared order parameter (top) and residual energy per spin (bottom) data exhibit close agreement between QPU and the ground truth across nearly three decades of $t_a$.  In the fast-quench limit $t_a\rightarrow 0$, correlations and therefore $\langle q^2\rangle$ are small; in the slow-quench limit $t_a\rightarrow \infty$, the system evolves adiabatically and produces classical ground states with vanishing residual energy. Approaching this limit, the problem crosses over into the much easier problem of classical ground-state sampling.  ({\bf B}) Spin-spin correlations $c_{ij}$ compared against ground-truth values $\tilde c_{ij}$ for $t_a{=}\SI{7}{ns}$.  Left, QPU.  Right, MPS at a bond dimension producing output of similar quality.  ({\bf C}) Comparison of individual state probabilities for the same simulations as {\bf B}; classical fidelities $\mathcal F$ estimated from $10^6$ samples are comparable.  ({\bf D}--{\bf E}) Ensembles of $20$ $L{\times} L$ instances (disorder realizations). ({\bf D}) Median correlation error $\epsilon_c$ remains roughly constant across system sizes for QPU (left), but matching the QPU error level in MPS requires an increasing QPU-equivalent bond dimension $\chi_{\text{Q}}$.  ({\bf E}) $\chi_{\text{Q}}$ exhibits exponential growth in $L$ for two quench rates.  For $t_a{=}\SI{20}{ns}$, the superior processor, ADV2, shows lower errors, and therefore steeper scaling of $\chi_{\text{Q}}$.  All error bars represent 95\% bootstrap confidence intervals of the ensemble median.}\label{fig:2}
\end{figure*}

NQS can also serve as ans\"atze for approximate wave functions. A variety of results have been demonstrated for calculation of ground states, but with few demonstrations relating to time evolution of disordered systems~\cite{carleo2012localization, medvidovic2021classical, Schmitt2022, Donatella2023}. 
We examine various time-integration techniques and architectures, including complex restricted Boltzmann machines (cRBMs) and autoregressive models such as transformers and recurrent neural networks (RNNs). We conclude that these approaches cannot produce results with acceptable accuracy beyond the shortest time scales and smallest models presented in this paper\cite{SM}.

\section*{Two-dimensional systems}

To make the case for beyond-classical computation in quantum simulation, we must first establish accurate and well-understood behavior on nontrivial problems that can still be simulated classically. We start with 2D spin glasses on cylindrical $L{\times} L$ square lattices, for which we calculate ground truths for up to $L{=}8$. Square lattices, relative to higher-dimensional topologies, are amenable to classical simulations: MPS and PEPS have been applied to related 2D lattice simulation problems 
\cite{Ebadi2020,Schmitt2022,Scholl2021}.

The computational task at hand is to fairly sample from the quenched state projected in the computational basis. We start with a single exemplary $6{\times} 6$ spin glass and compare QPU results with an MPS-based ground truth using two statistics:
a spin-glass order parameter
\begin{equation}
   \langle q^2 \rangle = \frac{2}{N(N-1)}\sum_{i<j}c_{ij}^2,  \label{eq:q2}
\end{equation}
and residual energy 
$E_{\rm res} {=} \langle \mathcal H_P \rangle - E_0$.
Here, $c_{ij} {=} \langle \sigma^z_i \sigma^z_j\rangle$ is the two-point correlation function and $E_0$ is the ground-state energy of $\mathcal H_P$.
As shown in Fig.~\ref{fig:2}A, the close agreement between the QPU data (symbols) and the ground truth (solid lines) extends across nearly three orders of magnitude of $t_a$.

For evaluating the estimated samples in comparison to ground-truth MPS results, our main figure of merit is based on spin-spin correlations.  Fig.~\ref{fig:2}B shows all $c_{ij}$ in the same problem. We compare the ground truth (converged at $\tilde\chi{=}256$) with QPU data (left) and approximate MPS output using a smaller $\chi{=}64$ (right). Agreement can be summarized as a correlation error 
\begin{equation}
\epsilon_c = \left(\frac{\sum_{i,j}(c_{ij}-\tilde c_{ij})^2}{\sum_{i,j} \tilde c_{ij}^2}\right)^{1/2},\label{eq:ec}
\end{equation}
where ``$^\sim$'' denotes a ground-truth value.

Various other distributional-divergence measures exist, such as Kullback–Leibler divergence, total variation distance, and classical fidelity (Bhattacharyya distance), but they have the disadvantage that when evaluated by sampling, the number of samples must scale exponentially with the distribution entropy, which grows with $N$ in our application. Nevertheless, we use classical fidelity $\mathcal F{=}\sum_z \sqrt{p_z\tilde p_z}$, where $p_z$ and $\tilde p_z$ are method and ground-truth probabilities of state $z$, to corroborate the main metric $\epsilon_c$ at small sizes.
Fig.~\ref{fig:2}C scatters $p_z$ against $\tilde p_z$ based on $10^6$ samples, again using QPU and MPS ($\chi{=}64$); resulting estimates of $\mathcal F$ are similar. 
In both Figs.~\ref{fig:2}B and \ref{fig:2}C, it is evident that the approximate MPS yields data closely resembling that of the QPU, as manifested in both error measures $\epsilon_c$ and $\mathcal F$.  In the following sections, the bond dimension that achieves comparable accuracy to the QPU, the “QPU-equivalent bond dimension”~\cite{Shaw2023}, is denoted as $\chi_Q$ ($\approx 64$ in this example).

Next, we address scaling behavior of $L{\times} L$ lattices. Fig.~\ref{fig:2}D (left) shows flat scaling of $\epsilon_c$ for QPU in $L$ (median over $20$ random disorder realizations) using both ADV1 and ADV2 processors.  Interpolating MPS \mbox{$\chi$-dependent} error (right) allows us to estimate  $\chi_Q$.  Fig.~\ref{fig:2}E shows $\chi_Q$ as a function of $L$ and $t_a$.  Crucially, we observe exponential dependence on $L$, which is consistent with area-law scaling~\cite{Vidal2003}.  For $t_a{=}\SI{20}{ns}$, ADV2 has lower error, therefore higher $\chi_Q$, than ADV1---henceforth all QPU results shown in the main text are from ADV2.

\begin{figure}
  \includegraphics[scale=1]{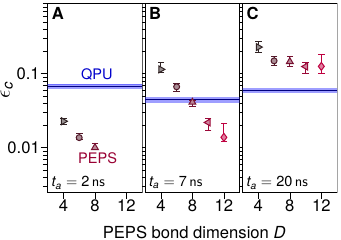}
  \caption{{\bf Breakdown of PEPS in 8${\times}$8 systems.} We show correlation error as a function of PEPS bond dimension $D$ for three annealing times. The best-performing PEPS variant, of many evaluated\cite{SM}, was ``plqt-NN+''. Computation time increases sharply with bond dimension $D$.  PEPS easily beats QPU error at $t_a{=}\SI{2}{ns}$, but degrades for slower quenches. At $\SI{20}{ns}$, all PEPS experiments have higher error than QPU, and PEPS appears to be neighborhood-limited: increasing $D$ does not significantly reduce error.}\label{fig:3}
\end{figure}

MPS can produce QPU-quality results---and better---for all annealing times of interest up to $L{=}8$.  
In contrast, we found that PEPS and NQS methods struggle for slower quenches ($t_a{=}\SI{20}{ns}$), even for those relatively small systems. Many classical methods, including PEPS, are especially effective in short-time regimes, where locality can be exploited with relative insensitivity to system size~\cite{Tindall2023}.

We have exhaustively probed PEPS approaches to continuous time-evolution~\cite{SM}, starting with a simple update scheme~\cite{Orus2014}, where only the immediate neighborhood is accounted for during the application of each gate of Trotterized evolution---a strategy that proved successful for the kicked Ising model~\cite{Tindall2023,beguvsic2023fast,liao2023simulation} despite its simplicity. We then systematically enlarged the size of the neighborhood---with quickly increasing numerical cost and heuristic compromises---to a scale approaching the entire lattice for an $8{\times}8$ example. 

Fig.~\ref{fig:3} highlights the results of the best-performing approach (``plaquette-nearest-neighbor+'' [plqt-NN+], see SM~\cite{SM}), showing errors generated during the PEPS evolution. All procedures perform well for $t_a{=}\SI{2}{ns}$, even with small PEPS bond dimension $D{=}4$. A combination of beyond-nearest-neighbor environment and increasing $D$ allows for systematic improvement at $\SI{7}{ns}$. Finally, neighborhood-limited errors saturate well above the QPU level for $t_a{=}\SI{20}{ns}$. This is consistent with observations from spatial decomposition methods we evaluated~(see SM~Sec.~XI.B~\cite{SM}), which indicate that correlations span the system for $L=8$. In contrast, the largest square-lattice simulations ($L=18$) may be amenable to spatial decomposition methods.

We have also attempted PEPS simulations and spatial decomposition of 3D lattices~\cite{SM}. For the largest 3D simulations considered in the next section, correlations are system spanning at $t_a{=}\SI{20}{ns}$, preventing PEPS from exploiting locality. The biclique problems have no such locality to exploit.

While NQS ans\"atze do not rely explicitly on locality, we arrive at qualitatively similar conclusions~\cite{SM}. Accurate results are easily reproduced for fast quenches ($t_a{=}\SI{2}{ns}$). However, modeling dynamics even at $t_a{=}\SI{7}{ns}$ requires numerical effort that grows prohibitively with system size, owing to the increase in required sampling precision at each step of the time evolution.

We conclude that these various PEPS and NQS methods, which we detail in the SM~\cite{SM}, do not match the utility of MPS as a competitive solver for the higher-dimensional topologies, which we now consider for the remainder of the paper.

\section*{Higher-dimensional systems}

\begin{figure}
\includegraphics[scale=1]{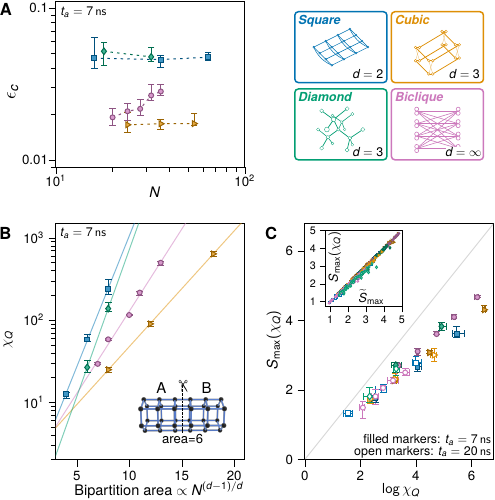}
  \caption{{\bf Entanglement and QPU-equivalent bond dimension.} ({\bf A}) QPU median correlation error for the four topologies as color coded in the right panels. ({\bf B}) For all four problem sets, the bond dimension required by MPS to match QPU simulation quality, $\chi_Q$, exhibits exponential dependence on the bipartition area (asymptotically $N^{(d-1)/d}$, see inset example), consistent with the area-law scaling of entanglement entropy (two-point fit is trivial for diamond at $t_a{=}\SI{7}{ns}$). ({\bf C}) Median $\log \chi_Q$ against maximum entanglement entropy $S_\text{max}(\chi_Q)$ for MPS with bond dimension $\chi_Q$. All data points for the four topologies and two annealing times fall roughly on the same line. Inset: $S_\text{max}(\chi_Q)$ versus ground-truth $\tilde S_\text{max}$ estimates for each input.}\label{fig:4}
\end{figure}

We now apply the approach outlined in the preceding section to evaluate sampling on more complex topologies: cubic, diamond and biclique lattices.  The cubic and biclique lattices are dimerized, with each standard lattice site represented by a pair of spins rather than a single spin~\cite{SM}.

As in square lattices (Fig.~\ref{fig:2}D), QPU error appears nearly constant for 
cubic and diamond lattices, as seen in Fig,~\ref{fig:4}A.  For biclique problems, $\epsilon_c$ increases with system size, which we believe to be a finite-size effect.  Nonetheless, the increase is accounted for in the calculation of $\chi_Q$, which is shown for all topologies in Fig.~\ref{fig:4}B. The $x$-axis is the MPS bipartition area, which scales to leading order as $N^{(d-1)/d}$.  Just as we observed in the square lattice behavior shown in Fig.~\ref{fig:2}E, each of these higher-dimensional topologies exhibits exponential dependence on area, again compatible with area-law scaling.

This scaling of bond dimension is a direct consequence of entanglement in the system: In Fig.~\ref{fig:4}C we plot the maximum entanglement entropy $S_\text{max}$ (maximum throughout the quench and over all MPS bipartitions) against $\log\chi_Q$, and observe a linear relation for multiple topologies, sizes, and quench rates. This underscores the intimate connection between MPS representation complexity and entanglement. The inset demonstrates close agreement between $S_{\text{max}}$ at bond dimension $\chi_Q$ and the ground-truth entanglement $\tilde S_{\text{max}}$.  Together, this tells us two things: First, MPS methods must capture almost all the entanglement in the system to match QPU quality.  Second, the MPS that achieves this is efficient, in the sense that its entanglement is close to the maximum possible value, $\log\chi_Q$, in an MPS of bond dimension $\chi_Q$.  We further note that for all topologies, $\chi_Q$ decreases as $t_a$ increases from $\SI{7}{ns}$ to $\SI{20}{ns}$.  While the bipartite entropy is expected to grow with the correlation length and annealing time, we also expect contributions from local fluctuations and finite system size.

\begin{figure*}
\includegraphics[scale=1.05]{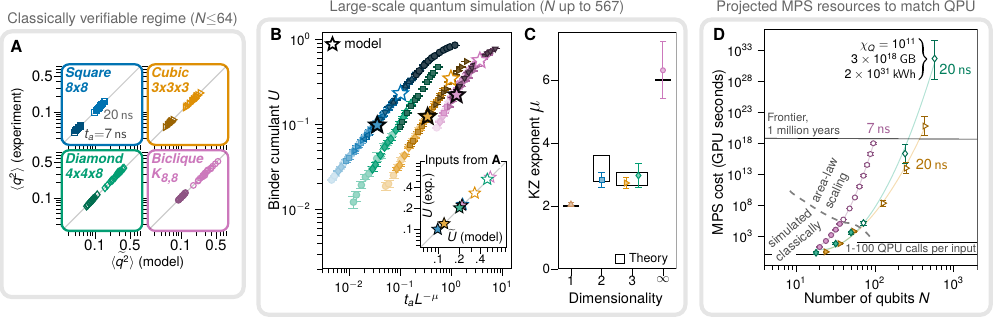}
  \caption{{\bf Dynamical scaling and resource estimates for large-scale quantum simulation.} ({\bf A}) We compare $\langle q^2\rangle$ in QPU output and the ground truth for 20 instances of each topology at $t_a=\SI{7}{ns}$ and $\SI{20}{ns}$, establishing close agreement (median relative error $<1\%$) for the largest MPS ground truths. ({\bf B}) Using a range of system sizes that we cannot simulate classically, we collapse Binder cumulants $U$ using a dynamic finite-size scaling ansatz, yielding best-fit KZ exponents $\mu$.  Curve in collapsed data indicates crossover from power-law (KZ) to system-spanning correlations.  Data points are computed from ensembles of 300 to 20,400 simulated spin-glass realizations (collapsed sizes $N$, indicated by different shades and markers, are 16--324 for square, 54--432 for cubic, 72--567 for diamond, and 32--96 for biclique).  Inset: Estimates of $U$ using simulations shown in {\bf A} for $t_a=\SI{7}{ns}$ and $\SI{20}{ns}$ (dark and light stars respectively, also in main plot) show close agreement with the ground truth at simulable sizes, bridging the gap between the classically-simulable and large-scale regimes.
  ({\bf C}) Extracted KZ exponents for one-, two-, three-, and infinite-dimensional systems agree with previous estimates for the corresponding QPT universality classes.
  ({\bf D}) Area-law scaling of $\chi_Q$ (Fig.~\ref{fig:4}B) allows extrapolation of required classical time based on $N\chi_Q^3$ scaling of MPS methods; all other methods failed at smaller sizes. Filled markers indicate classically-simulated 20-instance ensembles; larger problems are extrapolations.}\label{fig:5}
\end{figure*}

Our claim of beyond-classical computation requires evidence of QPU accuracy both within and beyond the classically simulable regime.  For small systems, in addition to the small distributional distance quantified by $\epsilon_c$ (and $\mathcal{F}$), we provide evidence in Fig.~\ref{fig:5}A with an instance-wise comparison of the spin-glass order parameter $\langle q^2\rangle$ on the largest lattice of each topology for which the MPS ground truth is converged at both $t_a=\SI{7}{ns}$ and $t_a=\SI{20}{ns}$.  The resulting quantities have a median deviation of only $1\%$ between quantum and classical simulation.

Beyond the classically verifiable regime, we show agreement with quantum mechanics up to the largest size using the theory of critical phenomena.  As shown in Fig.~\ref{fig:5}B,  we verify a dynamic finite-size scaling ansatz\cite{Liu2015,King2023} in which the  Binder cumulant
\begin{equation}
U = \frac 12 \left(3-\frac{[\langle q^4\rangle]}{[\langle q^2\rangle]^2} \right),
\end{equation}
collapses as a function of $t_aL^{-\mu}$, where $\mu$ is the universal Kibble-Zurek (KZ) exponent corresponding to the topology's universality class.  Here, $[\cdot]$ indicates an average over 300 to 20,400 disorder realizations.  The inset shows estimates of $U$ for the 20-instance ensembles for which we have a ground truth (Fig.~\ref{fig:5}A), with excellent agreement between experiment and model; inset $\tilde U$ values are marked as stars in the main plot (no ground truth could be computed for the smallest instances in the diamond collapse, $6{\times}6{\times}8$).  The crossover of the collapse functions from power-law to saturation near $U{=}1$ is an indication of system-spanning correlations, preventing PEPS or other approaches from exploiting locality.  QPU estimates of $\mu$ for square, cubic, diamond, and biclique ensembles---extracted from experimental data collapse---are in agreement with previous estimates\cite{Guo1994,King2023,Bernaschi2023} and theory\cite{Read1995}, as shown in Fig.~\ref{fig:5}C (agreement for 1D and ``cubic no-dimer'' lattices is also confirmed\cite{SM}).  These exponents, along with the stars in Fig.~\ref{fig:5}B, show that the QPU captures both universal quantum critical scaling and quantitative details of the non-universal collapse functions.

The area-law scaling observed in Fig.~\ref{fig:4}B provides a foundation for extrapolating $\chi_Q$ to larger sizes, as shown in Fig.~\ref{fig:5}D.  Based on classically simulable systems (Fig.~\ref{fig:4}B), we can estimate the computational resources required to match QPU simulation quality with MPS, using leading scaling forms of $N\chi_Q^2$ for memory and $N\chi_Q^3$ for time (and energy) requirements.  We generously assume perfect parallelization, and extrapolate GPU time required to generate MPS results\cite{SM}, ignoring practical scaling challenges.
Extrapolation (open symbols in Fig.~\ref{fig:5}D) indicates that MPS requires infeasible resources to accurately reproduce the measurements in Fig.~\ref{fig:5}B: On the largest problems, MPS would take millions of years on the Frontier supercomputer per input to match QPU quality.  Memory requirements would exceed its $\SI{700}{PB}$ storage, and electricity requirements would exceed annual global consumption~\cite{SM}.  We emphasize that this scaling analysis applies to MPS---the only method with which we can match QPU quality for all considered quench times---and is not an intrinsic lower bound for all classical methods.

\section*{Summary and conclusions}

We have demonstrated beyond-classical computation in the quantum simulation of nonequilibrium magnetic spin dynamics.  We simulated square, cubic, diamond, and biclique topologies that are relevant to applications in materials science and AI, and are amenable to scaling analysis through the area law of entanglement and universal quantum critical scaling.  More challenging and irregular topologies---for example those corresponding to deep neural networks\cite{Biamonte2017}---can also be simulated with the same QPUs.

Two distinct QPU generations yielded consistent solutions in agreement with quantum theory.  At small scales this is verified by simulating the Schr\"{o}dinger equation. At larger scales, where solving the Schr\"odinger equation is infeasible, QPU observables follow the anticipated universal quantum critical dynamic scaling, with critical exponents in close agreement with independent estimates.

To evaluate classical computation costs, we examined tensor networks and neural networks as the leading frameworks capable of providing approximate solutions to the Schr\"{o}dinger equation, seeking QPU-quality output.  Among all algorithms, we could only extract a reliable scaling from MPS; PEPS and NQS approaches failed even on modestly-sized 2D problems.

For MPS, we demonstrated a stretched-exponential dependence of the required bond dimension $\chi_Q$ on the number of qubits, consistent with area-law entanglement scaling.  This enabled a reliable extrapolation of $\chi_Q$, thus computational resources, to larger systems. Extrapolation beyond the simulable scale to a few hundred qubits, far below the maximum experimental scale, indicates a hypothetical runtime on the Frontier supercomputer surpassing millions of years with infeasible memory and energy requirements.

The QPU has been shown to accurately sample from a variety of random models with correct predictions of microscopic, macroscopic and scaling statistics without {\it a priori} knowledge of the ground truth.  We hope and expect that our results will inspire novel numerical techniques for quantum simulation. However, competing classical methods not only have to match QPU-quality results for all models at the scales where ground truths were obtained, but also should reproduce both universal and non-universal details of quantum critical scaling up to the largest experimentally achievable sizes.  Similar methods can be used to study dissipative dynamics in open quantum systems~\cite{Bando2020}, another problem of interest that is intractable for classical computers.

The nonequilibrium quantum dynamics studied here are responsible for a scaling advantage over classical dynamics in optimization~\cite{King2023}. This impracticability of classical simulation opens the door to quantum advantage in optimization and AI, addressing scientific questions that may otherwise remain unanswered, and applications that may be classically impossible.

\let\oldaddcontentsline\addcontentsline
\renewcommand{\addcontentsline}[3]{}
\bibliography{main}
\let\addcontentsline\oldaddcontentsline

\section*{Data Availability}

Code and data are available in the Zenodo online repository \url{https://doi.org/10.5281/zenodo.14063693}.  MPS and PEPS experiments were performed using the ITensor and \texttt{yastn} libraries respectively\cite{itensor, yastn}.

\section*{Acknowledgments}

We are grateful to the technical and non-technical staff at D-Wave, without whom this work would not have been possible.  We thank \L.~Cincio, W.~Zurek, V.~Martin-Mayor, S.~Boixo, A.~Potter, D.~Huerga for insightful and helpful discussions.  We also thank the ITensor community---especially M.~Fishman and K.~Pierce---for providing flexible and efficient tensor-network methods with support for GPU acceleration.  
Work at UBC was supported by Natural Sciences and Engineering Research Council of Canada (NSERC) Alliance Quantum Program (Grant ALLRP-578555), CIFAR and the Canada First Research Excellence Fund, Quantum Materials and Future Technologies Program.
A.N. acknowledges the support of computational resources from the Advanced Research Computing at the University of British Columbia.
This research used resources of the Oak Ridge Leadership Computing Facility, which is a DOE Office of Science User Facility supported under Contract DE-AC05-00OR22725. 
G.A. acknowledges support from the U.S. Department of Energy, Office of Science, National Quantum Information Science Research Centers, Quantum Science Center.
This research was supported in part by the National Science Centre (NCN), Poland under projects 2019/35/B/ST3/01028 (J.D.) and 2020/38/E/ST3/00150 (M.M.R.). 
R.G.M. was supported by NSERC.  Research at Perimeter Institute is supported in part by the Government of Canada through the Department of Innovation, Science and Economic Development Canada and by the Province of Ontario through the Ministry of Economic Development, Job Creation and Trade.
J.~Carrasquilla acknowledges support of NSERC, Compute Canada, and the Canadian Institute for Advanced Research (CIFAR) AI chair program. Resources used in preparing this research were provided, in part, by the Province of Ontario, the Government of Canada through CIFAR, and companies sponsoring the Vector Institute \url{www.vectorinstitute.ai/#partners}.  A.~Sandvik acknowledges support from the Simons Foundation, Grant No.~511064.

\clearpage
\widetext
\begin{center}
\textbf{\large Supplementary Materials:\\ \mytitle}
\end{center}

\tableofcontents
\setcounter{equation}{0}
\setcounter{figure}{0}
\renewcommand{\figurename}{FIG.}
\renewcommand{\thefigure}{S\arabic{figure}}
\renewcommand{\theequation}{S\arabic{equation}}
\renewcommand{\theHfigure}{S\arabic{figure}}
\renewcommand{\thefootnote}{{\Roman{footnote}}}
\makeatother

\input{sm_quantum_annealing.tex}

\clearpage

\input{sm_testbed.tex}

\clearpage
\input{sm_tensor_based_methods}

\clearpage
\input{sm_quality.tex}

\clearpage
\input{./sm_entanglement.tex}

\clearpage
\input{sm_neural_net_methods}

\clearpage
\input{./sm_classical_compute.tex}

\clearpage
\input{./sm_critical_exponents.tex}

\clearpage
\input{./sm_monte_carlo_methods.tex}

\clearpage
\input{./sm_spatial_decomposition.tex}

\end{document}

%% file: sm_quantum_annealing.tex
\section{Quantum annealing}\label{sec:qa}
\label{app:QA}

\subsection{Quantum annealing processors}

\begin{figure*}
  \includegraphics[width=\textwidth]{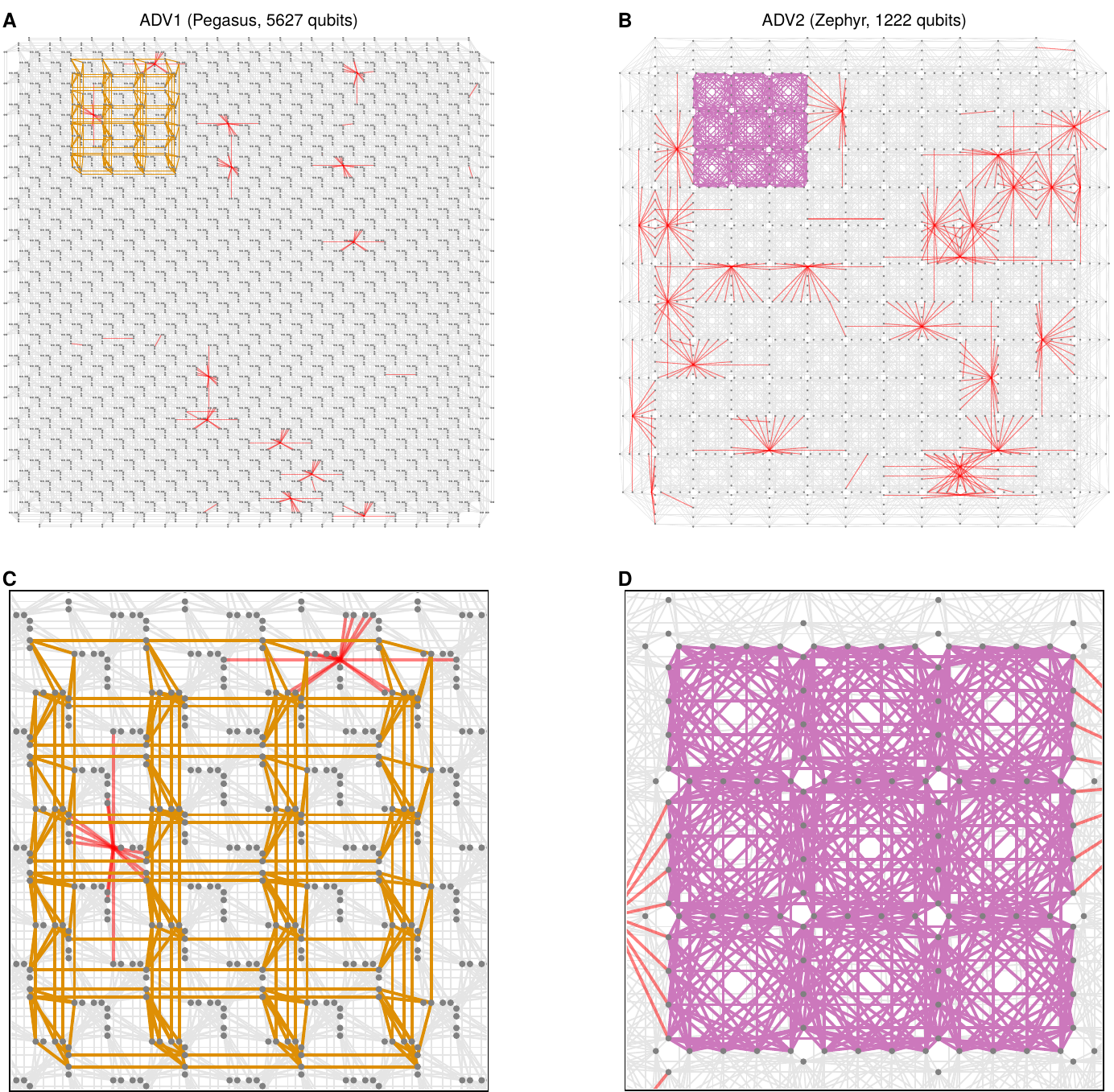}
  \caption{{\bf Qubit connectivity graphs.}  {\bf A}, ADV1 and {\bf B}, ADV2 qubit graphs.  Gray points and lines indicate operable qubits and couplers respectively; red points and lines indicate inoperable qubits and couplers.  Embeddings of example systems are shown (zoomed, {\bf C} and {\bf D}): cubic dimer, $4{\times} 4 {\times} 4$ dimers in orange, biclique $K_{24,24}$ with 48 dimers in purple.  Diagonal couplings are inter-dimer; horizontal and vertical couplings are intra-dimer.}\label{fig:qubitgraph}
\end{figure*}

Experiments were performed on two models of D-Wave quantum annealing processor: a prototype Advantage2\texttrademark\ processor (ADV2) and an Advantage\texttrademark\ processor (ADV1, online for general access as {``\texttt{Advantage\_system4.1}''} during experiments).  Both processors were maintained at a cryostat set point of $\SI{12}{mK}$. ADV1 was used in a recent demonstration of coherent annealing of 3D spin glasses\cite{King2023}.  ADV2 was fabricated using a lower-noise process and a different qubit connectivity family (Zephyr \cite{Boothby2021}) than ADV1 (Pegasus \cite{Boothby2020}).  ADV2 and ADV1 had, respectively, 1222 and 5627 qubits; the qubit connectivity graphs, which dictate the available programmable qubits and couplers, are shown in Fig.~\ref{fig:qubitgraph}.

\begin{figure*}
    \includegraphics[scale=0.8]{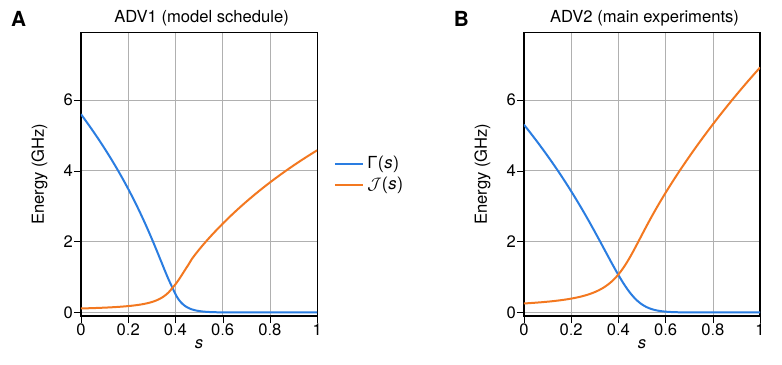}
  \caption{{\bf Annealing schedules for ADV1 and ADV2.}  {\bf A} and {\bf B} show effective TFIM schedules for Advantage and Advantage2 systems, respectively.  ADV1 annealing schedule is used to characterize the modeled quench, but ADV2 is used for main measurements, since the Advantage2 processor has lower noise and higher energy scale.  The annealing parameter $s$ is a normalized time: $s=t/t_a$.}\label{fig:schedules}
\end{figure*}

\subsection{Annealing schedules}
\label{spp:annsched}
The QPUs use radio-frequency superconducting quantum interference device (rf-SQUID) flux qubits to approximately realize ideal spin-1/2 particles in a TFIM\cite{Harris2010a}.  Each QPU has its own annealing schedule consisting of time-dependent transverse ($x$-basis) and longitudinal ($z$-basis) energy scales $\Gamma(s)$ and $\mathcal J(s)$.  In a previous work \cite{King2023} we performed detailed modeling of the effective many-body qubit Hamiltonian and mapped it to an effective TFIM schedule for ADV1, shown in Fig.~\ref{fig:schedules}A and Fig.~\ref{fig:1}.  We use this ADV1 schedule as the model quench schedule, which defines the parameters of the classical methods used.  We use both ADV1 and ADV2 to simulate this quench.

The ADV2 schedule (Fig.~\ref{fig:schedules}b) differs significantly, most importantly by having much higher energy scales than ADV1 in the region where $\Gamma(s)\approx \mathcal J(s)$.  Simulating the ADV1 quench with the ADV2 QPU therefore requires parametric adjustments as described below.

\subsubsection{Annealing time as a function of energy scale}
\label{app:ates}
\begin{figure*}
  \includegraphics{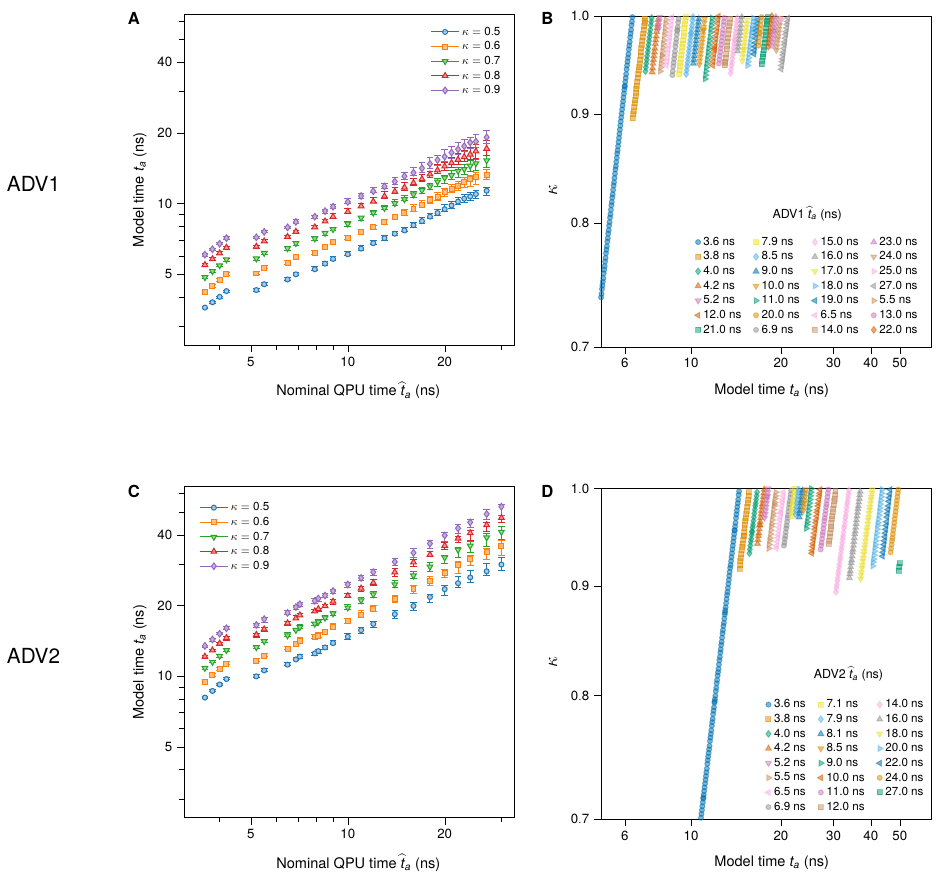}
  \caption{{\bf Calibrating QPU parameters by quench time for ADV1 (top) and ADV2 (bottom) QPUs.} {\bf A}, {\bf C}, effective model quench time $t_a$ is calculated for a range of energy scales $\kappa$ and nominal quench times $\hat t_a$ (legends); $t_a$ is taken as the median effective quench rate (matching $\langle q^2\rangle$) over 
  a $4{\times} 6$ 2D spin glass ensemble.  
    {\bf B}, {\bf D}, for each $t_a$ we choose an appropriate $\hat t_a$ and determine the appropriate $\kappa$ based on a power-law fit in $\kappa$ based on the data shown in the left-hand panels.}\label{fig:calibrate_time}
\end{figure*}

The energy scales in the annealing schedule dictate the time scales relevant to Schr\"odinger evolution. 
QPUs are calibrated to perform an anneal on a requested timescale, $\hat t_a$ (an input to the control electronics), which is well controlled in the generally available operational regime of the device ($\hat t_a>\SI{500}{ns}$). However, deviations from this nominal time are significant for the model quench times $t_a$ studied here, and we make a refined mapping of requested time to model time as discussed in this section. Synchronization of $s$---the 
normalized time $t/t_a$---for all qubits also becomes challenging in the small anneal-time limit, for which correction methods are discussed in Appendix \ref{sec:annealoffsets}.  We remark that any future availability of $t_a<\SI{500}{ns}$ will have properly synchronized annealing lines and translated $t_a$, making this section unnecessary for future users of these QPUs.

We construct a mapping from a given model quench time, $t_a$, to a pair of QPU parameters: a nominal QPU quench time $\hat t_a$ and an energy scale $\kappa$ (a multiplicative prefactor on $\mathcal H_P$ in Eq.~\ref{eq:ham}).  This approach is preferred over the more obvious approach of changing only $\hat t_a$ because of a technicality in this regime: due to the specifics of controlling waveforms at this speed, the effective quench time $t_a$ is not a smooth or monotone function of the nominal (i.e., requested) quench time $\hat t_a$.  It is, however, a smooth and monotone function of $\kappa$.

To calibrate time, we simulate an ensemble of 20 2D spin glasses on $4{\times} 6$ qubits, which is small enough to routinely determine accurate ground-truth correlations for many quench times.  We determine effective quench times $t_a$ for a range of $\hat t_a$ and $\kappa$ pairs by lining up a statistic that increases with $t_a$: the squared spin-glass order parameter $\langle q^2 \rangle$, Eq. \ref{eq:q2}.

We measure $\langle q^2 \rangle$ for $\kappa \in \{0.5,0.6,0.7,0.8,0.9\}$ and a dense selection of nominal times $\hat t_a \in \mathcal T$, along with smaller values of $\kappa$ down to $0.01$ for the fastest $\hat t_a$, $\SI{3.6}{ns}$.  For each $\hat t_a\in \mathcal T$, we perform a polynomial fit between $\log\kappa$ and $\log t_a$.  This gives us, for each $\hat t_a$, an invertible mapping $f_{\hat t_a}$ from $\kappa$ to effective $t_a$.  It only remains to choose, for $t_a$, a canonical nominal QPU $\hat t_a$ to use.  We choose the fastest $\hat t_a$ in $\mathcal T$ such that the appropriate energy scale $\kappa = f_{\hat t_a}^{-1}(t_a) \leq 1$.  The resulting time-calibration mappings are shown in Fig.~\ref{fig:calibrate_time}.

\subsection{Calibration refinement}

As mentioned, the QPUs used in this work are subject to a baseline calibration appropriate to the operational regime $\hat t_a > \SI{500}{ns}$.  Thus we do some additional work to refine the calibration for the relevant experimental regime $\SI{3.6}{ns} \leq \hat t_a \leq \SI{27}{ns}$.  All these methods have been applied effectively in previous works, especially Refs.~\onlinecite{King2022,King2023}, and as detailed in Ref.~\onlinecite{Chern2023}. 

To correct the calibration we use only symmetries of the input models.  For example, due to a lack of symmetry-breaking longitudinal field in the models studied, each qubit should have average magnetization $0$ in all experiments.  Further, the sampling distribution should be independent of the programming batch and the embedding; i.e., placement on the processor. This allows us to refine parameters (flux offsets and anneal offsets) by self-consistent criteria. 

Calibration refinement of effective time $t_a$ and anneal offsets are independent of the programmed spin-glass topology and coupling distribution. Calibrations of flux offsets restore trivial model symmetries. Hence these calibrations do not create any model-dependent computational advantage or overfitting relative to classical simulation techniques where such calibrations also exist by definition.

\subsubsection{Flux offsets}

All inputs studied in this work have no linear biases in the nominal Hamiltonian, and as a consequence $\langle \sigma^z_i \rangle=0$ for well-calibrated experiments. This symmetry is respected in ideal flux qubits, but is broken by flux drift and cross-talks that can be corrected on a per-qubit basis with compensating applied fields (flux offsets). Magnetization is smooth and monotonic in the flux bias for our models and dynamics. We calibrate flux offsets to achieve approximately zero magnetization on every qubit by an iterative method for each experiment (ensemble, $\hat t_a$, $\kappa$)~\cite{Chern2023}.

\subsubsection{Anneal offsets}
\label{sec:annealoffsets}

The QA Hamiltonian is controlled using several annealing lines (eight in ADV1, six in ADV2) that must be well synchronized. Calibrated synchronization has a typical tolerance of $\approx \SI{1}{ns}$, which is suited to anneals longer than $\SI{500}{ns}$ but too large to ignore in our experimental regime.  We therefore offset the qubits on each line, either advancing or delaying them depending on observed spin-spin correlations in a 512-qubit chain, as 
in Ref.~\onlinecite{King2022}. We determine these per-line anneal offsets for each $\hat t_a$ and $\kappa$ corresponding to the studied model quench times $t_a$, and use them throughout all experiments across all problem ensembles.

\subsubsection{Embedding and automorphism averaging}

When possible, we embed multiple copies (up to 200) of each problem into the QPU in parallel.  Typically around half of the qubits are used, in total.  For example, in ADV1 and ADV2 we use 87 and 23 parallel embeddings, respectively, of $6{\times} 6$ square lattices.

The unweighted graphs corresponding to the lattices studied in this work have rich automorphism groups.  On each programming of the QPU, we apply a random automorphism to the graph, which changes the mapping of model spins to physical qubits.  This allows averaging out of calibration imperfections even when only one embedding is available.

\subsection{Embedding in qubit graph}

Any problem must be mapped to the available coupling topology of the quantum annealing processor.  The problem of mapping an input to the qubit graph is an instance of subgraph isomorphism, rather than minor embedding.  We use the powerful Glasgow Subgraph Solver~\cite{McCreesh2020} for this mapping.  Examples of embedding with Glasgow Subgraph Solver, with code, can be found in Ref.~\onlinecite{Chern2023}.

%% file: sm_testbed.tex
\section{Experimental testbed}
\label{app:testbed}

\subsection{Definition of inputs}

The Ising spin glasses studied are defined for 2D, 3D and high-dimensional graphs. These can be subdivided into models with dimers  (biclique, cubic dimer), and those without. In the dimerized cases a subset of edges defined by a perfect matching are strongly ferromagnetically coupled ($J=-2$). To all non-dimer edges we assign uniformly and identically distributed random couplings $J_{ij}$, subject to a technicality in the cubic dimer lattice, described below.
The inputs we study are ``low precision'' and ``high precision'' with respect to the random $J$ values.

Low-precision inputs have random couplings drawn from $\{-1,1\}$, while high-precision instances have couplings drawn from $\{\tfrac{-128}{128},\ldots,\tfrac{-1}{128},\tfrac{1}{128},\ldots\tfrac{128}{128}\}$.  In the special case of bicliques, these couplings are further subject to a size-dependent rescaling, described below.  All edges in the graph are assigned couplings from the same distribution.  

Note that low- and high-precision ensembles differ not only in precision, but also effective energy scale---a lower energy scale alone (in the high-precision problem) can account in part for a delayed rate of correlation-length growth, but is not a sufficient explanation for the variation between the two ensembles.

Here we define five input topologies: Square, cubic dimer, cubic no-dimer, diamond, and biclique. All models, dimerized or not, are subgraphs of the QPU architecture and embed as a one-to-one mapping of lattice sites to qubits.  
Cubic no-dimer, as well as all low precision problems, are omitted from the main text for brevity.

\subsubsection{Square}

\begin{figure}
\includegraphics[scale=.8]{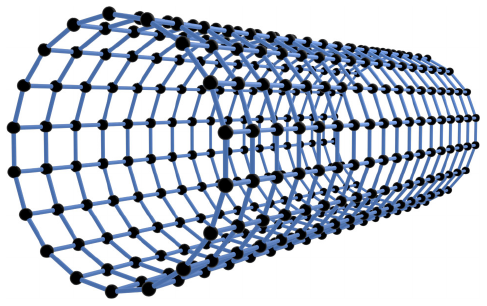}
\caption{{\bf Square lattice.} Largest cylinder embedded on ADV2, $18 {\times} 18$, shown.}\label{fig:2dtestbed}
\end{figure}

The 2D lattices we study are $L_x {\times} L_y$, periodic in the $x$-dimension (Fig.~\ref{fig:2dtestbed}). There are no dimers.

\subsubsection{Cubic dimer}

\begin{figure}
\includegraphics[scale=.8]{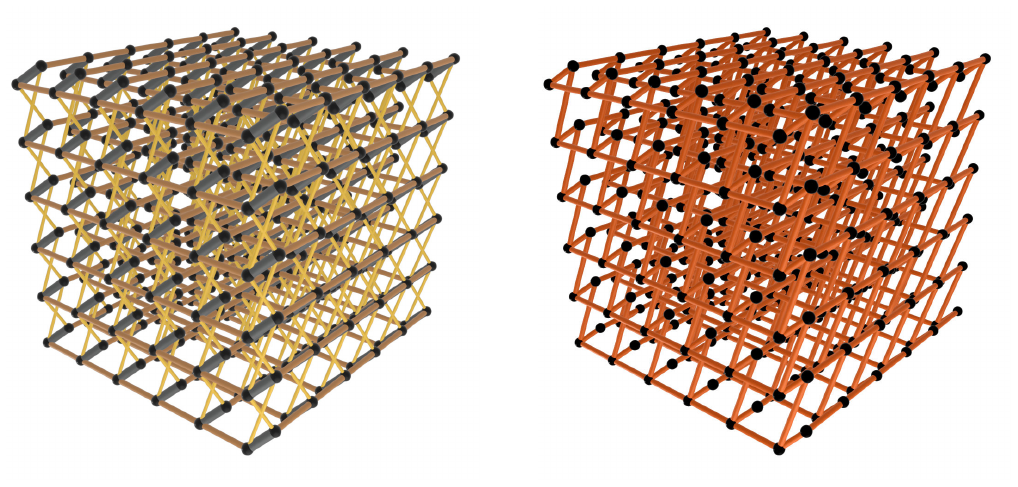}
\caption{{\bf Cubic dimer and no-dimer lattices.} Left, cubic dimer, and right, cubic no-dimer lattices with $(L_x,L_y,L_z)$ = $(6,6,6)$ on 424 qubits.  In cubic dimer problems, thick gray lines indicate dimer couplings ($J=-2$), dark lines indicate $x$- and $y$-dimension couplings, and light yellow lines indicate $z$-dimension couplings, in which a coupling term is distributed equally between two couplers joining two dimers.  In cubic no-dimer inputs (right), all couplers are drawn from the same distribution, and half of the $z$-couplings from the cubic dimer connectivity are discarded.  The disordered ``no-dimer'' couplings prevent the spins from being paired into approximately two-level objects.  Periodic couplings between top and bottom not shown.}\label{fig:3dtestbed}
\end{figure}

Cubic dimer inputs (simply called ``cubic'' in the main text), which were studied extensively in previous work~\cite{King2023}, are constructed starting with a spin glass on a simple cubic lattice of size $L_x {\times} L_y {\times} L_z$.  Each cubic lattice site, with coordinates $x,y,z$, is expanded into a dimer of two spins $s_{x,y,z}$ and $s'_{x,y,z}$.  Couplings in the $x$-dimension are placed incident to $s_{x,y,z}$, and couplings in the $y$-dimension are placed incident to $s'_{x,y,z}$.  In the $z$-dimension, the coupling is divided equally across two couplers; for example, if the cubic lattice spin glass has a coupling of $-1$ between sites $x,y,z$ and $x,y,z+1$, the cubic dimer spin glass will accordingly have a coupling of $-1/2$ between $s_{x,y,z}$ and $s'_{x,y,z+1}$ and a coupling of $-1/2$ between $s'_{x,y,z}$ and $s_{x,y,z+1}$.

All cubic dimer inputs have open $x$- and $y$-boundaries, and are periodic in the $z$ dimension if and only if $L_z>2$.  The $6 {\times} 6 {\times} 6$-dimer system is shown (without periodic $z$-couplings) in Fig.~\ref{fig:3dtestbed} (left).

\subsubsection{Cubic no-dimer}

Cubic no-dimer inputs are derived from cubic dimer inputs by making two changes:
\begin{enumerate}
    \item one coupling is omitted in the $z$-dimension: the one between $s'_{x,y,z}$ and $s_{x,y,z+1}$.
    \item the coupling between $s_{x,y,z}$ and $s'_{x,y,z}$ is no longer a dimer.
\end{enumerate}
The $6 {\times} 6 {\times} 6$-dimer system is shown (without periodic $z$-couplings) in Fig.~\ref{fig:3dtestbed} (right).

\subsubsection{Diamond}

\begin{figure}
\includegraphics[scale=.8]{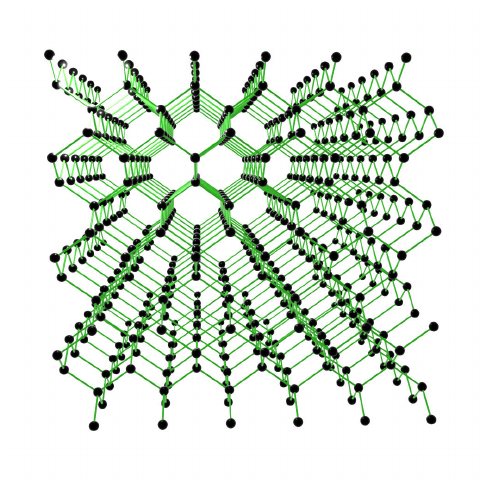}
\caption{{\bf Diamond lattice.} The largest input studied in this work is a diamond lattice with dimensional parameters $(12,12,16)$, in which 567 of 576 qubits (98.4\%) are operable.  Inoperable qubits not shown.  Periodic couplings between top and bottom not shown.}\label{fig:diamond}
\end{figure}

The diamond lattice, like the square lattice, is 4-regular. We parameterize these inputs with dimensions $L_x$, $L_y$, and $L_z$, which do not correspond exactly to Euclidean coordinates of diamond lattice sites.  Rather, assuming an edge length of 1 and in the limit of large systems, the Euclidean dimensions of an input of size $L_x{\times}L_y{\times}L_z$ are $\frac{L_x}{\sqrt{3/2}}$, $\frac{L_y}{\sqrt{3/2}}$, $\frac{L_z}{\sqrt{3}}$, and such an instance has only $N=L_xL_yL_z/4$ spins, not $L_xL_yL_z$ spins.  For the purposes of finite-size scaling, we take $L$ to be $N^{1/3}$.  The inputs studied here are open in the $x$- and $y$-dimensions and periodic in the $z$-dimension; this requires $L_z$ to be divisible by 4.  The maximum values of $L_z$ for inputs embeddable in ADV1 and ADV2 are 8 and 16, respectively.  The largest system we study has $(L_x,L_y,L_z) = (12,12,16)$ (Fig.~\ref{fig:diamond}), embedded in ADV2 with 9 defects (site vacancies) out of 576.
The largest diamond lattice is the only case for which we permit defects; due to their small number these are expected to have no significant impact on algorithmic complexity of simulation.

\subsubsection{Biclique}

\begin{figure}
\includegraphics[scale=.8]{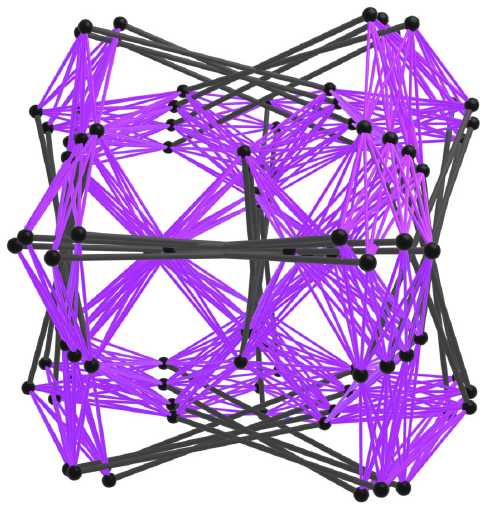}
\caption{{\bf Biclique lattice.} Shown is a $K_{24,24}$ topology on $96$ spins, where gray lines indicate $J=-2$ coupled dimers, and purple lines indicate inter-dimer couplings, forming a complete bipartite structure.}\label{fig:biclique}
\end{figure}

Biclique inputs in this work are based on complete bipartite graphs $K_{M,M}$ embedded using two-spin dimers for each biclique site.  We may call the bipartition of dimers $A$ and $B$, and the dimers in this bipartition $A_1$, $\ldots$, $A_M$ and $B_1$, $\ldots$, $B_M$.  Between any $A_i$ and $B_j$ there exists exactly one coupler.  Since the number of random couplings scales as $M^2$ it is necessary to rescale the coupling strength to ensure extensive energy (correct scaling behavior)~\cite{Venturelli2015}; we rescale couplers as 
$J_{ij}\rightarrow \frac{J_{ij}}{\sqrt{M/4}}$ for every $i$ in $A$ and $j$ in $B$.  The largest input studied is $K_{24,24}$ (Fig.~\ref{fig:biclique}); an embedding is shown in Fig.~\ref{fig:qubitgraph}.

%% file: sm_tensor_based_methods.tex


\section{Tensor networks}
\label{app:TN}
\begin{figure*}[htb!]
\includegraphics[scale=0.8]{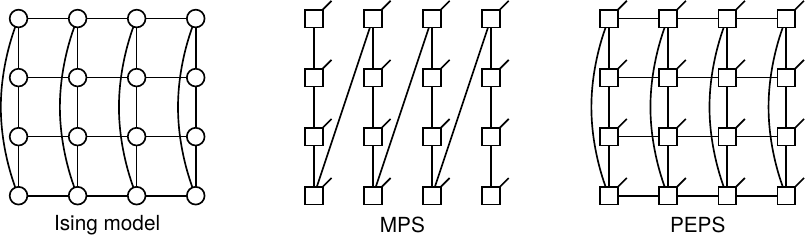}
  \caption{{\bf MPS and PEPS for a small cylindrical TFIM.} A $4 {\times} 4$ cylindrical transverse-field Ising model may be represented diagrammatically with spins or qubits depicted by circles and couplers depicted by lines (left). To represent a quantum state associated with such a lattice we may use a tensor network (MPS or PEPS). The above diagrams depict contracted networks for purposes of amplitude calculation, as in Eq.~\eqref{eq:mps}. Tensors are represented by squares, lines adjoining two tensors imply a contraction, and free lines depict unsummed indices (in this example, the qubit states). In an MPS (center), the tensors form a path with three indices (lines) per tensor (except on the first and last tensor where there are two), one index or line being of dimension 2 in the depicted case (matching the state-space size of a qubit). In our canonical ordering on a square lattice the tensor contraction circles around the lattice in the periodic dimension, minimizing pathwidth on the target TFIM. Local entanglement is captured indirectly through paths with longer-range connections, but contraction is efficient. In a PEPS (right), one dimension (bond) can be assigned for each non-zero coupling in the TFIM, capturing local full entanglement, but exact contraction is not efficient. In practice, tensor contractions for purposes of time evolution and estimation can be performed iteratively over local subsets, or effective approximate tensor environments, avoiding creation of large objects (many unpaired dimensions).}
  \label{fig:TN}
\end{figure*}

Tensor networks are considered the state of the art for simulation of quantum states and dynamics in low-dimensional quantum systems~\cite{Verstraete2008, 10.1016/j.aop.2010.09.012, Stoudenmire2012,  Orus2014, Ran2020, Scholl2021, Schmitt2022}. Several forms of tensor network state exist, suited to different applications. Sections~\ref{app:MPS} and~\ref{app:PEPS} respectively describe matrix product states (MPS)---and certain generalizations---and projected entangled pair states (PEPS), and discuss details of our implementations and their applicability to our problems.

The area law of entanglement and Kibble-Zurek scaling of the correlation length limit the performance of tensor network methods on lattices. In MPS, the area law of entanglement manifests as exponential scaling of the bond dimension $\chi \propto \exp\left({N^{(d-1)/d}}\right)$, where $N$ is the number of qubits and $d$ is the dimension. Memory, time, and energy requirements scale polynomially in this parameter. In PEPS, certain types of area law are simulable with finite bond dimension, but the numerical requirements can grow prohibitively with an increasing range of correlations.

\subsection{Matrix Product States (MPS)}
\label{app:MPS}

MPS or ``tensor train'' networks are ans\"atze for efficient factorization of high-dimensional tensors, including quantum states of many-qubit systems\cite{fannes1992finitely,White92,White93,10.1016/j.aop.2010.09.012}.

For an $N$-qubit system, an MPS may be defined in terms of $N$ tensors of the form $T^{i}_{l_i,z_i,l_{i+1}}$. The indices $\{l_i\}$ label virtual indices (bonds), and $z_i$ physical ones, with dimensions dim($l_i$)$\leq \chi\; \forall i$ and $\text{dim}(z_i)=p$. In general $p$ scales exponentially in the number of qubits associated with the tensor, which in all our experiments is constant; in the standard representation with one tensor per qubit, $p=2$. For a wave function $\ket{\psi} = \sum_z a_z \ket{z}$, amplitudes in the computational basis can be represented as
\begin{equation}
    a_{z_1, \ldots, z_N} = \sum_{\{l\}}\prod_{i=1}^N T^{\pi(i)}_{l_{p(i)}, z_i, l_{\pi(i)+1}} \label{eq:mps}
\end{equation}
where bond indices $\{l_*\}$ are summed over the corresponding bond dimension\footnote{States near the boundary can be represented with reduced bond dimension, for example $\chi_1=\chi_{N}=1$ without loss of generality.} and $\pi(i)$ is the ordering of the qubits in the MPS. A graphical representation of the MPS contraction, Eq.~\eqref{eq:mps}, is shown in Fig.~\ref{fig:TN}. It is important to point out the efficiency of the MPS representation: while the space of amplitudes is of size $2^N$, the space of tensors is of size $O(N p \chi^2)$, controlled by the maximum bond-dimension parameter $\chi$.

The quality of an MPS (ordering) can be understood in terms of entanglement entropy.
Consider a system divided into two subsystems, $A$ and $B$, at some bond. If $\rho_A$ and $\rho_B$ denote the reduced density matrices for these subsystems, the entanglement entropy is defined as 
\begin{equation}
    S_A = - {\rm Tr}_A (\rho_A \log \rho_A) 
\end{equation}
The reduced density matrix $\rho_A$ possesses $2^{N_A}$ diagonal elements, where $N_A$ is the number of qubits in subsystem A. The maximum possible entropy corresponds to the extreme case of a uniform probability distribution: $\max (S_A) = N_A \log 2$. This means it scales with volume of the subsystem. For a random state in the Hilbert space this scaling is also expected for any subsystem and is sometimes referred to as the volume law of entanglement entropy. However, in systems with local connectivity and a gapped ground state, the entanglement entropy of the ground state exhibits a linear scaling with the number of qubits at the boundary between the two subsystems. This phenomenon is commonly referred to as the area law of entanglement~\cite{Eisert2010}: $S_A \propto {\rm bipartition \ area}$. As discussed in Section~\ref{app:entanglement}, and in agreement with previous observations~\cite{Canovi2014}, we show that quenched quantum dynamics deviating from the ground state also show area-law scaling.

In open-boundary one-dimensional systems, optimal MPS ordering (for purposes of minimizing bond dimension) is typically aligned with the 1D topology. Such MPS states naturally encode the area law of entanglement~\cite{hastings2007}: ground states of gapped Hamiltonians of $N$ qubits can be approximated to arbitrarily high precision with an MPS of bond dimension $\chi$, which does not grow with the size of the system. Therefore, the entanglement entropy of a block of $L$ qubits constructed out of an MPS state, $S(L)=-\textrm{Tr}[\rho_L \log(\rho_L)]=O(\log \chi)$, with $\rho_L$ being the density matrix of the block, is bounded by a constant $S(L)\propto~$const for $L\gg 1$.

Choosing an optimal MPS ordering for higher-dimensional systems is ambiguous. In spite of this, geometrically intuitive orderings can be defined that are optimal from a scaling perspective, as discussed in Section~\ref{app:MPSordering}. 
For these orderings, area-law compliant states of $d$-dimensional systems have $S(N)\propto N^{(d-1)/d}$, requiring the MPS bond-dimension $\chi \sim \exp(S(N))$ to grow exponentially in the area of the lattice. Our experiments present evidence that coherently-annealed quantum states obey an area law of entanglement scaling. Hence, although the MPS representation is more concise than the full representation, it cannot be reduced further without incurring systematic errors.

MPS representations can also be modified by representing multiple physical qubits per tensor. Qubits composing dimers in our models exhibit stronger correlations than other qubit pairs, and it can be efficient to represent them in a single tensor---in this case the prescription Eq.~\eqref{eq:mps} applies but with $p=2^2$, reflecting the expanded state space. By extension, we might collapse a $d$-dimensional system to a $d-1$-dimensional system by collecting together $O(N^{1/d})$ qubits into a single state. Since $p$ grows exponentially in the number of qubits collected together, such a method cannot be used to bypass area-law scaling constraints, and typically leads to scaling inefficiencies.  These collection methods may lead to some modest practical speedups beyond the benefit of simply processing dimer pairs consecutively in the tensor train. Aside from dimers, a further example of multi-qubit tensor efficiency is examined in Section~\ref{app:PEPS2D} for PEPS.

\subsubsection{Initial state preparation}
\label{app:isp}
If the dynamics were initialized with the ground state of only the driver Hamiltonian $\mathcal H_D$, this would trivially be a product state with $\chi=1$.
Time-evolving an MPS with $\chi=1$ with the time-dependent variational principle (TDVP, see next subsection for more details) is ill-defined, as many of the long-range interactions are lost during the projection step and the resulting time evolution would be wrong~\cite{Haegeman2011,Haegeman2016,paeckel2019}. To avoid this problem, we run a ground-state Density Matrix Renormalization Group (DMRG) algorithm on the full system Hamiltonian at $s=0$ to estimate the initial state~\cite{itensor}. 
This is performed by setting the truncation error threshold to zero and keeping the state at a bond dimension $\chi$ larger than one.

For initialization and dynamics of MPS using the TDVP method we rely primarily on the {\tt ITensor}\cite{itensor} library. We also tested the dynamics using the local-Krylov approach for time-evolution as implemented in the {\tt DMRG++}\cite{dmrg++} library, but found the TDVP superior in terms of running time and precision for larger time steps. For the GPU implementation of the TDVP method, we have mainly used the {\tt ITensor} library and adapted the GPU backend to both Nvidia and AMD GPUs as used in the Summit and Frontier clusters, respectively. Our primary platform for PEPS experiments is the {\tt yastn} library~\cite{yastn}.

\subsubsection{Time-dependent variational principle (TDVP)}

For MPS time-evolution we employed a standard time-dependent variational principle algorithm with a two-site update, with second-order Trotterization error~\cite{itensor, dmrg++, Haegeman2016}. Unless stated otherwise, our presented results rely upon a time-step of $dt=\SI{0.01}{ns}$, and a discarded-weight threshold for the singular value decomposition (SVD) of $10^{-12}$. We have checked that this value of the time-step gives converged ground-truth results for all the different topologies studied. 

An important comment is in order about our use of the two-site versus one-site update for the TDVP algorithm. Refs.~\cite{Yang2020,LiDelft2022} proposed a scheme to dynamically enlarge the bond dimension within a one-site update, which is shown to perform better than a standard two-site TDVP algorithm. We expect a factor of 2-3 speedup when this refinement is implemented. 
Nevertheless, it is clear that these improvements won't change the exponential scaling of computing costs, which will eventually become classically prohibitive for the system sizes currently achievable with the QPU (see also Section~\ref{app:classicalcompplatform}). 

Continuing the discussion about details of our implementation of the TDVP method: for the solution of the forward and backward time-dependent Schr\"{o}dinger equations for the local MPS and center matrix tensors, we keep up to 100 Krylov vectors, allowing a Krylov truncation error of $10^{-12}$. Computational time also scales in proportion to $n_t = t_a/dt$, which is for practical purposes significant, since we consider evolutions on time scales of order $\SI{10}{ns}$; values of $dt$ larger than $\SI{0.01}{ns}$ quickly lead to large systematic errors in many simulations. The truncation threshold has little impact on the statistics or running time in the small-value limit (small $\chi$ and large $N$) in which we operate; increasing the truncation threshold can allow efficiencies through the early-time evolution of states where tensors need not saturate the bond-dimension bound, but this phase of simulation is not a computational bottleneck. We found that more careful optimization of these quantities did not impact our conclusions qualitatively---in other words, their impact is insignificant compared to the choice of $\chi$.

The dominant contribution to computing time is the choice of maximum bond dimension $\chi$, which is applied uniformly to all bonds and all times, and is saturated for a large finite fraction of tensors in space and time, during our simulation. Memory requirements scale as $O(N \chi^2)$, and time requirements scale as $O(N n_t \chi^3)$, owing to SVD and Krylov iterations that must be applied to all bonds and at every time step of the algorithm. 

MPS errors are systematically controlled by bond dimension $\chi$ and dynamical discretization $dt$. Our statistics converge smoothly at large $\chi$ and small enough $dt$, and it is possible to infer a ground truth. By reducing $\chi$, it is possible to obtain a computationally less expensive, but less accurate distribution. 

\subsubsection{Classical computational platform limitations}
\label{app:classicalcompplatform}

Computational platforms used for MPS simulations include CPU, GPU and tensor processing unit (TPU)~\cite{DMRGtpu} implementations.
Our walltimes and energy consumption estimates of Figures \ref{fig:4}C--\ref{fig:4}D of the main text are based on an implementation of TDVP algorithm on a single GPU node with a two-site update, which by definition necessitates an SVD of joined consecutive MPS tensor pairs to implement the MPS tensor truncations. 

We performed this SVD using the divide and conquer algorithm, which is notoriously difficult to implement efficiently on GPU architectures. 
In this regard, Li et al.~\cite{Unfried2023} recently proposed a QR decomposition to eliminate the SVD step of the time-evolving block decimation (TEBD) algorithm (not for the TDVP method), demonstrating substantial speedups on single GPU architectures.
Furthermore, a real-space parallelized version of the TDVP algorithm has been recently proposed in Ref.~\onlinecite{Secular2020}, which might offer further performance speedups.
We believe that combining all these approaches together with a multi-GPU or TPU~\cite{DMRGtpu} implementation of the two-site (or better one-site update with bond controlled expansion) TDVP scheme will give the possibility of benchmarking the QPU against even larger system sizes both in 3D and biclique lattices, and that our time and energy estimates could likely be reduced by one order of magnitude. Nevertheless, even including this renormalization factor, we stress that current TPUs are optimized to fit at most a $\chi=2^{16}=65536$ bond dimension~\cite{DMRGtpu}, while for a 3D or biclique lattice our work estimates that a bond dimension of $\chi_{Q}\simeq 10^7$ is required to match the QPU error.  For diamond lattices, the estimated requirement grows to $\chi_Q\simeq 10^{11}$.

We conclude that quantum simulations of coherent anneal quenches with sufficiently large annealing time $t_a$ on systems with several hundreds or thousands of qubits are prohibitive for current MPS-TDVP approaches on existing GPU or TPU architectures.

\subsubsection{MPS ordering}
\label{app:MPSordering}
The efficiency of an MPS representation, and its manipulation in TDVP, depends critically upon tensor ordering.  For one-dimensional systems with open boundaries, the optimal path matches the topology.
In higher dimensions, it is generally hard to find the optimal MPS ordering to represent a generic quantum state. In this work, we resort instead to geometric intuition, and have tested this intuition against some heuristic methods. Standard orderings that have proven successful for 2D align with our heuristic method~\cite{Stoudenmire2012, Scholl2021, DaLiao2021, Schmitt2022}.

Ordering in an MPS becomes more efficient when strongly-correlated sites are close in ordering. Directly coupled qubits are more entangled (on average), as are qubits coupled by short coupling paths (short geometric distance), in line with the intuition that there will be a path length beyond which one might assume weak correlation (a correlation length).

We can formalize this intuition with a heuristic objective function to be minimized, a quadratic assignment problem with a cost 
\begin{equation}
    Q(p) = \sum_{i,j} F_{ij} R(\pi(i),\pi(j))\;,
\end{equation}
where $R$ is a distance between qubits under a particular ordering choice, and $F$ is a flow (`correlation strength') between those qubits.

Minimizing the cost under various intuitive definitions of flow and distance remains hard, but an approximate solution can be efficiently obtained and add value relative to an uninformed MPS ordering. This problem can be mapped to an unconstrained quadratic model, and naturally, we find this is amenable to heuristic solution by the QPU through a Leap\texttrademark~hybrid solver as one option. 

A special intuitive case of the measure takes $R(x,y)=||x-y||_\infty$ (maximum distance between tensors) and $F_{ij} = 1 - \delta(J_{ij},0)$ (a penalty for coupled qubits, no penalty for uncoupled qubits). I.e., we minimize how far apart coupled qubits are in the tensor train. The objective is then equivalent to the graph-theoretic problem of pathwidth minimization, which is also relevant to tensor decomposition in other contexts\cite{10.4230/LIPIcs.TQC.2019.10}.

For results presented in this paper we have used orderings that aim to minimize pathwidth\footnote{A formal proof of minimal pathwidth is challenging in some topologies; we settle for simple scalable patterns that are provably optimal to within $O(1)$ corrections (as $N$ becomes large), in some cases provably optimal.}. The MPS ordering used for the square lattice covers first the periodic dimension and then the open dimension, creating a spiral path along the cylinder (an example is shown in Fig.~\ref{fig:TN}). Similar ``one-dimension at a time'' patterns apply to the diamond and cubic lattice cases, proceeding first over dimer pairs where relevant. For bicliques we proceed first over half of one partition, then the entire other partition, then the remaining half of the first partition.  A feature of these orderings is that the partitions define lattice interfaces that scale in worst case as $N^{(d-1)/d}$. I.e., bonds define partitions, and these partitions typically define $d{-}1$-dimensional planes with respect to the lattice topology. As such, we do the best job possible in minimizing the entanglement entropy that a bond must support (in a scaling sense), since we cannot do better than $d{-}1$-dimensional partitions (but can do worse).

It is expected that we might obtain efficiencies by working harder to determine an efficient MPS ordering. To test this we have used several forms of $R$ (1,2, and infinity norms) combined with several definitions of $F$ ($1-\delta(J_{ij})$, $|J_{ij}|^2$, and mutual information based on simulated states) incorporating disorder-realization specific information. The computational savings made by such optimizations were not reliable, and did not show clear evidence of scaling advantage. We also investigated the use of a Hilbert-curve ordering on the square lattice. This ordering has been used to improve MPS efficiency to solve the quantum Ising model on square lattices, but while the ordering better preserves locality and has a lower average neighbor distance than the pathwidth-minimizing order, it comes at the cost of a maximum $O(N)$ neighbor distance (as opposed to the pathwidth-minimizing case of $N^{1/2}$)~\cite{cataldi2021}. We found the Hilbert-curve ordering used with TDVP produced higher correlation error across all disorder realizations tested, as well as worse scaling. In our final analysis of time we used the simpler pathwidth-minimizing (lattice specific, but disorder-realization independent) MPS orderings. 

\subsubsection{Tree tensor networks and multi-scale entanglement}

Tree tensor networks~(TTN) are a natural generalization of MPS, maintaining qualitatively similar properties in our application~\cite{Shi2006,Murg2010}. Virtual bonds define a tree rather than a path. Canonical forms can be constructed by iterating from the leaves, at a cost scaling as $\chi^{k+1}$, where k is the maximum connectivity of any node ($k\geq 3$ in contrast to $k=2$ for MPS). The space of TTN is a superset of the space of MPS (as defined by orderings), thus we might expect some gain in optimization of the TTN to the problem (lattice and disorder realization) along the lines explored in Sec.~\ref{app:MPSordering}. 

It is possible that the optimal TTN choice is not an MPS and that performance gains are sufficient at $k>2$ to overcome the polynomial disadvantage in $\chi$ scaling (plus any implementation-specific complications). However, given that there is no obvious tree-like substructure to a homogeneous spin-glass lattice problem, there is no obvious choice for a TTN.
As in MPS, bonds in TTN define partitions, and it is not possible to define trees with finite $k$ such that the worst partition scales better than $N^{(d-1)/d}$. Thus given area-law entanglement scaling, a tree tensor network cannot provide a scalable advantage relative to an MPS.  
Empirical investigations fall beyond the scope of this work.

The multi-scale entanglement renormalization ansatz~(MERA) is a means to expand the scale of quantum entanglement captured without exponential growth in the contraction complexity~\cite{Vidal2007}. This is achieved by combining a TTN with disentanglers constrained to an isometric (canonical) form for efficient contraction. This has been shown to effectively capture scalable correlation structures that describe one-dimensional critical states, in line with renormalization-group scale invariance that is present by construction. MERAs have been specialized for use in the analysis of ground states for 1D systems with translational invariance in the large-system limit.  Although a time-dependent MERA algorithm has been proposed and successfully applied to translationally invariant 1D systems\cite{Rizzi2008}, its unfavorable computational scaling $O(\chi^9)$ makes a generalization to our finite, higher dimensional disordered models impractical.

\subsection{Projected entangled pair states (PEPS)}
\label{app:PEPS}

Projected entangled pair states (PEPS), also called tensor product states, are a generalization of matrix product states to arbitrary graphical structure~\cite{Verstraete2004, Nishino_2DvarTN_04, Verstraete2008, Orus2014, Ran2020}. For some states subject to geometrically constrained patterns of entanglement and correlation, in particular a subset of those exhibiting area-law entanglement, PEPS can provide a memory-efficient representation relative to MPS when virtual indices (bonds) are chosen to match the target topology~\cite{Ge2016, Verstraete2006}. Unfortunately the contraction of PEPS in 2D and higher is $\#P$-complete~\cite{Schuch2007}, and in practically interesting cases it relies on heuristic contraction schemes (lattice contraction is described in Sec.~\ref{app:latticestructuredpeps}). Contraction is required for all inference purposes: the calculation of bond environments for purposes of (polynomial-scaling) Trotterized evolution, the evaluation of local expectations, and the calculation of state probabilities. 

Efficient contractions have been achieved for some interesting lattice problems appearing in the context of claims of beyond-classical performance that are practically inaccessible to conventional MPS methods~\cite{Tindall2023, Patra2024}. 
Despite being a \#P-complete problem in worst case, it is understood that structured states, in particular those controlled by a finite correlation length, $\xi$, can allow efficiencies in scaling\footnote{The Kibble-Zurek mechanism anticipates a correlation length $\hat \xi$ growing with quench duration, and as a function of the problem class and energy scale; we have demonstrated various experimental results aligned with this intuition. Here the correlation length might be associated with the Kibble-Zurek mechanism, but other phenomena can contribute to correlation length; the use is more generic.}.
Supposing a weak dependence of some environment (local observables) on the state (tensors) beyond some neighborhood scale ($\xi$), we can propose to contract a neighborhood of volume $O(\xi^d)$ surrounding the environment. The vertex congestion (and tree width) associated with contraction on such a neighborhood scales as $\xi^{d-1}$, and computational complexity scales exponentially in this quantity. In so far as the contraction neighborhood scale is linked to the Kibble-Zurek scaling of correlation length we might estimate scalable bounds on such approximations. Thus we can anticipate the failure of PEPS neighborhood approximations owing to a combination of dimensionality and correlation length.

In line with this intuition we show that PEPS is accurate for lattice evolution on very short time scales, where short correlation lengths apply. 
At longer times and for larger problem energy scales (correlated with connectivity and reduced precision), $\xi$ grows and larger neighborhoods are required. We investigate a variety of practical heuristics and conclude that the systematic errors for PEPS are larger than those of the QPU in our more challenging experiments. 

We present the fundamental problem of PEPS contraction for lattices in Section~\ref{app:latticestructuredpeps}. A $d+1$ dimensional PEPS representation, compatible with circuit models, is examined in Section~\ref{app:dplus1PEPS}. In Sections~\ref{app:PEPSltl} and~\ref{app:PEPSltl2} we examine locally tree-like approximations and an associated simple update (and simple observable) algorithm. In Sections~\ref{app:PEPS2D}-\ref{app:PEPS3D} we consider more complex neighborhood update methods tailored to 2D square and 3D cubic lattices. The interaction of correlation length and neighborhood contraction scale is examined in the well-controlled context of 1D models in Section~\ref{app:PEPS1D}. 

\subsubsection{Contraction of lattice structured PEPS}
\label{app:latticestructuredpeps}
Lattice PEPS are a generalization of MPS in which tensor virtual bonds are in alignment with lattice edges; an example is shown in Figure~\ref{fig:TN} for a square lattice.
In this section we consider primarily the standard form of PEPS with one tensor per qubit (state dimension $p=2$) fully entangled with neighboring tensors, and with a uniform bound on bond dimension ($D$). For the square and cubic lattices, we consider also tensors in alignment with plaquette and cubic states ($p=2^4$, $2^8$ respectively), as shown in Figure~\ref{fig:ntu_env_2d}G. Use of finite spatially-local (plaquette or cubic base) tensors does not change qualitative dimensionality considerations, but can be practically useful as shown later. All tensors are of size $\leq p D^k$, $k$ being the lattice connectivity.

The process of contraction involves taking inner products of tensors (contracting tensor-pairs to single tensors), creating sequences of intermediate tensors. Contractions are required for controlled truncation (compression) of PEPS, and for the calculation of observables. Successive multiplications can be represented by a binary tree in which one bond (adjoining two tensors) is eliminated to create a new tensor at each node proceeding inwards from the leaves to a root. Assuming inner products are the dominant computational operation, the complexity (memory and time) of the contraction is tied to the (bond-dimension-weighted) vertex and edge congestion of the best possible binary tree~\cite{10.4230/LIPIcs.TQC.2019.10}. If bond dimension is homogeneous, the scaling of floating-point operations (and memory) is equivalently understood as being exponential in the treewidth of the line graph~\cite{10.1137/050644756}. Line graphs of $d$-dimensional lattices are also (different) $d$-dimensional lattices, and have treewidth that scales as $N^{(d-1)/d}$. Therefore, even at very modest scales, heuristics are required for contraction. 

A heuristic of primary interest to us involves a neighborhood approximation: rather than contracting over the full volume we might consider a symmetrical neighborhood of volume $O(\xi^d)$, so that contraction only scales as $\xi^{(d-1)/d}$\cite{10.1103/PhysRevB.104.094411}. Another heuristic of interest is boundary-MPS (or boundary-PEPS), where a $d{-}1$-dimensional PEPS (MPS in the case $d=2$), defined at some boundary, is approximately contracted by a sequence of matrix-product operations with truncation~\cite{Orus2014}. Whether for full, neighborhood or boundary-MPS/PEPS contractions, the details of the contraction process can be limited not just by dimensionality, but also the specifics of the lattice. It should be noted that practical (small finite $D$, $\chi_B$ and $\xi$) implementations can be dominated by costs other than the inner products, and many methods require iterative procedures that do not guarantee convergence and have a high propensity toward numerical instability. 

Mirroring our MPS method, our PEPS methods separate the time evolution of a lattice-structured PEPS from the calculation of observables on the time-evolved PEPS. As we are interested in $z$-basis sampling, our more-costly numerical experiments are terminated at $s=0.6$, where the transverse field becomes negligible (see Figs.~\ref{fig:schedules}A and \ref{fig:s_cutoff}). 

Section~\ref{app:dplus1PEPS} discusses an interpretation of our PEPS contraction in terms of a $d+1$ (space plus time) dimensional PEPS problem; from this perspective, the time-evolution element of the contraction can be interpreted as a boundary-PEPS method ($\chi_B=D$), with the boundary proceeding from time $t=0$ to our final state at $t_a$. 
An initial state is efficiently prepared (as discussed in Section~\ref{app:isp}, the ground state of a driver Hamiltonian is already sufficient) and Trotterized time evolution applied. Coupled tensors are updated via action of a rank-2 (or 4) matrix product operator, as illustrated in Figure~\ref{fig:Circuit2ndOrder}. This enlarged tensor pair is then compressed via heuristic methods to prevent exponential growth of the rank (memory requirements). 
Estimation of observables is required following the time evolution. We can again contract on some neighborhood for this purpose. The neighborhood might match that used in time evolution, for example in Section~\ref{app:PEPSltl2} a single-edge neighborhood is used (simple observable matching the simple update). In most places we consider a larger neighborhood---the full lattice, exploiting efficiencies (such as MCMC) for approximate contraction.
Errors can arise either in the dynamics or in the final contraction of the time-evolved state. We find that the former results in more practically significant computational constraints. 

We consider two types of heuristic truncation for time evolution: simple update (SU) and neighborhood tensor updates (NTU) at various scales. 
For final state measurement of observables, we consider boundary-MPS, exploiting where necessary the efficiencies of Markov-chain Monte Carlo sampling methods, as well as contractions limited to small-neighborhood scales.

\begin{table}[t!]
\centering
\begin{tabular}{||c c c||} 
 \hline
 Method & Lattice type and dimensions & Vertex congestion (complexity)\\ [0.5ex] 
 \hline\hline
 Tree-like (SU,SVDU) & single bond environment & $O(p D^{\text{connectivity}+1})$~(Ref.~\onlinecite{PhysRevLett.101.090603})\\
 NTU(NN, NN+) & $2{\times}3$ core, on square lattice; Fig.~\ref{fig:ntu_env_2d}A,B  & $O(D^8 + p D^7)$ (NN in Ref.~\onlinecite{10.1103/PhysRevB.104.094411})\\
 NTU(NNN, NNN+) & $3{\times}4$ core, on square lattice; Fig.~\ref{fig:ntu_env_2d}C,D & $O(p D^{10})$ \\ 
 NTU(3DNN) & 4 nearest loops, on simple cubic lattice; Fig.~\ref{fig:3D_delta_v_t}C & $O(D^{12}+p D^{11})$\\
 NTU(3Dcube) & 3x3x4 core, on simple cubic lattice & $O(p D^{29})$ \\
 Normalization & square &  $O(p D^{2L+6})$  [$O(p \chi_B^2 D^{8} + \chi_B^3 D^6)$] \\
 MCMC & square &  $O(D^{L+4})$ $[O(\chi_B^2 D^{5} + \chi_B^3 D^3)]$\\
 MCMC & cubic &  $O(D^{L^2+L+5})$\\ 
 MCMC & biclique $K_{h,v}$ &  $O(D^{h v/2 + 1})$ (Ref.~\onlinecite{10.1016/j.jctb.2018.03.007})\\ [1ex] 
 \hline
\end{tabular}
\caption{{\bf Vertex congestion (time complexity) scaling as a function of bond dimension $D$ at large $L$}. Scaling is enumerated for: local environment approximations (NTU, SU) for time evolution; full contraction for amplitude estimation (MCMC sampling); and full bra-ket contraction (normalization, exact environment estimation or FU). Cubic and square lattice are assumed periodic in one dimension per main-matter experiments.  
Scaling of boundary-MPS methods for the square lattice are shown, for some choice of heuristic MPS bond dimension $\chi_B$, in square brackets. Prefactors that scale as volume (and time) also apply to a practical usage, just as with MPS. Neighborhood-update costs, beyond simple update, have severe scaling as local loops are incorporated in higher-dimensional lattices~\cite{VlaarCorboz3D}. Tree-like decorations, or one-dimensional NTU-extensions~\cite{10.1103/PhysRevB.90.064425}, on neighborhoods can be added, raising only the prefactor scaling of contractions; the limited power of such decorations is explored in experiments.}
\label{table:PEPS}
\end{table} 

A summary of contraction complexity for several lattices and contraction schemes is provided in Table~\ref{table:PEPS}. Calculation of environments for purposes of heuristic truncation require a bra-ket contraction, and the complexity of these operations is a function of the neighborhood structure. For our neighborhoods, we can determine an optimal ordering, and denote the vertex congestion assuming homogeneous bond dimension.
Whereas many methods for calculation of observables also require bra-ket local (or global) contraction, we 
rely primarily upon Markov-chain Monte Carlo (MCMC) sampling~\cite{10.1103/PhysRevB.104.235141}. An amplitude can be calculated with respect to only a bra (or ket) network by fixing the state legs; taking the absolute square of the amplitude yields an unnormalized probability. This achieves approximately quadratic speedup in contraction as shown in Table~\ref{table:PEPS}.  Provided correlation strengths are not too large, approximated probability ratios can be used as rejection criteria for an MCMC method for $z$-basis sampling, sufficient for estimation of the observables of interest. 

While methods are well developed for square-lattice models, particularly ground-state estimation in translationally invariant models~\cite{Liao2019,Liu2022}, the performance of practical approximations in three and higher dimensions are poorly understood in application to disordered models and diabatic time evolution. Even on the 3D simple cubic lattice, scaling of basic neighborhood bra-ket contractions (beyond locally tree-like approximations) can be as severe as $O(D^{29})$, thereby limiting bond dimension to $D \leq 2$ and hence reducing the support for entanglement\cite{VlaarCorboz3D}.

\subsubsection{Circuit model and {\it d}+1-dimensional PEPS contraction}\label{app:dplus1PEPS}

\begin{figure*}
\includegraphics[scale=0.7]{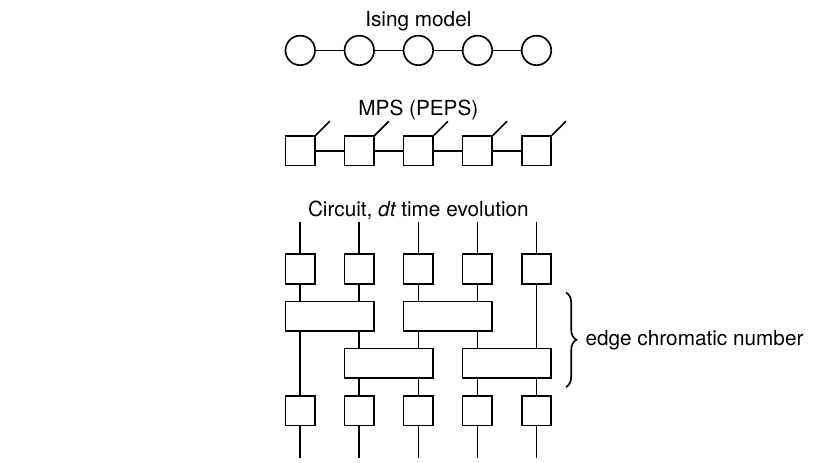}
  \caption{{\bf A PEPS (circuit-model equivalent) representing a 2nd-order Trotterized approximation to time evolution of duration $dt$.} Repeated action of a $dt$-operator on a PEPS enables a discretized simulation of dynamics with error controlled by $dt$. The $dt$ operator can be decomposed into $2+\kappa$ layers of rank-2 unitaries (gates), where $\kappa$ is the edge chromatic number of the lattice being simulated ($\kappa=2=\text{site connectivity}$ for the case of a 1D chain shown). 2-site operators are $\exp(-i 2 \pi dt \mathcal{J}(t/t_a) J_{ij}\sigma^z_i\sigma^z_j)$ and single-site operators are $\exp(-i \pi dt \Gamma(t/t_a) \sigma^x_i)$. The depth of 2-site operators is controlled by $\kappa$, and the total number of 2-gates matches the number of edges in the target lattice. To compress the circuit, 1-site operators between time steps can be combined. With care in the order of operations (to ensure 2nd-order Trotterization error), depth can be reduced to $\kappa$ per unit $dt$ by use of 2-site (rank 4) gates of the form $\exp\left(-i 2 \pi dt [\mathcal{J}(t/t_a) J_{ij}\sigma^z_i\sigma^z_j - \Gamma(t/t_a) (\frac{\sigma^x_i}{k_i} + \frac{\sigma^x_j}{k_j}) ]\right)$, $k_*$ being site connectivity, eliminating the need for 1-site gates.}
  \label{fig:Circuit2ndOrder}
\end{figure*}

We formulate the 2nd-order Trotterized model of dynamics as a circuit model ($d+1$-dimensional PEPS). We show that the scale of such circuits is well beyond the capability of any current circuit (gate)-model computer. Given a circuit model, we might also evaluate the complexity of contraction by circuit-model-inspired PEPS methods that have proved successful in undermining some claims of beyond-classical performance~\cite{PhysRevLett.129.090502, Patra2024, Tindall2023, liao2023simulation, Kechedzhi2024}. The PEPS formulation motivates boundary-PEPS (time-evolved lattice-PEPS) formulations, which we evaluate in detail in other sections. 

Simulation of the time-dependent Schr\"odinger equation is a challenging problem that has been tackled by a variety of methods from the fields of physics and ordinary differential equations~\cite{Ikeda2023minimum, Low2019hamiltonian, arXiv:2402.17667}. Trotterization methods are commonly employed, and allow discretization errors that are controlled polynomially in $dt$, with a number of unitary operations that scales linearly in the number of Hamiltonian terms and simulation time\footnote{State-of-the-art (in scaling) qubitization approximations to time-dependent evolution, based on 2-qubit unitaries (gates), have been proposed that allow for exponential reduction in the error as a function of $dt$, with an overhead relative to Trotterization that does not scale with simulation time $t_a$~\cite{Low2019hamiltonian}. However, no such circuits have been tested in practice. In our view the most practical means to simulate our results, assuming availability of a large digital universal quantum computer, would be a 2nd order Trotterization.}. Trotterized dynamics can be formulated as a PEPS with one tensor per unitary operator (gate); these gates are coupled in a discretized time dimension, so that a lattice-structured PEPS analysis becomes a problem of $d+1$ dimension. The challenging task is to contract to the final time layer and to measure $z$-basis statistics. The PEPS can be interpreted as a circuit model. A high-fidelity circuit-model computer with an amenable structure, programmable gates, and $z$-basis measurement also suffices for dynamical simulation. Our classical methods, based upon the mid-point rule method in PEPS (2nd-order Trotterization) and 2-site 2nd-order TDVP methods in MPS, show that Trotterization with $dt \sim \SI{0.01}{ns}$ is sufficient for accurate simulation. It is therefore interesting to evaluate the circuit-model formulation with a view to either implementation on a hypothetical circuit-model quantum computer, or by PEPS heuristics.

Figure~\ref{fig:Circuit2ndOrder} demonstrates a PEPS formulation of 2nd-order Trotterized dynamics applied to a 1D model. Higher-dimensional models require additional 2-site unitary layers aligned with the edge chromatic number $\kappa$ (in our lattices, matches the lattice connectivity). To perform a single time step requires at least $\kappa$ layers of $|E|$ 2-qubit gates, where $|E|$ is the number of non-zero couplers in the problem Hamiltonian. Repeating the action of these operations $n_t=t_a/dt$ times creates a PEPS sufficient for simulation of dynamics.

Implementation of this PEPS, assuming suitable programmability of 2-gates matched to the target lattice architecture, could require as few as $|E|t_a/dt$ high-fidelity 2-site gates with depth $k t_a/dt$. 
Assuming a measurement at $s=0.5$ and a step size $dt=\SI{0.1}{ns}$ are sufficient, as shown in the context of second-order TDVP dynamics (Figs.~\ref{fig:s_cutoff} and \ref{fig:timestep})\footnote{Limited experiments indicate that 2nd-order Trotterization (Fig. \ref{fig:Circuit2ndOrder}, PEPS dynamics) may be more sensitive to time discretization than 2nd-order TDVP (MPS dynamics), requiring smaller $dt$; $dt \sim \SI{0.01}{ns}$ appears to provide sufficient accuracy in both cases.}, this would modify our lower-bounds on depth/gates by a factor 20 compared to our analysis with $dt=\SI{0.01}{ns}$ evolution to $s=1$; it does not change the conclusions.  

In the context of simulating to the end of the anneal ($s=1$) with $dt=\SI{0.01}{ns}$, per our work, lower bounds on gate-model resources can be considered, assuming no constraints on coherence, gate tuning or processor connectivity.
For our smallest experimental results---a $4 {\times} 4$ cylindrical lattice (28 couplers, edge chromatic number $4$) at $\SI{7}{ns}$---this would require depth of 2800 and 39,200 2-gates. Bounding the resources for our more challenging experiments, the Advantage processor used in this study has an edge-chromatic number of $15$, and $40279$ programmable couplers, so that a 2nd-order Trotterized circuit simulation at $\SI{20}{ns}$ requires depth 30,000 and 80,558,000 2-gates. Recent state-of-the-art simulations with time-independent coarsely-grained Trotterization on $N=127$ qubits ($|E|<1.25N$, mean connectivity $\lesssim 2.5$) heavy-hex lattices have depth $60$ and 2880 CNOT gates by contrast. Simulation by state-of-the-art circuit-model quantum computers is limited by gate type, scale, topology and fidelity. Since we see many orders of magnitude separation between the scale of digital simulation required for QPU emulation and published state-of-the-art performance, we can conclude that near-term circuit-model quantum computers are impractical for this purpose. Similar conclusions have been drawn in more detailed analyses of related problems~\cite{arXiv:2402.17667}.

We can also consider the implications for PEPS and MPS simulation in the context of the $d+1$-dimensional PEPS that is the circuit model. Full contraction can be interpreted as a problem of $d+1$-dimensional PEPS contraction, but special care must be taken, relative to Section~\ref{app:latticestructuredpeps}, owing to the asymmetry of the time dimension (and reduced bond dimension approaching $t=0$). 
Complexity arises in the contraction of the network, in line with standard vertex congestion considerations~\cite{10.4230/LIPIcs.TQC.2019.10}.  In the regime of deep circuits where $t_a/dt > N$ (our experimental regime), an exact contraction requires resources scaling exponentially in $N$, worse than for contraction of the $d$-dimensional PEPS models. Our time-evolved PEPS methods (and also time-evolved MPS methods) are attempts to leverage the improved scaling understood for PEPS-lattice models at fixed time. They can be interpreted as boundary-PEPS heuristics for the circuit model: we approximate the boundary $t=0$ as an (effectively factorized) PEPS and contract in time with compression after each operation to maintain a maximum bond dimension. The shortcomings of various standard and bespoke boundary-PEPS heuristics are evaluated in the following sections.

Many methods have been developed to accelerate the contraction of circuit-model PEPS. These can exploit inhomogeneity in the circuit structure, small circuit depth, rank-reduction of gates, Clifford-gate methods, an assumed fidelity threshold, or locally tree-like structure (or many of the above)~\cite{PhysRevLett.129.090502, Patra2024, Tindall2023, liao2023simulation, Kechedzhi2024}. Considering our ensemble of rank-2 gates, $\exp(-i \pi/2 [4 dt \mathcal{J}(t/t_a) J_{ij}]\sigma^z_i\sigma^z_j)$ and $\exp(-i \pi/2 [4 dt \Gamma(t/t_a)] \sigma^x_i)$, we can note that these are close to Clifford gates when angles $\theta = [\cdot]$ are integer: if $dt$ could be chosen to satisfy this criterion for a significant subset of gates throughout the anneal, efficiencies become possible. Since small $dt$ is required and $\{\mathcal{J}, \Gamma\}$ vary across a broad range of values, we view discretization using Clifford gates as infeasible. To achieve a sufficiently refined time evolution (faithful high-fidelity $J_{ij}$ representation, and small Trotterization error [$dt$], at the evolved energy scales throughout the anneal) a large variety of angles $\theta$ is required; ideally a continuous range, which is impractical for calibration purposes in current circuit-model quantum computer platforms. We evaluate locally tree-like methods in sections \ref{app:PEPSltl} and \ref{app:PEPSltl2} without success. The demonstrated growth of correlations and entanglement through the quantum phase transitions lead us to anticipate, more generally, a failure of perturbative expansion methods. Homogeneity of gates in space in time, a large number of short loops, and large depth, leaves few opportunities to leverage slicing and contraction-order optimization beyond those exploited in our boundary-MPS and boundary-PEPS methods. 

\subsubsection{Locally tree-like approximations}
\label{app:PEPSltl}

The success of MPS methods is contingent on the ability to efficiently manipulate isometries and canonical forms, which allow efficient full-scale contraction. Heuristic methods are required for the manipulation of isometries in PEPS and have only been well studied, with limitations, in square lattices~\cite{PhysRevLett.124.037201}. Computationally useful canonical forms have not been defined for disordered lattice models. Foregoing global consistency of isometries and gauges, a local consistency of gauges and/or isometries can be sought, and such approximations have had much success. 
Uniqueness of a locally self-consistent gauge, in agreement with full contraction on environments up to small corrections, is expected for ``locally tree-like lattices'' with short correlation lengths~\cite{10.1103/PhysRevB.98.085155}.

When such gauges are stable, two methods are commonly used: belief propagation (BP) and simple update (SU)~\cite{PhysRevLett.101.090603, 10.1103/PhysRevResearch.3.023073, Tindall2023}.
The gauges resulting from a locally consistent criterion can be defined based only on tensors (tensor pairs) and their neighbors, matching those on a tree tensor network. These can be iterated from some initial condition, but unlike a tree tensor network it is not possible to systematically contract from leaves to produce a canonical form. Convergence of the iterated local criteria is not guaranteed, and systematic errors associated with converged gauges are typically difficult to quantify. However, if correlation length is short compared to the length of loops, it follows intuitively that such methods can converge on a locally tree-like environment with reasonable accuracy.

For these types of approximations to be efficient, a tensor network needs to look ``tree-like'', which means that the network can locally be represented by a tree with a diameter bounded by the smallest loop. For example, in the aforementioned simulation of the kicked Ising experiment on the heavy-hex lattice, the shortest loops are length $12$, and tree-tensor networks proved to be very fast and economical in computational resources since correlation lengths are relatively short~\cite{Tindall2023}.

For a random $k$-regular graph, it is easy to show that the expected average loop size goes to infinity with the number of nodes\cite{random_graphs}, and hence a tree-tensor approach is expected to efficiently perform quantum annealing protocols on such graphs for any finite correlation length. 
Our $k$-regular lattice models, in contrast, have many short loops incident on every site: length 2 for dimerized bicliques, length 4 for dimerized-cubic and square lattices, and length 6 in diamond lattices and non-dimerized 3D lattices. As shown in Section~\ref{app:PEPSltl2}, we can use a locally tree-like approximation efficiently in all these lattices for both time updates and estimation of observables---agnostic to the requirement of gauge convergence---but find that for even the shortest experimental time scales and simplest lattices, large systematic errors result.

\subsubsection{Simple update}
\label{app:PEPSltl2}

Simple update assumes a locally tree-like approximation for purposes of truncation during time evolution~\cite{PhysRevLett.101.090603}. The initial condition of factorized spins meets the locally tree-like gauge conditions. In our annealing context, SU with fine-grained Trotterization ought to track a gauge consistent with BP where such a fixed point is stable\cite{10.1103/PhysRevResearch.3.023073}, and so we omit an independent evaluation of BP. 

The impact of an unconverged, or converged (but poor), gauge during time evolution is that inferred truncations are uncontrolled. The systematic errors associated with uncontrolled truncation can nevertheless represent a good trade-off when errors in the final state are permissable. 
SU methods have been successfully applied for the simulation of a kicked Ising model on a heavy-hex lattice for example, which despite finite dimensionality (2) has a paucity of short loops\cite{Tindall2023, Patra2024}.

Simple update is based upon an approximation of the bond environment. This environment can be used not only for PEPS truncation in time evolution, but also for estimation of expectation values. We call this approximation SO (simple observable). The self-consistent combination of SO with SU has been successfully applied in related PEPS applications~\cite{Patra2024}, and this method is feasible in all of our lattices. 

\begin{figure*}
  \includegraphics[scale=1]{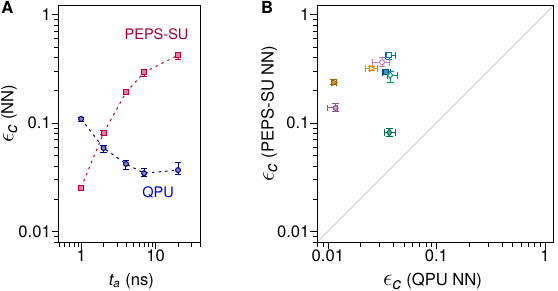}
  \caption{{\bf SU (evolution) plus SO (correlation estimation) versus QPU.} ({\bf A}) Ensemble-median $8{\times}8$ square-lattice (high-precision) correlation error, with numerator and denominator taken only over nearest-neighbor pairs, for SU and QPU.  ({\bf B}) Nearest-neighbor correlation error for high-precision square ($8{\times }8$), cubic ($3{\times }3{\times}3$), diamond ($4{\times }4{\times}8$), and biclique ($K_{6,6}$) ensembles at $t_a=\SI{7}{ns}$ (colored markers) and $\SI{20}{ns}$ (white markers).  PEPS SU+SO errors are smallest on shorter time scales and in models with fewer short loops (like the diamond ensemble), as expected.
  }\label{fig:PEPS1}
\end{figure*}

In application of SU and SO to the experimental regimes many results were well converged, with a weak dependence of $\epsilon_c$ on bond dimension $D$ (for $D\geq 2$). A weak dependence of errors on $D$ might result from the insufficient neighborhood scale. In some more strongly correlated regimes SU was found to be unstable, with sensitivity to $dt$ and the minimum singular value threshold. This is not unexpected since inversion of small singular values necessary in the algorithm can amplify errors quickly, and locally-tree-like assumptions fail as correlation length grows. We present in Fig.~\ref{fig:PEPS1} results with $dt = \{2^{-7},2^{-8}\}\,\SI{}{ns}$ and $D=\{2,3,4\}$ optimized for minimal error on a per instance basis, and with a small ($10^{-10}$) cutoff in the singular values supported. Correlation error $\epsilon_c$ (NN) is taken only over nearest-neighbor pairs; this quantity correlates strongly with $\epsilon_c$ in~\eqref{eq:ec}, but SO environments are local and hence more suited for estimation only of nearest neighbor correlations, so we use the nearest-neighbor variant for convenience.  Errors are significantly larger than those obtained from the QPU in our experimental regime (lattices and time scales).

\begin{figure*}
  \includegraphics[scale=0.8]{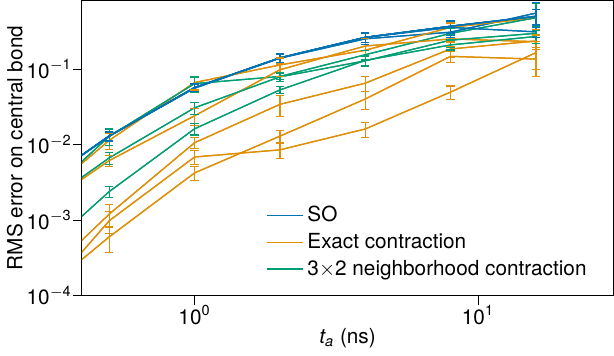}
  \caption{{\bf Absolute correlation error on the (unique) central bond in an open-boundary $3{\times}4$ $J=\pm 1$ square lattice as a function of quench time.} Exact diagonalization methods are applicable at this scale. We present a root mean square error on SU dynamics with respect to 17 disorder realizations. Three different contraction methods for calculation of observables are used, indicated by color, and are combined with SU at various bond dimensions: SO($D=2,\ldots,6$), $3{\times}2$ neighbor contraction ($D=2,\ldots,4$), and exact contraction($D=2,\ldots,6$). As simulation time approaches zero, all methods approach zero error for bond dimension $\geq 2$. At the shortest time shown, error decreases monotonically with bond dimension for all methods.
  For time $\ll \SI{1}{ns}$ all methods converge to zero error at small bond dimension. Approaching the short-time limit, errors go to zero much quicker (in absolute terms shown, and also relative to $c_{ij}$), and SO is sufficient. At intermediate time scales estimation quality improves both as a function of neighborhood scale and bond dimension (albeit SO has a very weak dependence on bond dimension).  For time $\gtrsim \SI{4}{ns}$ all methods approach larger errors, and without a monotonic dependence of error on bond dimension and neighborhood size. Even with no error in observable calculations we see the systematic errors at experimental time scales---larger problems with additional loops can be anticipated to have larger systematic errors, explaining the trends in Figure~\ref{fig:PEPS1}.}
  \label{fig:PEPS2}
\end{figure*}

Sources of errors might be attributed to SU or SO. A spin-glass ($\pm J$, low precision) ensemble for small square open-boundary lattices provides a controlled context for this evaluation. SU might be combined with one of: SO; contractions on $2{\times}3$ neighborhoods (combined with Frobenius norm outside the sublattice, matching the standard NTU update for time-evolution); or exact full contraction. As shown in Fig.~\ref{fig:PEPS2}, SU at small bond dimension combined with SO is accurate for short time scales ($t_a<\SI{1}{ns}$). $3{\times}2$ and full contractions improve slightly on errors in these regimes, in line with perturbative loop corrections. SO degrades at longer time scales in a manner that is almost independent of bond dimension, whereas $3{\times}2$ and full contraction respond to variation in bond dimension, with full contraction superior. For contraction methods, on time scales $\gtrsim \SI{7}{ns}$, both full and $3{\times}2$ contractions fail in a manner that cannot be reduced systematically by increasing bond dimension. In this regime the accumulation of errors in SU prevents attainment of small errors by any contraction method for observables. This aligns with our intuition that as correlation length grows more complex, neighborhoods must be contracted~\cite{10.1103/PhysRevB.90.064425, 10.1103/PhysRevB.104.094411}. We examine this in the context of square and cubic lattices in Sections~\ref{app:PEPS2D} and~\ref{app:PEPS3D}.

\subsubsection{Non-canonical MPS in 1D}  
\label{app:PEPS1D}

\begin{figure}[b]
\includegraphics[width=0.75\columnwidth]{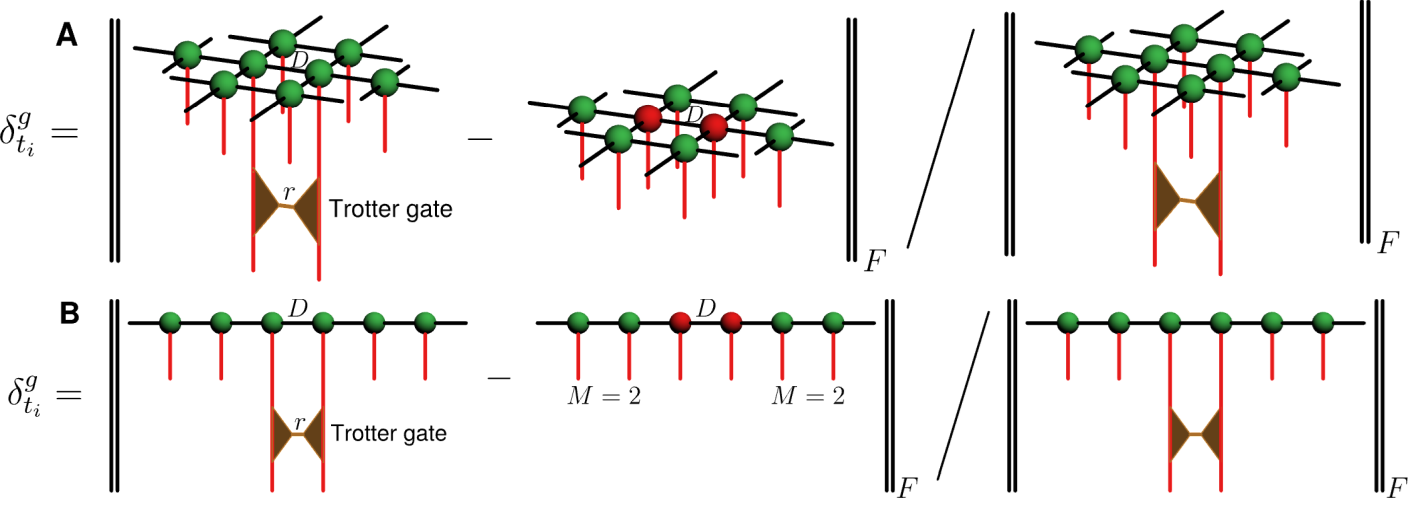}
  \caption{{\bf Optimizing bond dimension truncation in 2D PEPS and non-canonical 1D MPS.} 
  In {\bf A}, for non-canonical PEPS in 2D, the two central tensors are applied with a Trotter gate $g$ with SVD rank $r=2$. When the gate tensors are absorbed into the central ones, the bond dimension increases from $D$ to $Dr$. It is truncated back to $D$ by approximating the central tensors with a pair of new tensors with bond dimension $D$. The new tensors minimize the Frobenius norm $\delta_{t_i}^g$ of the relative difference between the left and the right diagram, where $t_i$ index time steps of the Trotterized evolution. Here, in addition to the two central sites, the 2D cluster includes their six nearest neighbor sites. The total cluster size during such NTU algorithm~\cite{10.1103/PhysRevB.104.094411} update is $8$ sites.
  In {\bf B}, we consider a non-canonical MPS in 1D.
  Here, in addition to the two central sites, the diagrams include $M=2$ sites to the left and $M=2$ sites to the right. A total cluster size is $2+2M$ sites.
  }  
  \label{fig:ntu12D}
\end{figure}

In this work we use the standard PEPS techniques that do not refer to canonical forms.
Potential advantages of the isometric PEPS~\cite{PhysRevLett.124.037201} seem to be offset by the errors of Moses moves that attempt to change the canonical structure of the network.
In order to better understand problems arising from a non-canonical structure of PEPS in 2D, we simulated the Kibble-Zurek (KZ) quench in 1D for an infinite pure-ferromagnetic quantum Ising chain with an infinite MPS ansatz, but without enforcing any canonical gauge. The ansatz is a tensor train with two sublattice tensors: $...-A-B-A-B-...$. The time evolution is performed in a similar manner as for an infinite PEPS in 2D. Not counting local gates with a transverse field, in each time step there are two 2-site Suzuki-Trotter gates with a ferromagnetic interaction on bonds $AB$ and $BA$ that commute with each other. After the application of a gate, the bond dimension is doubled and has to be truncated back to the original bond dimension $D$ in a way that minimizes truncation error (see Fig.~\ref{fig:ntu12D}B). The error is measured on a cluster of $2+2M$ sites including the two central sites with the gate and two blocks of $M$ spins adjacent to the central sites to the left and the right. The error $\delta^g_{t_i}$ is the Frobenius norm of the relative difference between the clusters before and after the truncation (see Fig.~\ref{fig:ntu12D}B), following application of the gate at time step $t_i$ to a bond $g=AB$ or $BA$. In the worst-case scenario, the individual errors sum up to a (upper estimate) total truncation error $\Delta_g = \sum_i \delta^g_{t_i}$.
We define the error estimate as
\begin{equation}
    \Delta = \max_{g} \Delta_g = \max_{g} \sum_i \delta^g_{t_i}.
    \label{eq:integrated_truncation_error}
\end{equation}
For lack of canonical structure, \eqref{eq:integrated_truncation_error} serves as a rough estimate for a relative error inflicted on observables by the truncations. We will show below that it correlates well with the errors of the expectation values calculated at the end of the quench. Moreover, $\delta^g_{t_i}$ becomes proportional to $dt$ in the limit of small time steps, making the definition in Eq.~\eqref{eq:integrated_truncation_error}, and the quality of the solution, independent of the time-step in that limit. This is illustrated in Fig.~\ref{fig:CZZ_Delta_2D}B in a 2D setup.

\begin{figure}[t]
\includegraphics[width=0.98 \columnwidth]{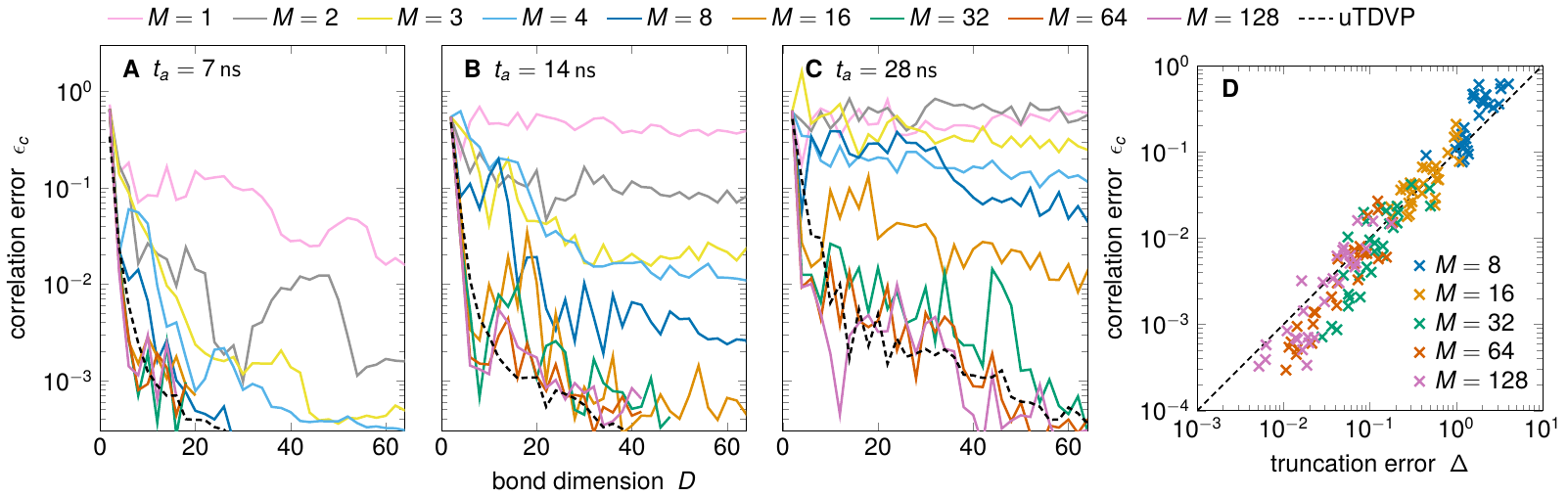}
  \caption{{\bf Correlation and truncation errors in non-canonical MPS.} 
  In {\bf A}, {\bf B}, and {\bf C}, we show the correlation error, $\epsilon_c$ in Eq.~\eqref{eq:ec}, as a function of the bond dimension $D$ for annealing times $t_a=7$, $14$, and $\SI{28}{ns}$, respectively, for a range of cluster sizes $M$. 
  For each $t_a$, the function $\epsilon_c(D)$ saturates above a critical value of $M$. We also show the results of an uniform TDVP (uTDVP) simulations with one-site unit cell~\cite{Vanderstraeten2019}, which roughly coincide with the results for the largest clusters in the presented range of parameters.
  In {\bf D}, we plot the correlation error as a function of the truncation error $\Delta=\max\left(\Delta^{AB},\Delta^{BA}\right)$, according to Eq.~\eqref{eq:integrated_truncation_error}. The two are roughly linearly correlated. The correlation error is an order of magnitude lower than the truncation error---the dashed line indicating $\Delta/10$ is a guide to the eye. In this panel, $t_a=\SI{28}{ns}$ and $D=4,\ldots,60$.
  }  
  \label{fig:delta_v_D}
\end{figure}

The correlation error, Eq.~\eqref{eq:ec} of the main text, after the quench with annealing times $7,14,\SI{28}{ns}$, simulated with various cluster sizes $M$, is shown in Fig.~\ref{fig:delta_v_D}. 
Overall, a longer annealing time results in a final state with a longer KZ correlation length $\hat\xi\propto\tau_Q^{1/2}$ and requires a higher bond dimension to reach the same level of accuracy\footnote{We use $\hat \xi$ to denote the Kibble-Zurek correlation length in line with literature notation. This is to be contrasted with the usage of a more generic correlation length ($\xi$) elsewhere in our paper (which is not assumed to follow KZ scaling).}. Here $\tau_Q=t_a/t_0$ is a dimensionless KZ quench time with $t_0$ being a characteristic system's time scale. 
As we can see in Figure \ref{fig:delta_v_D}, for shortest values of $M$ the error improves slowly, if at all, with increasing $D$. 
We can also see that for a given $D$, the error decreases with the cluster size until it saturates for long enough $M$. The minimal $M$ needed for convergence, $M_c$, depends on the annealing time. It might be natural to expect $M_c\propto\hat\xi\propto\sqrt{t_a}$, but it grows faster than $\sqrt{t_a}$, as $t_a$. For example at $t_a=\SI{7}{ns}$ we can estimate $M_c\approx16$ while for $t_a=\SI{28}{ns}$ it is four times longer, $M_c\approx 64$. As such, $M_c\propto t_a$ seems to be more consistent with the data.
This steeper scaling is not necessarily at odds with the KZ mechanism. After correlated pairs of quasiparticles with opposite quasimomenta $(k,-k)$ are excited near the critical point, during the following quench across the ferromagnetic phase, they are free to move along the chain and spread their pairwise entanglement. In this way, a second length scale, $l=\sqrt{\tau_Q}\ln\tau_Q$, develops~\cite{Cincio2007} that is longer than $\hat\xi$. In the limited range of $\tau_Q$ where we are operating, what might at first be dismissed as a mere logarithmic correction to $\hat\xi$, is actually substantial and makes $l$ appear to grow nearly linearly with $\tau_Q\propto t_a$.

Fig.~\ref{fig:delta_v_D}D shows the correlation between the correlation error and the total truncation error in Eq.~\eqref{eq:integrated_truncation_error}. The log-log plot is not inconsistent with a linear dependence. It is noteworthy that the correlation error is an order of magnitude lower than the truncation error. We observe that $\epsilon_c \approx \Delta/10$ is a good approximation to the data, allowing prediction of the correlation error based on the truncation error.

The 1D toy model shows how important it is for the cluster size to be at least comparable to the range of correlations in order to make efficient use of the limited bond dimension. The issue becomes more important in 2D and 3D where the limitations are more stringent. 

\subsubsection{2D PEPS}
\label{app:PEPS2D}

\begin{figure}[t]
\includegraphics[width=\columnwidth]
{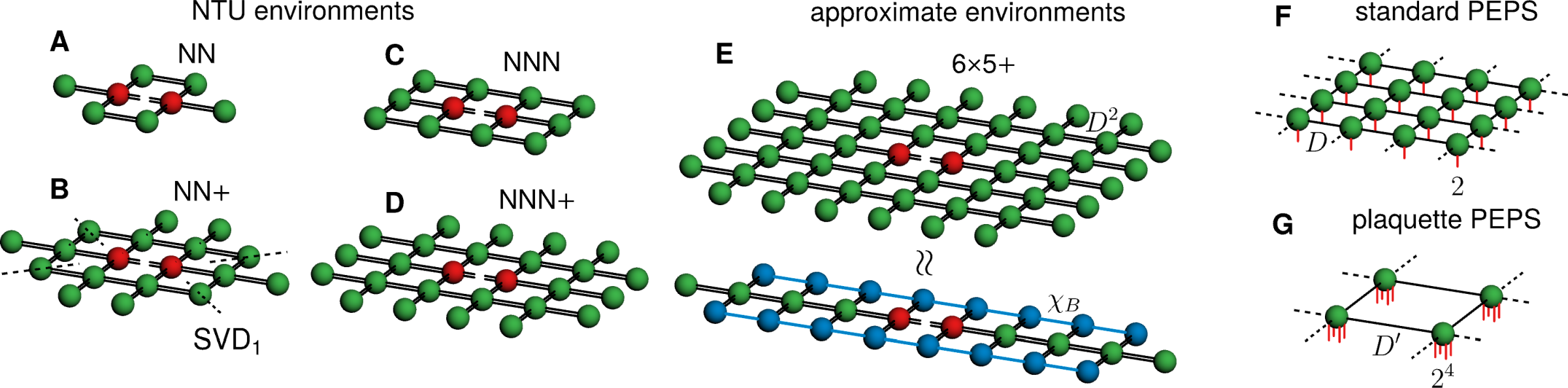}
  \caption{{\bf  2D PEPS: environments for truncating enlarged bonds in time-evolution simulations.}  The two-layer diagrams correspond to the calculation of the Frobenius norms depicted in Fig.~\ref{fig:ntu12D}A. In panel {\bf A}, we show the neighborhood tensor update environment and, in  {\bf B}--{\bf D}, its direct extensions that are contracted numerically exactly. In {\bf B}, SVD$_1$ indicates a product approximation, where we retain a single leading singular value for tensor decomposition along black dashed cuts. Larger environments, such as the one in {\bf E}, require approximate contraction, which, however, might disrupt the positivity of the depicted bond metric. Here, to that end, we employ boundary MPS. In {\bf F}, we indicate a standard PEPS representation with one tensor per spin and, in {\bf G}, a plaquette PEPS with one tensor for four spins. While a $2{\times}2$ plaquette of a standard PEPS with bond dimension $D$ can be contracted into a single plaquette PEPS tensor of dimension $D^2$, such bonds might not be used optimally. Indeed, we observe that a larger effective environment in plaquette PEPS might result in the compression of the required bond dimension, $D' < D^2$, for reaching similar accuracy.
  }  
  \label{fig:ntu_env_2d}
\end{figure}

In 2D, we start with the current state-of-the-art algorithm that can match QPU accuracy only for the fastest annealing times considered. We then systematically examine a family of extensions, which reduce errors below QPU level at intermediate time scales (though not for the slowest experimentally relevant quenches).  The \texttt{yastn} library implements the various neighborhood contractions discussed in these sections, with example code and further documentation \cite{yastn}.

As our baseline method, we employ the neighborhood tensor update (NTU)~\cite{10.1103/PhysRevB.104.094411, 10.1103/PhysRevB.107.144510} with one PEPS tensor assigned to each lattice site. In NTU, the cluster used for truncation includes the two central sites with the gate plus their six nearest neighbor sites, including two loops, as shown in Fig.~\ref{fig:ntu12D}A. Such a neighborhood is contracted numerically-exactly at a cost of ${\cal O}(D^8 + p D^7)$ operations, where $p$ is the physical dimension of PEPS tensor. This exactness preserves the desired positivity of the bond metric tensor in Fig.~\ref{fig:ntu_env_2d}A that gets inverted during optimization of the truncation.

We next systematically probe several methods that expand the neighborhood scale considered in the truncation. Within the paradigm of local bond environments that get contracted exactly (preserving positivity), we expand the nearest neighbor (NN) environment in Fig.~\ref{fig:ntu_env_2d}A by adding partial information from the next layer of tensors, treated tree-like (see Fig.~\ref{fig:ntu_env_2d}B). Calculating this metric has the same ${\cal O}(D^8 + p D^7)$ asymptotic numerical cost. Further extensions involve including extra loops in the exactly-contracted environment, with both next nearest neighbor (NNN) and NNN+ in Fig.~\ref{fig:ntu_env_2d}C,D coming at ${\cal O}(p D^{10})$ numerical cost and limiting feasible PEPS bond dimensions to approximately $D=8$, in practice. Finally, larger environments can be considered, going toward full update. They rely on approximate contraction schemes, such as boundary MPS shown in Fig.~\ref{fig:ntu_env_2d}E. Unlike exact contraction, such approximations do not preserve the non-negativity of the metric tensor.

Another strategy still uses the NTU cluster for truncation, but each PEPS tensor covers a plaquette of $2{\times}2$ sites, as shown in Fig.~\ref{fig:ntu_env_2d}G. In this way, the NTU cluster of plaquettes covers four times as many qubits as in the standard setup of Fig.~\ref{fig:ntu_env_2d}F. Surprisingly, for many considered protocols, the bond dimension required for the same truncation error is less than a square of the bond dimension in the one-site NTU. Intuitively, two bond indices with dimension $D$ extending in a given direction out of a $2\times2$ plaquette in the one-site PEPS should fuse into one bond index with dimension $D^2$. We observe, however, that the dimension needed to achieve equivalent accuracy is lower than $D^2$.  Such compression is possible because a plaquette represented by a contraction of 4 one-site PEPS tensors is more constrained than the same plaquette covered by a single plaquette tensor (see Fig.~\ref{fig:ntu_env_2d}F and G). This compression also enables a numerically exact contraction of the (four-times bigger) plaquette NTU cluster. This larger effective cluster allows making more efficient use of the truncated bond dimension, contributing to the compression. 

In all the above cases, expectation values follow from the approximate contraction of the entire network. We employ a boundary MPS approach adapted to a cylinder. Larger bond dimensions require exploiting Monte Carlo sampling operating in the $z$-basis. Here, we follow the approach of Ref.~\onlinecite{10.1103/PhysRevB.104.235141}, but with a local trial probability distribution. As compared to the deterministic evaluation of correlators, which requires a double-layer PEPS (PEPS contracted with its conjugate), Monte Carlo sampling requires only the contraction of a single (projected) PEPS layer. It is worth noting, that we produce on the order of one MC sample per second for an $8 \times 8$ system at $D=10$ (running on a single CPU core), which is roughly three orders of magnitude slower than the QPU sampling rate.

\begin{figure}[t!]
\includegraphics[width=0.7\columnwidth]{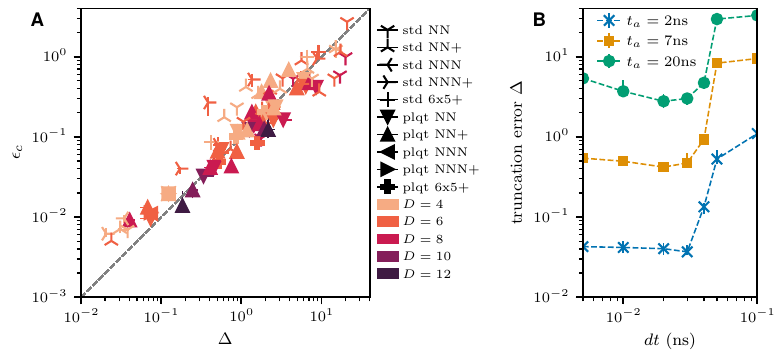}
  \caption{{\bf 2D PEPS:  Correlation errors vs truncation error.} 
In A, we compare the correlation errors $\epsilon_c$ with the truncation error in Eq.~\eqref{eq:integrated_truncation_error}. The two quantities are roughly linearly correlated for a broad range of simulation parameters, with $\epsilon_c$ being an order of magnitude smaller than $\Delta$. The dashed line showing $\epsilon_c = \Delta/10$ is a guide for an eye. Each point corresponds to a different PEPS bond dimension (color), the environment used for truncation (symbols, see Fig.~\ref{fig:ntu_env_2d}), annealing time, and coupling precision (the latter two are not distinguished by color nor symbol). Data for $8 {\times} 8$ cylindric lattice. In B, we show the convergence of $\Delta$ with $dt \to 0$, where the data are for plaquette PEPS with NN+ environments and $D=10$. 
}  
\label{fig:CZZ_Delta_2D}
\end{figure}

\begin{figure}[b!]
\includegraphics[width=0.95\columnwidth]{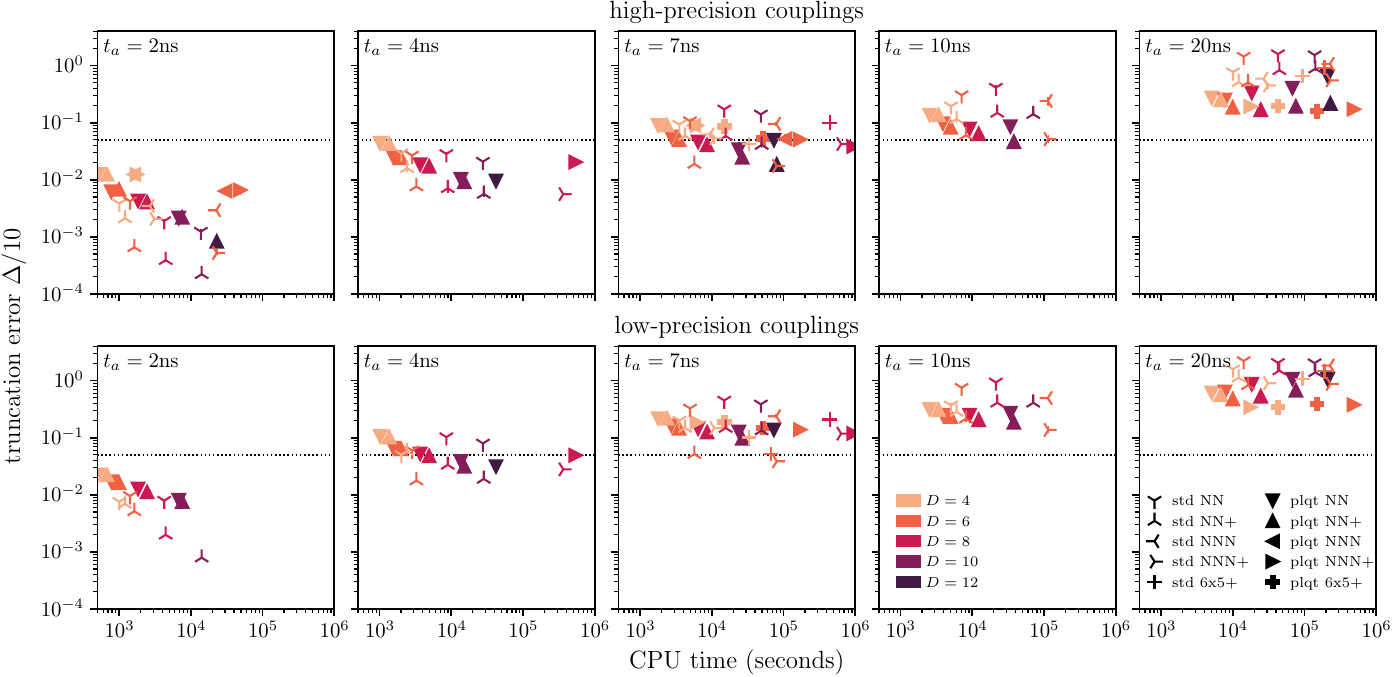}
\caption{{\bf 2D PEPS: errors vs numerical costs.}
We show estimates of PEPS accuracy, employing truncation errors $\Delta / 10$,
as a function of simulation times for problems defined on $8 {\times} 8$ cylinders. Each panel corresponds to a fixed annealing time and type of random couplings (high- or low-precision, see Sec.~\ref{app:testbed}). 
Points in each panel represent PEPS attempts to simulate a given problem for a selection of PEPS bond dimensions and type of environment used in truncation (see Fig.~\ref{fig:ntu_env_2d}). Those plots complement Fig.~\ref{fig:3} of the main text, which shows the correlation errors for $t_a = 2$, $7$, and $\SI{20}{ns}$ for high-precision couplings. We see systematic accuracy improvements with the bond dimension for shorter annealing times. This trend systematically stagnates for slower evolutions, where the necessity to employ larger environments becomes apparent. The low-precision couplings prove more challenging for PEPS to simulate than the high-precision. Indeed, energy degeneracies related to discrete $J=\pm 1$ couplings are expected to facilitate entanglement spreading as compared to high-precision inputs, where such degeneracies are much less common. As CPU times, we report simulation wall times executed on a single CPU core.}
\label{fig:2D_errors_costs}
\end{figure}
We find the systematic errors associated with the calculation of the expectation values to be much smaller than those resulting from approximate truncation in the time evolution part of the simulation. In Fig.~\ref{fig:CZZ_Delta_2D}A, where we collect simulations done for $8 {\times} 8$ cylinder lattice, both for high and low inputs precision studied on the QPU, we show the relation between total truncation error in Eq.~\eqref{eq:integrated_truncation_error} and the correlation error $\epsilon_c$. The latter is roughly an order of magnitude smaller than the corresponding truncation error $\Delta$ among many tested environments for standard and plaquette PEPS and a range of bond dimensions used to simulate time evolution for several annealing times $t_a$. Interestingly, a rough estimate of the correlation error as $\Delta / 10$ is consistent with a similar estimate in the 1D toy example in Fig.~\ref{fig:delta_v_D}. 
This corroborates that time evolution is the main source of errors in the PEPS results and truncation error is a valid metric to estimate the quality of the approximation. We also show in Fig.~\ref{fig:CZZ_Delta_2D}B that the total truncation error becomes roughly independent of the time step in the limit of small time steps. 
In those simulations, we terminate the dynamics at $s=0.6$. This has insignificant impact on statistics in the case of exact simulation (see Fig.~\ref{fig:s_cutoff}A) but prevents PEPS from accumulating extra errors during the last stage of the evolution.

We use the truncation error metric in Fig.~\ref{fig:2D_errors_costs} to compare the simulation errors with the simulation times, where all times are for the time-evolution only; i.e., without sampling from the final state, executed on a single CPU core. While for $t_a = \SI{2}{ns}$ all tested simulation methods produce errors below QPU precision, already for $D=4$, longer annealing times require larger environments and bond dimensions. In particular, we identify a plaquette PEPS with NN+ environments as a leading method for $\SI{7}{ns}$ quenches. However, all tested methods give worse than QPU results for $\SI{20}{ns}$ quenches. In particular, low-precision inputs prove to be harder to simulate using PEPS (note that the data in Fig.~\ref{fig:3} of the main text shows high-precision results). Finally, we note that the simulated problem has a global $\mathbb{Z}_2$ symmetry. We observe that imposing this global symmetry directly in PEPS ansatz does not affect the quality of the results compared to using dense tensors with no symmetry imposed.

\subsubsection{3D PEPS}
\label{app:PEPS3D}

\begin{figure}[b]
\includegraphics[width=0.80\columnwidth]{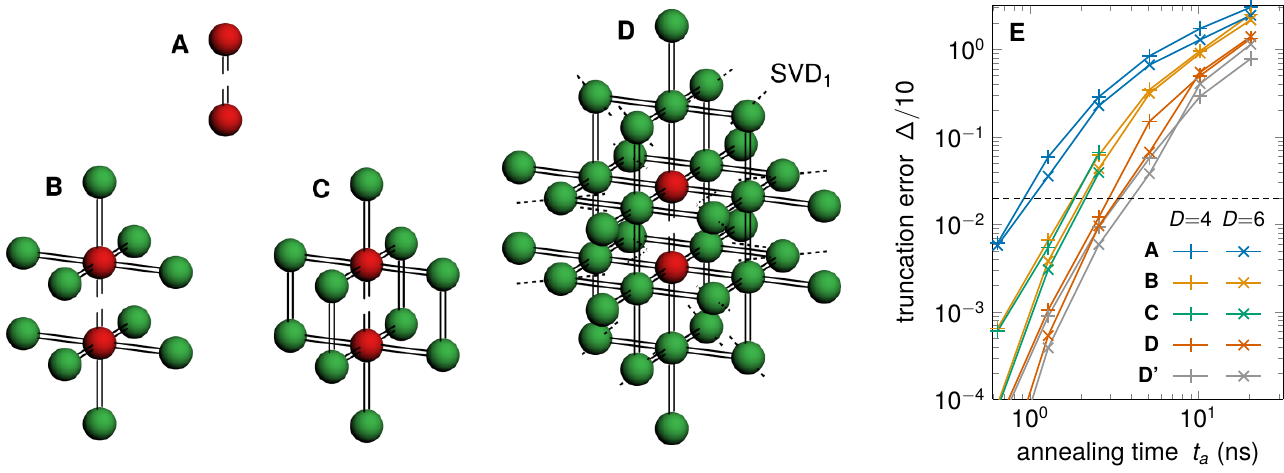}
  \caption{{\bf 3D PEPS.} 
  The left panel shows four out of five truncation clusters used in 3D PEPS time evolution. The diagrams represent Gramm-Schmidt metric tensors that measure the truncation error. With cluster {\bf A} the truncation is just an SVD, cluster {\bf B} includes whole nearest-neighbor tensors (more than the simple update), and {\bf C} includes vertical bonds between the nearest-neighbor tensors (like the 2D NTU, called 3DNN). Cluster {\bf D} reaches out to third nearest neighbors but is made tree-like by SVD truncations to just the single leading singular value (SVD$_1$) that are marked by the black dashed cuts. The biggest cluster {\bf D'} (not shown) is a $5\times5\times6$ rectangle made tree-like with a few SVD$_1$. 
  The right panel, {\bf E}, shows total truncation error as a function of the annealing time $t_a$ for clusters {\bf A}, {\bf B}, {\bf C}, {\bf D}, {\bf D'} and bond dimensions $D=4$ and $6$.
  Here $L=8$, $dt=\SI{0.01}{ns}$, and the error bars result from averaging over 8 realizations of the random couplings.
  A dashed line indicates a $2\%$ level of error obtained in QPU simulations for 3D problems with dimers.}  
  \label{fig:3D_delta_v_t}
\end{figure}

We performed simulations on a cubic $L {\times} L {\times} L$ lattice with open boundary conditions and random nearest-neighbor couplings $J_{ij}$ from a uniform distribution in $[-1, 1]$. This is a simplification of the dimerized cubic model used in experiments which has twice as many qubits, but the simplification is already challenging enough to highlight limitations in simulation.
Five truncation schemes,explained in Fig.~\ref{fig:3D_delta_v_t}, were applied. The simplest one is the SVD truncation in {\bf A} that ignores any tensor environment and requires only ${\cal O}(p D^{7})$ operations. Scheme {\bf B} includes NN tensors but still ignores contractions between them, keeping the numerical cost to calculate the Gramm-Schmidt metric tensor at the same level of ${\cal O}(p D^{7})$. Scheme {\bf C} includes the contractions, raising the cost to ${\cal O}(D^{12} + p D^{11})$ operations. Scheme {\bf D} includes up to third nearest neighbors. In order to keep the cost at the tolerable level of ${\cal O}(p D^{8})$ a sequence of SVD truncations to the single leading singular value (SVD$_1$) is applied that renders an approximate tree-like diagram. First the eight double-PEPS tensors at the corners of the diagram are SVD$_1$-truncated and then, after almost all contractions are completed, the four vertical double bonds are cut. Finally, scheme {\bf D'} extends {\bf D} to a $5\times5\times6$ cluster with more SVD$_1$ truncations to keep the cost at the same level of ${\cal O}(p D^{8})$. As in Section \ref{app:PEPS2D} we are able to truncate dynamics at $s=0.6$ with negligible impact on results. 

In 2D it was practical to combine neighborhood updates with approximate full contraction, the former being a computational bottleneck.
By contrast, the calculation of observables by comparable methods in 3D is a computational bottleneck. The deterministic scheme employing double-layer PEPS~\cite{VlaarCorboz3D} is not practical beyond $D=4$. We also tested code for a 3D generalization of the single-layer Monte Carlo~\cite{10.1103/PhysRevB.104.235141}, which partly trades the contraction cost for sampling, but it still does not extend much beyond $D=4$. We estimate that $D=6$ is the upper limit, if allowed a few months of computations, but $D$ is expected to grow exponentially with the Kibble-Zurek length and so does the computation cost. 

The right panel of Fig.~\ref{fig:3D_delta_v_t} shows the total truncation error as a function of the annealing time for $D=4,6$. For a given bond dimension, the errors of schemes {\bf B,C} are practically the same, while those of {\bf A/D} are much larger/smaller, suggesting that the cluster size matters more than the contractions between the nearest-neighbor tensors. For $D=4,6$ the error of scheme {\bf D} crosses the $2\%$ QPU benchmark around $\SI{4}{ns}$. 
In the regime of parameters of Fig.~\ref{fig:3D_delta_v_t}, the accuracy of the evolution is limited mainly by the cluster size: there is little improvement with increasing $D$ but much with the size. With approximate contraction, the size could be further increased but, for long-enough annealing time, a crossover to a $D$-limited regime would inevitably take place and the cost of time evolution would start to grow exponentially with $t_a$.

As a means to increase the exactly contractible cluster size, we also attempted a 3D cubic PEPS, similar to the plaquette one in 2D, where a single PEPS tensor covers a cube of $2^3$ sites. In 3D, a one-site PEPS has 4 bond indices with dimension $D$ extending in each direction out of the $2^3$ cube. In a cube PEPS tensor, the 4 indices can be fused into one index with dimension $D^4$. Therefore, the minimum required to simulate the evolution with only the classical nearest-neighbor Hamiltonian (the transverse field set to zero) is $D=2$ for the one-site PEPS, $D^2=4$ for the plaquette PEPS, and $D^4=16$ for the cubic one. The cubic PEPS has a much higher minimal bond dimension than the plaquette one. In 3D, $D\geq16$ may still be feasible for time evolution, despite an overhead due to the physical dimension, $2^8=256$, but the evaluation of expectation values within a controlled approximation seems out of reach. 

%% file: sm_quality.tex
\section{Measuring the quality of sampled distributions}\label{app:quality}

\begin{figure*}
  \includegraphics[scale=1]{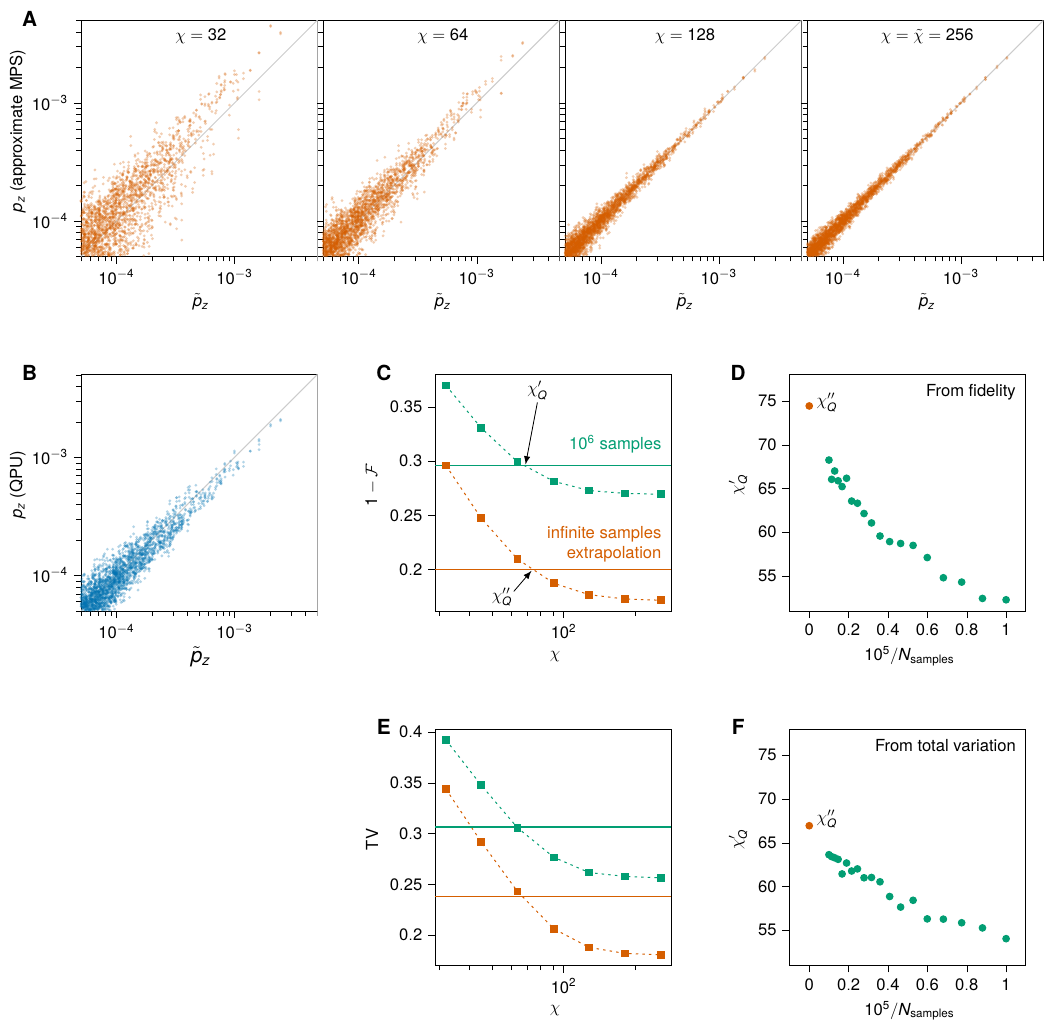}
  \caption{{\bf Estimating QPU-equivalent bond dimension from classical fidelity.}  ({\bf A}) Fig.~\ref{fig:2}C (right) is extended, with data for four MPS bond dimensions, up to and including the ground-truth bond dimension $\tilde\chi$, for which only sampling error remains.  ({\bf B}) Fig.~\ref{fig:2}C (left) is shown for comparison, qualitatively similar in spread to $\chi=64$.  ({\bf C}) We can estimate the QPU-equivalent bond dimension from $\mathcal F$ instead of from $\epsilon_c$, by using $10^6$ samples or by extrapolating to the limit of infinitely many samples drawn.  We call these quantities $\chi'_Q$ and $\chi''_Q$, similar to the estimate $\chi_Q$ derived from correlation error.  Analogously to $\chi_Q$, error ($1-\mathcal F$ in this case) is plotted for a range of MPS bond dimensions and the crossing point of QPU error (solid horizontal lines) gives the QPU-equivalent bond dimension. ({\bf D}) $\chi_Q'$ is shown for varying resampled sets of size $10^5$ to $10^6$, and plotted against the inverse sample size along with $\chi''_Q$.  ({\bf E}--{\bf F}) analysis of D--E is repeated using total variation (TV) in place of $1-\mathcal F$.}\label{fig:fidelity1}
\end{figure*}

\begin{figure*}
  \includegraphics[scale=1]{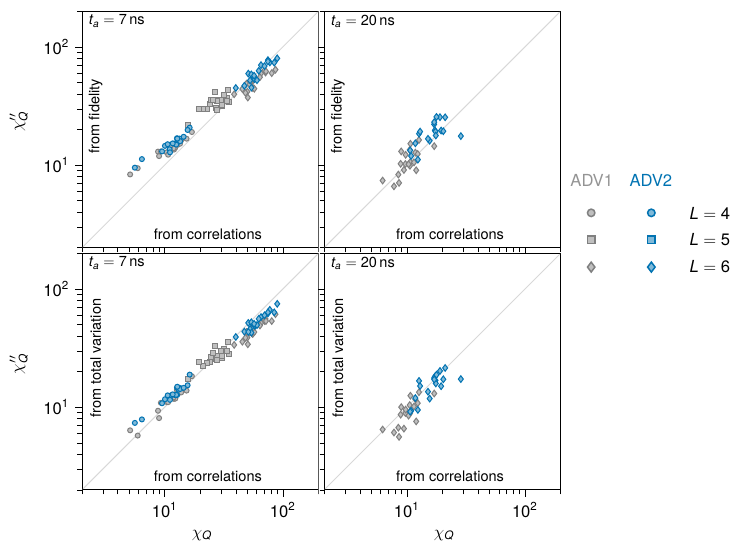}
  \caption{{\bf Comparing correlation-based and fidelity-based QPU-equivalent bond dimensions.}  Shown are $\chi_Q$ and $\chi''_Q$ (as in Fig.~\ref{fig:2}D and Fig.~\ref{fig:fidelity1}C respectively), with $\chi''_Q$ computed from fidelity (top) and total variation (bottom).  At a fixed maximum sample size $10^6$, we expect $\chi''_Q$ to deviate downwards from $\chi_Q$ for systems of increasing entropy, since inference requires exponentially many samples.}\label{fig:fidelity2}
\end{figure*}

The computational task we demonstrate is approximately independent and identically distributed sampling from a quenched quantum state projected in the computational basis.
The quality of approximation can be captured by various distributional divergence measures: a correlation-error metric $\epsilon_c$ and classical fidelity ($\mathcal F$, also called Bhattacharyya distance) are presented in the main text.

A popular measure of sampling quality relevant to random-circuit sampling is the cross entropy\cite{Boixo2018}; in that application, the target distribution approaches a Porter-Thomas distribution and a cross-entropy-based measure is known to approach 1 from below as a function of gate fidelity. The chaotic nature of the distribution makes spoofing of large values challenging. 

In this work, we might define the cross entropy as $-\langle \log {\tilde p}(z)\rangle_p$, where ${\tilde p}$ is the ground-truth probability of states projected in the $z$-basis and angled brackets denote an expectation taken by Monte-Carlo sampling of some approximating distribution $p$ (the QPU). Unfortunately such a measure appears not to be readily bounded in a manner informative for fidelity. Furthermore, and unlike random-circuit distributions, state probabilities in our application are correlated with efficiently calculated statistics (such as $\mathcal H_P$); the strength of correlation with $\mathcal H_P$ is controlled monotonically by $t_a$ and noise. It is easy to raise (lower) the cross-entropy measure by preferential sampling of states with low (high) classical energy. Thus the cross-entropy measure is not a useful quality measure.

The interpretation of a cross-entropy measure is made simpler by comparison with (subtraction of) the model entropy, $\textit{KL}(p,\tilde p) = \langle \log \frac{p(x)}{\tilde p}\rangle_{p}$, yielding a Kullback-Leibler divergence measure that approaches zero only when the distributions are in agreement. The problem with such a measure is twofold: first, it measures the ability of the ground truth to predict model samples; second, the calculation of entropy from sampling requires a number of samples that scales with the model entropy~\cite{Grassberger2003}, and the estimate can be subject to strong variance if the ground-truth distribution is narrower than the model's one. To solve the first issue we might consider the opposite form, $\textit{KL}({\tilde p},p)$, but the problem of scalability persists with this definition. Due to entropy scaling with system size, we could not formulate an unbiased estimator for either form of KL-divergence that was informative on larger problems. We found that several estimators exhibit related issues, including the classical fidelity. Because the standard techniques for dealing with scaling---coarse graining, binning, and projection of samples into a lower-dimensional space---are challenging to implement without simplifying the computational task, we selected the scalable measure $\epsilon_c$ as our primary metric. As we show, this correlates well with other measures of distribution divergence. 

Alongside $\epsilon_c$ we present classical fidelity, $\mathcal F(p,\tilde p)=\sum_x \sqrt{p(x){\tilde p}(x)}$, and total variation distance $\text{TV}(p,\tilde p)= \tfrac 12\sum_x | p(x)-\tilde p(x)|$ as complementary measures to test distribution divergence beyond that captured by two-body correlations. Fidelity is upper bounded by 1 (when distributions agree) and lower bounded by 0; likewise TV is 0 when distributions agree and in $(0,1]$ otherwise. We present data with $2^N$ bins (one per state), but limit usage to smaller scales where the measure is not dominated by sampling bias since, like the entropy-based measures already discussed, the number of samples must scale with entropy to avoid a dominant sampling bias. 

Distribution divergence measures, such as $\epsilon_c$ and $\mathcal{F}$, compress high-dimensional information to a number. These numbers may be skewed in some non-trivial manner; for example, our measure $\epsilon_c$ is not sensitive to sign-symmetry breaking $(p(s)\neq p(-s))$. A divergence of this type would yield a strong signal in classical fidelity by contrast. For this reason it is useful to consider several numbers and also contextualize them with additional distributional information where possible. It is also important to demonstrate that our core results, based upon $\epsilon_c$, are robust to a change in divergence measure. With this in mind, we provide visualizations of the full probability distribution to provide confidence in the measures, and show that if our QPU-equivalent bond-dimension analysis, based on $\epsilon_c$, is replaced by an analysis based on $\mathcal{F}$, no practically significant difference in results is apparent.

Fig.~\ref{fig:2}C scatters $p_z$ against $\tilde p_z$ for $10^6$ samples drawn from QPU and from approximate MPS at bond dimension $\chi=64$.  We first extend this data in Fig.~\ref{fig:fidelity1}A, showing the spread of $p_z$ versus ${\tilde p}_z$ for various values of $\chi$.  When $\chi=\tilde\chi =256$, the only error in the distribution is sampling error.  We compare this with QPU samples in Fig.~\ref{fig:fidelity1}B, noting, as in the main text, that results look roughly equivalent in quality to MPS at $\chi=64$.

To provide a more precise estimate of QPU-equivalent bond dimension, we first define $\chi'_Q$ as the intersection in infidelity $1-\mathcal F$ with the interpolating function of many evaluated bond dimensions---this is exactly how we define $\chi_Q$, based on correlation error.  This infidelity is a function of the sample-set size used, since we use an empirical estimate to the distribution $p$ based on sampling. We can suppress the effect of finite sample sizes by extrapolating infidelity to the infinite-sample limit; assuming that sampling error decreases as the square root of the number of samples is sufficient in lower-entropy use cases.  Using the infinite-sample limit gives another measure, $\chi''_Q$, which we analyze instead of $\chi'_Q$.  Sampling error in measurement of $\chi'_Q$ for one input is shown in Fig.~\ref{fig:fidelity1}D.

We compare $\chi_Q$ with $\chi''_Q$ in Fig.~\ref{fig:fidelity2}, for ensembles of 2D problems.  As previously noted, sampling error in estimating $\mathcal F$ is severe for distributions with high entropy (i.e., larger problems and faster anneals), which will affect not only direct estimates of $\chi'_Q$, but also extrapolations for $\chi''_Q$.  As seen in Fig.~\ref{fig:fidelity1}D, we can expect small sample sizes to result in underestimates of $\chi''_Q$, and therefore we should see a deviation downwards---compared to $\chi_Q$---for larger problems and faster anneals.  Indeed, in Fig.~\ref{fig:fidelity2} this is what we see, with strong quantitative agreement between $\chi_Q$ and $\chi''_Q$ beginning to decay for larger problems at $t_a=\SI{7}{ns}$.  Nonetheless, these results give confidence that $\chi_Q$---derived from pairwise correlation errors---is closely linked to the more expressive fidelity measure, and thus the low distributional distance seen for the QPU is not just limited to a plurality of two-body correlations, but rather is based on accurate reproduction of the amplitudes in the final wave function.

%% file: sm_entanglement.tex
\section{Dynamics of entanglement entropy}\label{app:entanglement}

The dynamics of entanglement in a quantum quench have been widely studied in quantum many-body systems, both in the context of instantaneous quench and in the context, relevant to this work, of a continuous quench across a QPT\cite{Calabrese2005,Cincio2007,Canovi2014,RevModPhys.80.517}.
In ``anneal'' quenches, the state of an $N$-qubit system, prepared initially in the ground state, is evolved in time with \emph{slowly} varying time-dependent Hamiltonian parameters. In such a finite system, the evolution across a continuous critical point can remain adiabatic for sufficiently slow quenches due to a finite-size gap at the critical point. Consequently, at least in one dimension the entanglement entropy of bipartition follows instantaneous ground-state behavior in this scenario. With faster parameter variation, the system gets excited upon crossing the critical point. Entanglement dynamics in such a case has been studied in a homogeneous 1D TFIM in Refs.~\onlinecite{Cincio2007, Canovi2014}, showing that at the end of the anneal it grows logarithmically as a function of annealing time $t_a$, $S(t=t_a)\propto\log(t_a)$. Focusing on a finite system~\cite{Canovi2014}, there is a crossover between those two regimes, where, in the adiabatic regime, the entanglement entropy at the critical point is $S\propto\log(N)$~\cite{Calabrese2004}. Much less is known about entanglement scaling as a function of number of qubits and quench time $t_a$ in anneal quenches in dimensions $d>1$.

Accordingly, this work relies mostly on MPS to reveal the scaling behavior of entanglement entropy in anneal quenches for lattice topologies with dimensions $d>1$. Section~\ref{QPUMPSexp} analyzes the behavior of the QPU-equivalent bond dimension, $\chi_Q$, in terms of the correlation error, $\epsilon_c$, and ground-truth bond dimension. We further show how we extrapolated our MPS results in bond dimension in order to to establish the ground truths.

\section{Comparing and evaluating QPU and MPS experiments}\label{QPUMPSexp}

In our experiments, we estimate bipartite von Neumann entanglement entropy $S_A$ using MPS, with our main practical constraint being the exponential growth of $\chi_Q$ as a function of MPS area.  The main quantity of interest is the peak entanglement $S_\text{max}$, which is the maximum both over MPS cuts $(A,B)$ and over annealing parameter $s=t/t_a$.

\subsection{Estimating entanglement entropy}

\begin{figure*}
    \includegraphics[scale=1]{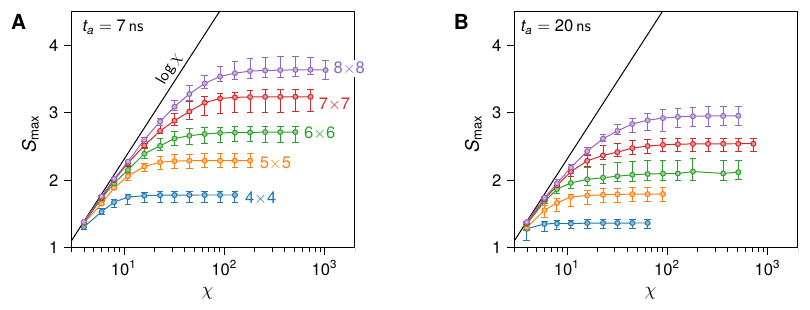}
  \caption{{\bf Convergence of peak entanglement entropy.} Entanglement $S_\text{max}$ grows as a function of MPS bond dimension $\chi$, converging in the $\chi\rightarrow\infty$ limit.  {\bf A} and {\bf B} show this convergence in 2D models for $\SI{7}{ns}$ and $\SI{20}{ns}$ quenches, respectively.  The line $\log(\chi)$ indicates the upper bound of entanglement that can be expressed in an MPS of bond dimension $\chi$.}\label{fig:entanglement_convergence}
\end{figure*}

For both $S_{max}$ and $\chi_Q$, our approach is to grow $\chi$ in factors of $\sqrt{2}$ until quantities are converged.  Fig.~\ref{fig:entanglement_convergence} shows convergence of $S_\text{max}$ as a function of MPS bond dimension $\chi$ for 2D models with $t_a=\SI{7}{ns}$ and $t_a=\SI{20}{ns}$.  In general we find that $S_\text{max}$ converges to the $\chi\rightarrow \infty$ limit faster than $\chi_Q$ does.

\subsection{Evaluating ground-truth convergence}
\label{app:groundtruthconvergence1}

\begin{figure*}
    \includegraphics[scale=1]{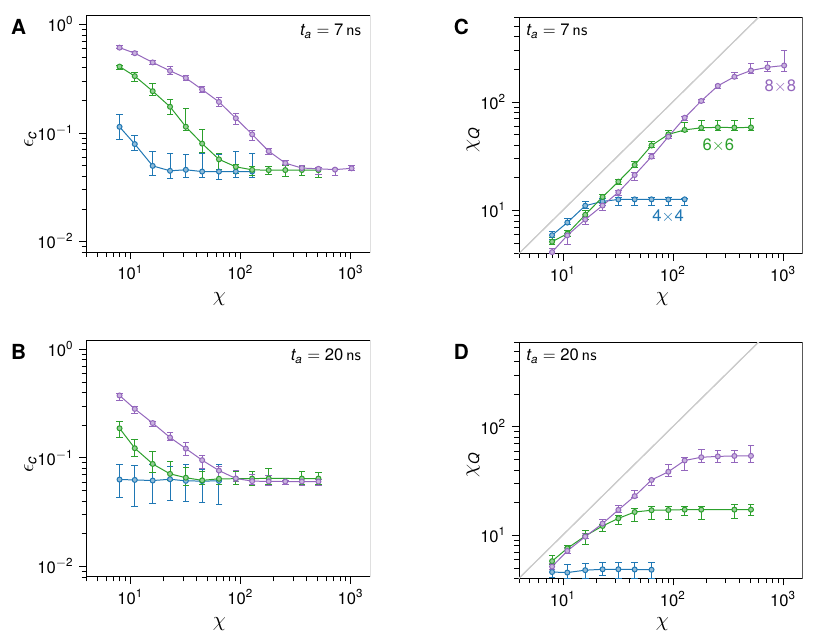}
  \caption{{\bf Convergence of error and equivalent bond dimension as a function of ground-truth bond dimension.}  As in Fig.~\ref{fig:entanglement_convergence}, we study square lattices at $t_a=\SI{7}{ns}$ and $t_a=\SI{20}{ns}$.  QPU error $\epsilon_c$  ({\bf A}--{\bf B}) and QPU-equivalent bond dimension $\chi_Q$ ({\bf C}--{\bf D}) are shown as a function of MPS bond dimension $\chi$.  The MPS used for the ground truth must have sufficient bond dimension such that these quantities are converged.  Gray lines indicate the upper bound $\chi_Q\leq \chi$.}\label{fig:chiQ_convergence}
\end{figure*}

The ground truth is ideally obtained in our experiment by taking the limit $\chi\rightarrow \infty$ in our MPS simulations. Our distribution accuracy measures---$\epsilon_c$, $\mathcal F$, and $\chi_Q$---can be shown to converge as a function of increasing $\chi$ approaching this limit much in the same way as entanglement entropy (see Fig.~\ref{fig:chiQ_convergence}).  

Statistics converge smoothly to an asymptote as shown, consistent with an exponential tail of truncated singular values~\cite{Vidal2003}.
Ideally we might observe a clear and sustained asymptote across a large range of $\chi$, indicating that singular value truncation is a negligible source of error.  For the hardest problems, this is not practical due to constraints on computational resources.  We therefore consider the possibility of extending the reach of MPS simulations by extrapolating in $\chi$.

\subsection{Extrapolation of bond dimension}

\label{app:bonddimensionextrapolation}
\begin{figure}
\includegraphics{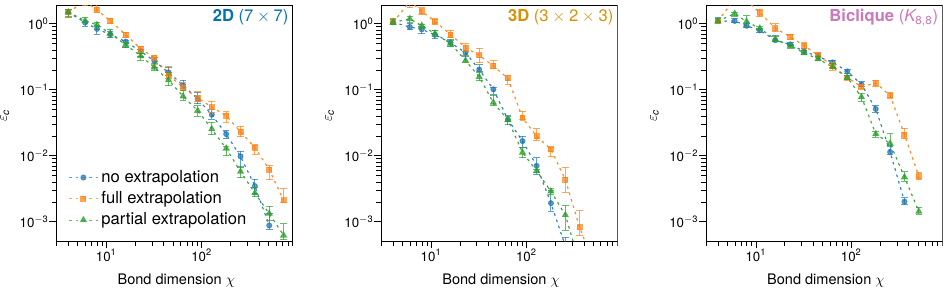}
    \caption{{\bf Effect of extrapolation in MPS bond dimension on correlation errors.}  MPS experiments of varying $\chi$ are evaluated under three types of two-point extrapolation of spin-spin correlations, in terms of their error against a well-converged ground truth with $t_a=\SI{7}{ns}$.}\label{fig:extrapolation}
\end{figure}

Since observables in MPS simulations converge as a function of $\chi$, it is often helpful to extrapolate to the limit of unbounded bond dimension ($\chi=\infty$).  In this work, we study problems that are near the edge of our computational capacity; we are therefore sufficiently far from convergence that extrapolation must be approached carefully.  Extrapolating using too-small bond dimensions is unreliable, so we restrict extrapolation to linear extrapolation of spin-spin correlations in $1/\chi$ based on the largest two bond dimensions available.  With this restriction, extrapolation to $\chi=\infty$ (``full extrapolation'') tends to increase errors, but we find that extrapolating from $\chi_\text{max}/\sqrt{2}$ and $\chi_\text{max}$ to $\sqrt{2}\chi_\text{max}$ (``partial extrapolation'') generally reduces errors, particularly in the important region $\epsilon_c \approx 0.03$, where errors are the same order of magnitude as QA errors.  Fig.~\ref{fig:extrapolation} shows examples of 2D, 3D, and biclique ensemble errors for no extrapolation, full extrapolation, and partial extrapolation.  Errors are evaluated against the ground truth of maximum available bond dimension $\chi_\text{max}$ with no extrapolation.  In all cases, partial extrapolation gains at best a factor $\sqrt{2}$ in bond dimension.  For this reason, although partial extrapolation is useful for refining ground-truth statistical estimates, it does not change the competitiveness of approximate MPS methods; we therefore restrict the use of extrapolation to ground-truth estimation.

\subsection{Effects of time step and stopping point}

\begin{figure*}
    \includegraphics[scale=1]{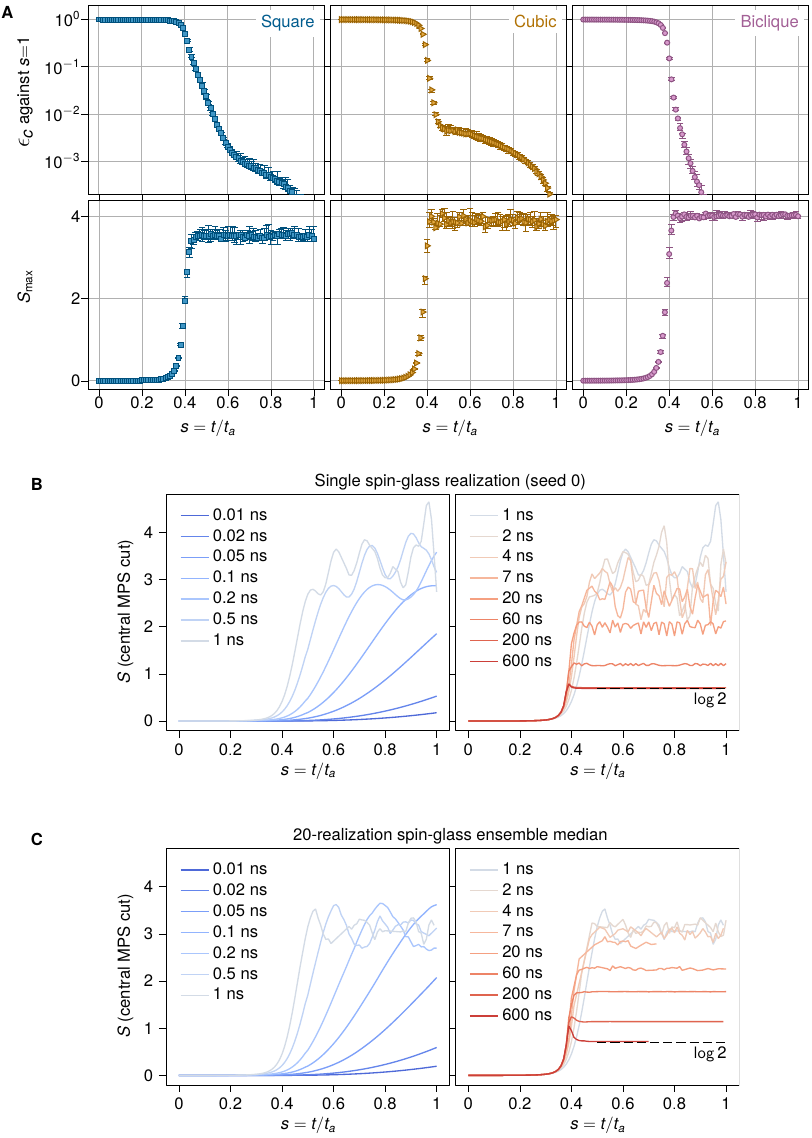}
  \caption{{\bf MPS error from stopping at $s<1$, compared with growth of entanglement entropy.} ({\bf A}) (Top) We plot median correlation error of MPS correlations throughout the quench, using a reference ground truth at $s=1$.  Square ($8{\times}8$, left), cubic ($3{\times}3{\times}3$, middle), and biclique ($K_{8,8}$, right) ensembles ($t_a=\SI{7}{ns}$, $\chi=1024$) approach the $s\rightarrow 1$ limit in varying manner. (Bottom) Growth of ensemble-median $S_\text{max}$ (maximum over MPS cuts for each instance) saturates when  $\epsilon_c$ is near $0.1$, higher than QPU error, consistent with Fig.~\ref{fig:4}C (inset).  ({\bf B}) Bipartite entanglement entropy in a single $8{\times}8$ spin-glass instance, taken across the central MPS cut (32 spins on each side) as a function of $s$ for varying $t_a$. Even $t_a=\SI{1}{ns}$ is far slower than the speed-limited regime in which entanglement grows monotonically as a function of $s$. ({\bf C}) Central-cut entanglement entropy, median over 20 $8{\times}8$ spin-glass instances.  All 20 inputs have only two classical ground states, related by spin-flip symmetry, so the ground state as $s\rightarrow 1$ is the GHZ state between them, with $S=\log{2}$.}\label{fig:s_cutoff}
\end{figure*}

\begin{figure*}
    \includegraphics[scale=1]{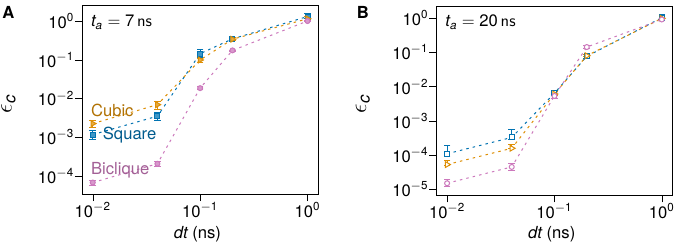}
  \caption{{\bf MPS error by time step $dt$.} ({\bf A}) $t_a=\SI{7}{ns}$ and ({\bf B}) $t_a=\SI{20}{ns}$.  Correlation error is plotted as a function of $dt$, against extrapolation to $dt\rightarrow 0$.  System sizes are $8{\times}8$ square, $3{\times}3{\times}3$ cubic, and $K_{8,8}$ biclique.}\label{fig:timestep}
\end{figure*}

The main parameter we have tuned to trade speed against accuracy in MPS methods is $\chi$.  We found MPS incapable of matching QPU quality for the largest simulations due to both time and memory requirements, and showed that nearly all entanglement entropy must be captured in the MPS to match QPU quality.  In Fig.~\ref{fig:s_cutoff}A we show that the time evolution of the quench cannot be truncated early---before nearly full entanglement is reached, requiring explosion of the bond dimension---without increasing correlation errors above QPU levels.  Therefore some time can be saved by stopping the simulation early, sometimes at a small cost to accuracy, but it will not be useful in reducing memory requirements.  Fig.~\ref{fig:s_cutoff}B shows entanglement dynamics for a single MPS cut in a single instance as a function of $s$.  Oscillations in $S$ after the critical point are hidden when taking the ensemble average and maximum over all cuts.

Throughout the work we have generally used $dt=\SI{0.01}{ns}$; in Fig.~\ref{fig:timestep} we verify that this time step is not a leading source of error.

\subsection{Oblong 2D lattices}
\begin{figure*}
    \includegraphics[scale=1]{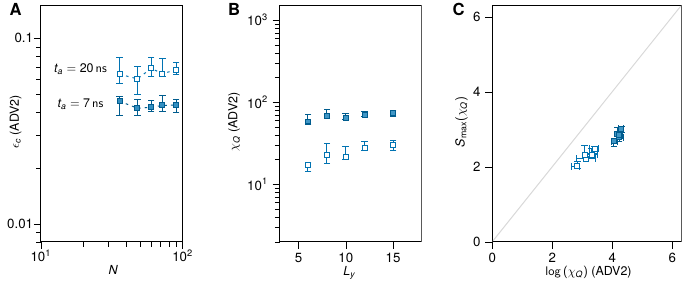}
  \caption{{\bf QPU error and entanglement in oblong cylinders.} We study $L_x{\times} L_y$ systems with $L_x=6$.  Data are analogous to Fig.~\ref{fig:4}; filled and empty markers indicate $t_a=\SI{7}{ns}$ and $\SI{20}{ns}$ respectively.  {\bf A}, QPU correlation error $\epsilon_c$ is flat as a function of $N = 6L_y$.  {\bf B}, $\chi_Q$ is nearly flat for $t_a=\SI{7}{ns}$, but shows a slight scaling for $t_a=\SI{20}{ns}$.  {\bf C}, $S_\text{max}$ and $\log(\chi_Q)$ are similarly related as in other ensembles, with $S_\text{max}$ increasing gently with $L_y$.  In the limit $L_y\rightarrow \infty$, we expect $S_\text{max}$ to increase as a function of $t_a$; larger systems are required in this range of $t_a$.}\label{fig:oblong}
\end{figure*}

We are interested in the scaling of QPU error and entanglement as a function of system size, but for the ensembles studied in the main text, this quickly becomes intractable.  One compromise is to study quasi-1D models, which should have favorable scaling in MPS bond dimension requirements.  Here we provide data on $6{\times} L_y$ cylinders for $6\leq L_y \leq 15$, whose MPS area remains constant at $6$.

As in the main text, we once again see flat scaling of QPU error $\epsilon_c$ as a function of system size (Fig.~\ref{fig:oblong}A); the bounded MPS width allows extending to larger systems than before.  As expected, QPU-equivalent bond dimension $\chi_Q$ remains almost constant, with a mild scaling evident at $t_a=\SI{20}{ns}$.  In all other systems we saw $\chi_Q$ (and $S_\text{max}$) decreasing as a function of $t_a$, suggesting that we are near the adiabatic, finite-size regime, as opposed to a large-system Kibble-Zurek regime.  The data in Fig.~\ref{fig:oblong}B hint that we may see a crossover for larger systems, as was previously observed in 1D chains~\cite{Canovi2014}.

\subsection{Area-law scaling for slower quenches}

\begin{figure*}
    \includegraphics[scale=1]{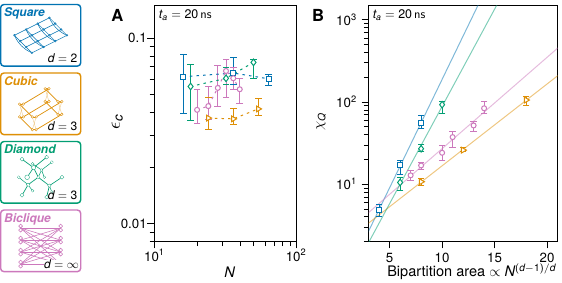} 
  \caption{{\bf Correlation error and area-law scaling of $\chi_Q$ at $t_a=\SI{20}{ns}$.} ({\bf A}) and ({\bf B}) are analogous to Fig.~\ref{fig:4}A and B respectively.}\label{fig:4_20ns}
\end{figure*}

Fig.~\ref{fig:4_20ns} shows data analogous to Fig.~\ref{fig:4} for $t_a=\SI{20}{ns}$ instead of $\SI{7}{ns}$.  We note that correlation errors $\epsilon_c$ are slightly higher for all topologies, and show less size-dependent growth for biclique problems than at $\SI{7}{ns}$.  In all topologies the longer quenches have lower $\chi_Q$, which allows us to evaluate larger lattices in some cases.

%% file: sm_neural_net_methods.tex
\section{Neural networks}
\label{app:NN}
Neural network quantum states (NQS) offer a powerful approach to constructing computationally efficient ans\"atze wave functions. 
Several types of neural networks, such as complex restricted Boltzmann machines (cRBMs)~\cite{carleo2017solving, czischek2018quenches}, feed-forward neural networks (FNNs)\cite{roth2021group, schmitt2020quantum,schmale2022efficient}, recurrent neural networks (RNNs)~\cite{hibat2020recurrent, hibatallah2022supplementing, hibatallah2023topological, lange2023neural}, and transformers~\cite{carrasquilla2021prob, sprague2023variational, rende2023simple}, have been employed to approximate ground states and out-of-equilibrium dynamics of quantum many-body systems. For the ground-state problem, the variational Monte Carlo method~\cite{BeccaSorella2017} provides a numerical method for ground-state estimation that is competitive with other state-of-the-art approaches such as tensor networks or quantum Monte Carlo~\cite{wu2023variational}. Crucial for the success of this method is the ability to sample from the NQS.  Whereas cRBMs and FNNs parameterize unnormalized quantum states and are sampled using Markov chain Monte Carlo, autoregressive models such as RNNs and transformers can be sampled using direct sampling. The latter come with a drawback, since recent results indicate that the autoregressive property of the models may restrict the expressiveness of the model~\cite{lange2023neural, bortone2023impact}.

Beyond ground states, it is possible to simulate the Schr\"odinger equation on the variational manifold induced by the neural network ansatz. There are currently two approaches to simulate real-time dynamics. The first method is called time-dependent variational Monte Carlo (tVMC)~\cite{carleo2012localization}. For an NQS $\ket{\psi_{\theta_t}}$ with $p$ parameters $\theta$, we calculate the force vector $F$ at time $t$, which depends on the variational energy of the Hamiltonian driving the dynamics. The parameter update to the network for a small time step is then given by
 \begin{align}
    \dot{\theta} = S^{-1} F,\label{eq:tVMC}
\end{align}
which is closely related to the second-order optimization method of natural gradient descent~\cite{amari1998natural}.
Here, $S$ is the metric on the variational manifold induced by the Fubini-Study metric, which is the natural metric on a Hilbert space~\cite{hackl2020geom}. As $S$ is a $p {\times} p$ matrix, its inversion takes time $O(p^3)$, which significantly limits the number of variational parameters that can be used. In addition, both $S$ and $F$ must be accurately estimated by sampling from the variational state. For a stable simulation, one typically requires $N_{\mathrm{s}}\sim 10^4$ samples per time step. In practice, the matrix $S$ is typically ill-conditioned and needs to be regularized, for which there exist several approaches~\cite{carleo2017solving, medvidovic2021classical,schmitt2020quantum}. The stability of the regularization depends on many factors, including the model architecture, the current variational state and the Hamiltonian. Regularization problems are particularly acute for the RNN architecture. The authors of Ref.~\onlinecite{Donatella2023} argue that RNNs fail for tVMC due to problems with regularizing the $S$ matrix and recent results indicate that the autoregressive property of the model may restrict its expressiveness~\cite{bortone2023impact}. Once the $S$ matrix is successfully regularized, we can solve the differential equation in Eq.~\ref{eq:tVMC} for a small time step $\tau$ with a second-order differential-equation solver (e.g., Heun's method or Runge-Kutta-Fehlberg~\cite{Press1992numrecipes}).

The second approach for time evolution of variational quantum states is to learn the effect of the time-evolution propagator $\mathcal{U}$ on our variational state~\cite{barison2021efficient,medvidovic2021classical,gutierrez2022real, Donatella2023, sinibaldi2023unbiasing}. We refer to this method as projected time-dependent variational Monte Carlo (ptVMC). Here, instead of solving a differential equation at each time step, we explicitly minimize the distance, or infidelity, between the variational state at time $t$ and time $t+\tau$:
\begin{align}\label{eq:infid}
    \min_{\theta_{t+\tau}} d(\ket{\psi_{\theta_{t+\tau}}}, e^{i\tau H}\ket{\psi_{\theta_t}}).
\end{align}
We see that there are two variational states: the state $\psi_{\theta_t}$ at time $t$ and the state  $\psi_{\theta_{t+\tau}}$. We ``freeze" the parameters $\theta_t$ and optimize the parameters $\theta_{t+\tau}$.
Since $e^{i\tau H}$ is not a local operator that we can easily evaluate on our variational wave function, in practice we have to approximate the full unitary propagator, which can be done in several ways~\cite{gutierrez2022real, Donatella2023, sinibaldi2023unbiasing}. In order to efficiently estimate the infidelity, the authors of Ref.~\onlinecite{sinibaldi2023unbiasing} propose a control-variate method to reduce the variance of the estimator similar to Ref.~\onlinecite{hibat2020recurrent}. This estimator is quite efficient and, in our experiments, requires only order $N_{\mathrm{s}}=10^2$ samples per optimization step. However, this comes at the cost of having to minimize the above distance, which could take on the order of $10^3$ steps in our experiments. Additionally, to achieve stable results at long times, a second-order approximation to the unitary propagator is required, which results in a quadratic increase in complexity that becomes prohibitive at larger system size.
In ptVMC, we have traded the sampling problem of tVMC for an optimization problem. 
The advantage of ptVMC over tVMC is that the former is unconstrained in terms of parameters, which opens up the possibility of using highly expressive models for the simulation of real-time quantum dynamics. 

For the problem at hand, which is the simulation of non-adiabatic critical dynamics of the TFIM, the Kibble-Zurek mechanism has been studied with an NQS approach that uses tVMC~\cite{Schmitt2022}. Additionally, Ref.~\onlinecite{Donatella2023} studied critical dynamics with ptVMC. There are two major differences between previous investigations of critical dynamics and the work presented here. First, the Ising model considered has no symmetries (apart from $\mathbb{Z}_2$), hence we cannot make use of translation or lattice symmetries, and we have to use a general variational ansatz. Second, the time scales considered here are beyond what has previously been investigated in the literature. In both Ref.~\onlinecite{Schmitt2022} and Ref.~\onlinecite{Donatella2023}, NQS approaches are only used close to the critical point. In particular, Ref.~\onlinecite{Schmitt2022} notes the NQS method was not stable in the ferromagnetic regime at long times. Finally, a recent work on the simulation of Rydberg arrays was unable to achieve stable results with NQS~\cite{Shaw2023}.

\begin{figure}[htb!]
    \centering
    \includegraphics[width=0.8\textwidth]{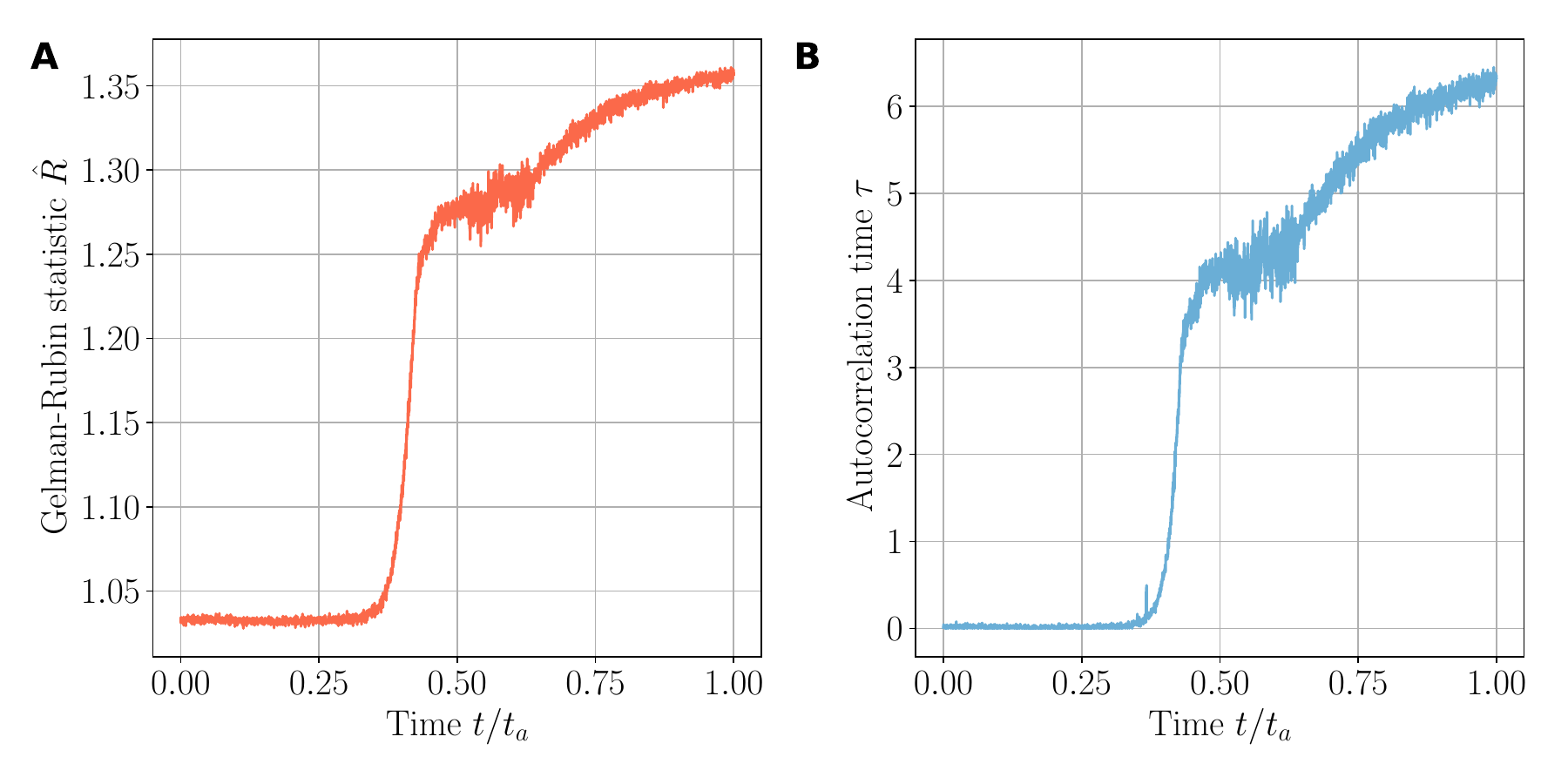}
    \caption{{\bf Monte Carlo convergence statistics for an $8{\times} 8$ instance at \SI{7}{ns}}. \textbf{(A)} Gelman-Rubin statistic ($\hat R$) plotted versus time~\cite{Gelman1992InferenceFI}. A value of $<1.1$ indicates that Markov chains are converging~\cite{gelman2013}. 
    \textbf{(B)} Autocorrelation time ($\tau$) of the Markov chain. Longer autocorrelation times mean that more sweeps are required to obtain uncorrelated samples.
    Both statistics indicate that as we pass through the critical point, the Monte Carlo sampling becomes more difficult, resulting in a higher number of samples required to obtain the correct dynamics.}
    \label{fig:sampling_hard}
\end{figure}
\begin{figure}[htb!]
    \centering
    \includegraphics[width=0.9\textwidth]{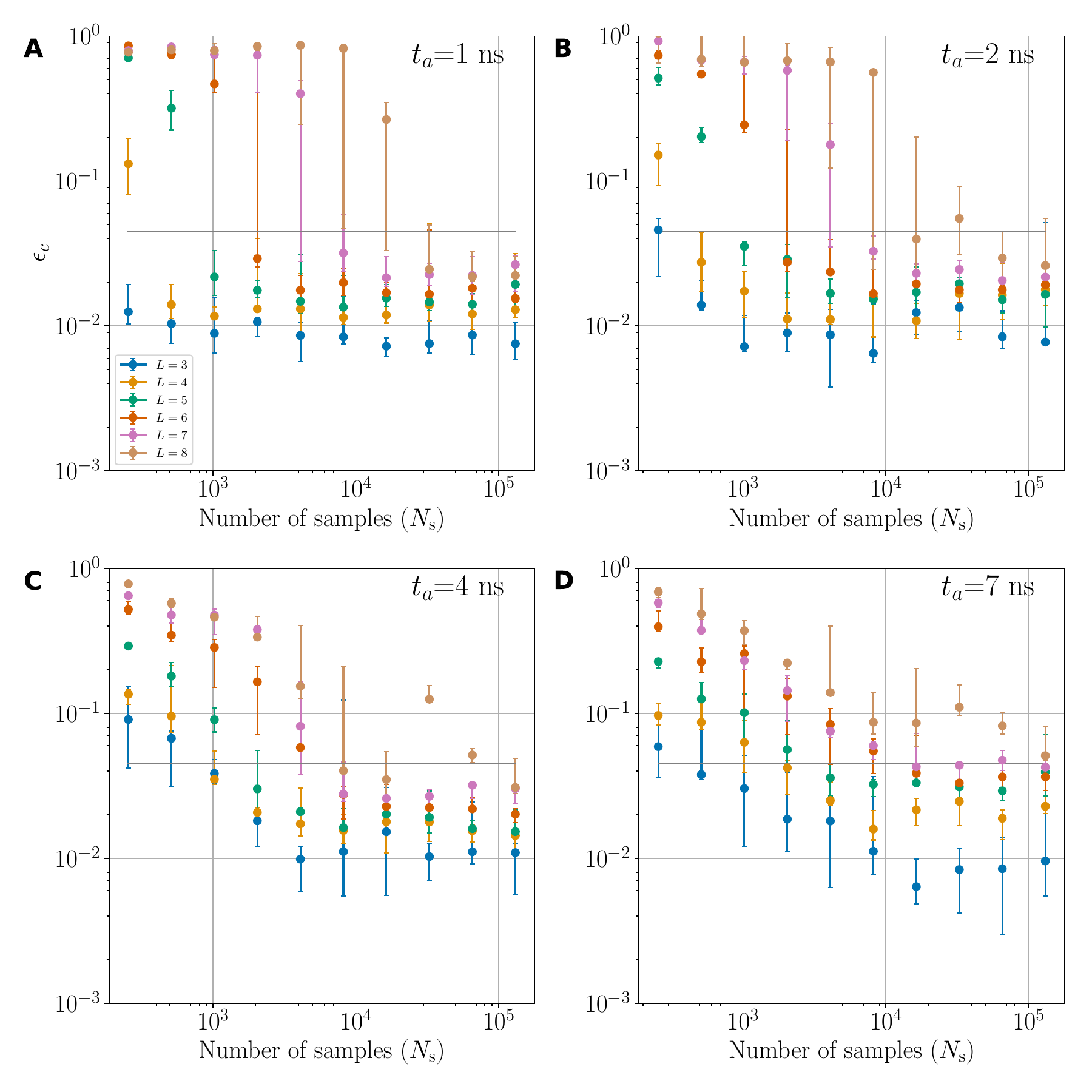}
    \caption{{\bf Average correlation error for a system of size $L{\times} L$  using the cRBM.} The correlations are estimated with a total of $10^6$ samples (equivalent to QPU). Error bars represent the standard deviation across 5 different disorder realizations. We focus on correlations at $s=0.6$ due to the smallest correlation error observed at this point (see also Figure~\ref{fig:s_cutoff}). Depending on the system size and number of samples $N_s$, we increased the number of GPUs from 1 to 8 and implemented a chunking scheme that splits up the total number of samples into batches. This reduces memory requirements at the expense of increased compute time. (\textbf{A-C}) For times up to $\SI{4}{ns}$, matching the QPU error is achievable for larger system sizes by increasing $N_s$. (\textbf{D}) As system sizes grow, matching the QPU error becomes more challenging. For all time scales, we observe that the error plateaus after reaching a certain sample count.}
    \label{fig:tvmc}
\end{figure}

We note that before we start the dynamics, the state of the neural network is brought to the ground state of $\mathcal{H}(0)$ via variational Monte Carlo~\cite{BeccaSorella2017}, which converges rapidly to the desired paramagnetic state. In our simulations, we consider only the high-precision square lattices (see Section~\ref{app:testbed}). 
Once the ground state is found, we can attempt to simulate the dynamics with either tVMC or ptVMC.  With tVMC, we have been able to simulate systems up to $8{\times} 8$ at a time scale of $\SI{7}{ns}$ at significant computational cost (3-4 days of compute time with multiple Nvidia T4 GPUs per instance).
We chose a fixed time step of $\tau = 10^{-2}$ in units of $\Gamma_c \cdot \pi$ ($\sim \SI{4.32}{ps}$), where $\Gamma_c$ is set to the point in the schedule where $\Gamma(s) = \mathcal{J}(s)$, which is where the dynamics are critical. Going beyond $8{\times} 8$ is possible for shorter annealing times $t_a < \SI{7}{ns}$ but would require a large number of GPUs to obtain the necessary samples in a reasonable time. We experimented with adaptive Runge-Kutta-Fehlberg schemes to adaptively set the time step $\tau$. This could significantly reduce the computational cost, since we could speed up the integration at the beginning and slow down as we cross the critical point. However, we were not able to reach stable results for large system sizes with such an approach. The most successful architecture was a cRBM model with the number of hidden units equal to the number of spins (see Figure~\ref{fig:tvmc}). Due to the $O(p^3)$ complexity of inverting $S$, we limited the number of hidden units to $N$ ($\alpha=1$), which was sufficient for the system sizes considered here. 
 
While a Jastrow ansatz~\cite{BeccaSorella2017} proved to be much more numerically stable than the cRBM, it was not expressive enough to capture the dynamics of the system, resulting in larger errors than the cRBM. Feed-forward neural networks with multiple layers showed promise for small system sizes ($L<5$), but failed at larger system sizes. Similarly, convolutional neural networks were ineffective, most likely due to the lack of translational symmetry of the spin-glass Hamiltonian considered here.  
Autoregressive architectures like the RNN and transformers came with $S$-matrix spectra that were hard to regularize, which resulted in the wrong dynamics.

The main bottleneck in the simulation of the dynamics is the estimation of the $S$ matrix in Eq.~\ref{eq:tVMC}. Here, we used a Metropolis-Hasting sampling method with local updates for the Markov chain, which can be parallelized over multiple GPU machines via Message Passing Interface (MPI). As we cross the critical point, the sampling becomes very challenging due to the glassy nature of the system (see Figure~\ref{fig:sampling_hard}). The long auto-correlation times result in an increase in variance of the estimators of the forces and $S$ matrix, necessitating a larger sampling budget to keep errors under control. For regularization of the $S$ matrix, we explored both the method by Schmitt et al.~\cite{schmitt2020quantum} and Medvidovi\'c~\cite{medvidovic2021classical}, and found that the latter was the most stable at $r_\mathrm{cond} = 10^{-5}$.

Autoregressive models such as the RNN and transformer fail for tVMC. Attempts to perform dynamics with these models with ptVMC indicate that this method has large (but practically feasible) costs, since the number of steps required to minimize the infidelity for each $\tau$ can be of order $10^3$. However, a major difficulty with this approach is that the infidelity optimization of Equation~\ref{eq:infid} can fail to converge, which leads to numerical instabilities. As a result of these convergence issues, we were not able to simulate system sizes larger than $L=5$. Experiments with the cRBM using ptVMC were much more stable, but not competitive with the tVMC simulations in terms of accuracy, especially at larger system sizes.

Our studies have been limited to square lattices at limited space and quench times; square lattice experiments at \SI{20}{ns} proved too costly to simulate. If we assume that simulation costs scale up as a function of the correlated volume (consistent with square lattice data from \SI{1}{ns} to \SI{7}{ns} in Figure \ref{fig:tvmc}), an NQS approach to more-connected topologies on experimental time scales may exhaust GPU resources at even shorter time scales.
All numerical simulations were performed with NetKet~\cite{netket}, a Python framework built on top of JAX~\cite{jax2018github}.

%% file: sm_classical_compute.tex
\section{Classical compute resources}\label{app:compute}

\begin{figure}
\includegraphics[scale=1.2]{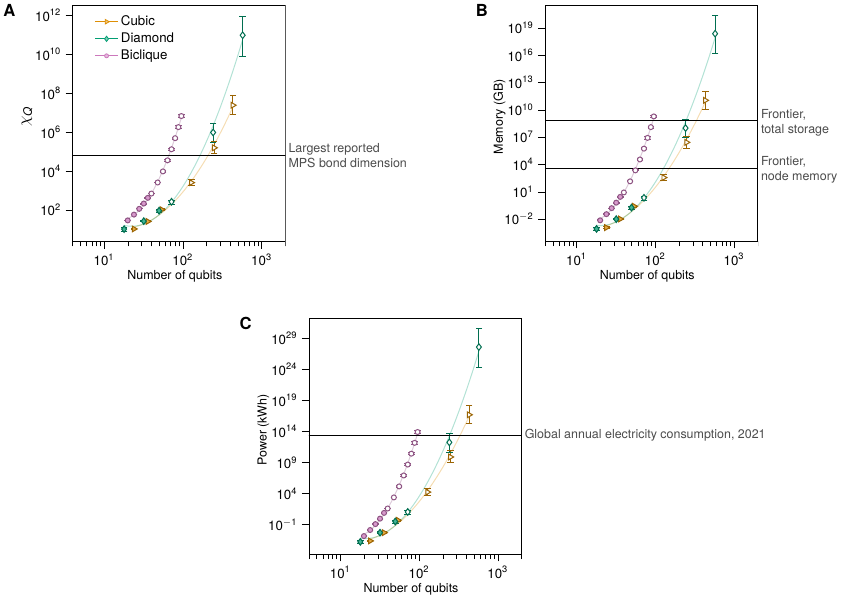}
  \caption{{\bf Resource requirements for MPS to match QPU quality.} ({\bf A}) Area-law extrapolation is used to estimate the QPU-equivalent bond dimension $\chi_Q$ required by MPS methods to achieve QPU-level accuracy.  Values far exceed the largest reported MPS bond dimension to date~\cite{DMRGtpu}. ({\bf B}) Dominant scaling of memory is $\propto N\chi_Q^2$; assuming this form we compute MPS memory requirements, which exceed both Frontier node flash memory ($\SI{4}{TB}$) and total storage of the Frontier filesystem ($\SI{700}{PB}$). ({\bf C}) Power, like computation time, scales $\propto t_aN\chi_Q^3$.  Here, GPU-time requirements (Fig.~\ref{fig:5}D) are converted to power requirements based on \SI{300}{W} draw of an Nvidia Tesla V100.  These requirements far exceed the total global electricity consumption throughout human history.  All plots use $t_a=\SI{7}{ns}$ for biclique problems and $t_a=\SI{20}{ns}$ for diamond and cubic problems.}\label{fig:memory}
\end{figure}

This work includes experiments run on a heterogeneous collection of classical CPU and GPU compute resources.  These include the Summit and Frontier supercomputers at the Oak Ridge Leadership Computing Facility, as well as clusters at the Vector Institute, University of British Columbia, Jagiellonian University, and D-Wave.

The main MPS experiments were run on Summit and Frontier, whose workhorse GPUs are, respectively, Nvidia Tesla V100 and AMD Radeon Instinct MI250X. The NQS experiments were performed on multiple Nvidia T4 cards.

Our extrapolation of compute resources is based on measurements from Summit for a 54-qubit system, estimated to require $\SI{12}{GB}$ of memory and \num{7e4} seconds at a bond dimension $\chi=724$ and $t_a=\SI{7}{ns}$.  These figures are extrapolated to larger systems and bond dimensions with the leading scaling forms
\begin{equation}
    \text{memory} \propto N\chi^2
\end{equation}
and
\begin{equation}
    \text{time} \propto Nt_a\chi^3.
\end{equation}
since we use a constant time step $dt=\SI{0.01}{ns}$ throughout (see Fig.~\ref{fig:timestep}).

The V100 GPUs advertise 14 teraFLOPS in single precision; we compare this with a peak performance of 1.7 exaFLOPS for Frontier to produce the comparison in Fig.~\ref{fig:5}D stating that matching QPU quality on the largest systems would exceed millions of years using all of Frontier.

This is of course an oversimplification in many respects, most importantly ignoring the practical constraints on such computations, which are far too large to perform naively on a GPU.  Each Frontier compute node has \SI{4}{TB} of memory, and the total storage is \SI{700}{PB}. These figures are marked in Fig.~\ref{fig:memory}.  We note that the largest bond dimension simulated to date\cite{DMRGtpu} is $2^{16}=\num{6.5e4}$.

%% file: sm_critical_exponents.tex

\section{Critical exponents and dynamic finite-size scaling ansatz}\label{sec:dfss}
\label{app:dfss}

\begin{table}[t!]
\centering\setlength{\tabcolsep}{12pt}
\begin{tabular}{r|ccc}
& Square & Cubic & Biclique\\\hline
$z$, literature &$2.46(17)$  (Ref.~\onlinecite{Bernaschi2023})&$1.3$ (Ref.~\onlinecite{Guo1994})& $2$ (Ref.~\onlinecite{Read1995})\\
$1/\nu$, literature&$0.71(24)$ (Ref.~\onlinecite{Bernaschi2023})&$1.55(11)$ (Ref.~\onlinecite{King2023})&$4$ (Ref.~\onlinecite{Read1995})\\
$\mu$, literature &{\bf 3.17}$(41)$ & {\bf 2.85}$(21)$  & {\bf 6}\\
$\mu$, experiment, all sizes, low precision  & $2.67(30)$ & $2.65(12)$ & $6.1(11)$\\
$\mu$, experiment, all sizes, high precision & $2.54(13)$& $2.77(12)$ & $6.5(7)$\\
$\mu$, experiment, largest sizes, low precision  & $2.99(19)$ & $2.65(7)$ & $6.9(79)$  \\
$\mu$, experiment, largest sizes, high precision  & $2.67(29)$ & $2.79(24)$ & $6.9(49)$\\
$\mu$, experiment  & {\bf 2.83}$(24)$& {\bf 2.72}$(16)$& {\bf 6.3}$(9)$\\
\end{tabular}
\caption{{\bf Critical exponents}. We compare experimental (QPU) measurements of $\mu$ based on collapse of the Binder cumulant $U$ against predictions from independent estimates based on classical Monte Carlo (square and cubic) and theoretical (biclique) for the three relevant universality classes.  We perform collapses using all sizes available (square: $L=4$ to $18$; cubic: $L=3$ to $6$; biclique: $M=8$ to $24$), and the three largest sizes available for each topology, to take into account small-size distortions.  For square and cubic our final estimate is the large-size average between high and low precision.  For biclique our final estimate is the all-size average between high and low precision, since large-size estimates have very large uncertainties.}
\label{table:exponents}
\end{table}

In this section we briefly explain the dynamic finite-size scaling analysis of Fig.~\ref{fig:5}B, which generalizes the approach of Ref.~\onlinecite{King2023} from cubic to multiple topologies.

A crucial element of any claim of beyond-classical performance is evidence that experimental results scale correctly beyond the classically verifiable regime.  Since we simulate traversals of quantum phase transitions (QPTs), we can appeal to the existing theoretical foundation of critical phenomena to provide this evidence.  For large problems, i.e., in the thermodynamic limit, the Kibble-Zurek mechanism~\cite{Zurek1985,Kibble1976,Dziarmaga2010,Polkovnikov2011,delCampo2014,Rossini2021} allows us to predict the scaling of correlations and related quantities---including the Binder cumulant---as a function of quench time.  More specifically, quantities will be related by the universal critical exponents reflecting the universality class of the QPT.  This approach has previously been used to compare quantum simulations with theoretical expectations in superconducting quantum annealers\cite{Weinberg2020,King2022,Bando2020,King2023} and annealed Rydberg arrays\cite{Keesling2019,Ebadi2020}.

Fig.~\ref{fig:5}B shows data collapses for the Binder cumulant 
\begin{equation}
U = \frac 12 \left(3-\frac{[\langle q^4\rangle]}{[\langle q^2\rangle]^2} \right),
\end{equation}
where $[ \cdot ]$ indicates an average over disorder realizations.
The Binder cumulant gives a statistical signature of phase transitions.  In this work, we use $U$ to estimate---via collapse---the Kibble-Zurek exponent
\begin{equation}
    \mu = z+1/\nu,
\end{equation}
where $z$ and $\nu$ are the dynamic exponent and the equilibrium correlation-length exponent pertaining to the QPT at hand\cite{Sachdev2011}.  That is, the paramagnetic-to-spin-glass QPT in the universality class for each topology.  These are 2D Ising, 3D Ising, 3D Ising, 3D Ising, and mean-field (fully connected) for square, cubic, cubic no-dimer, diamond, and biclique, respectively.

At the critical point in the quench, it is expected\cite{Liu2015} that both $\langle q^2\rangle$ and $U$ should collapse (separately) as a functions of $L$ and $t_a$.  While $\langle q^2 \rangle$ is less statistically sensitive, collapse of $U$ is preferable here for two reasons.  First, $\langle q^2\rangle$ was found\cite{King2023} to be susceptible to distortion from post-critical dynamics between the QPT and the end of the quench, where the output state is read.  Second, $\langle q^2 \rangle$ must be collapsed as a function of both $\mu$ and the anomalous dimension $\eta$; the collapse of $U$ is only in one parameter $\mu$, making the fit more straightforward in the face of limited system sizes and uncertainty in $\eta$.

Thus we measure $U$ for multiple linear system sizes $L$ and quench times $t_a$.  For each topology (and treating high-precision and low-precision ensembles separately), we expect data collapse when plotting $U$ against $t_aL^{-\mu}$.  We therefore minimize deviation between the collapsed data and a common collapse curve of the parametric form
\begin{equation}
    f(x) = a_0 + a_1 \log{(1+\exp{(a_2(x-a_3)}))}
\end{equation}
in the ($\log{(U)}$, $\log{(t_aL^{-\mu})}$) plane (as in~Ref.~\onlinecite{King2023}, Fig.~3c--d), where for bicliques we take $L=M^{1/8}=(N/4)^{1/8}$ based on the upper-critical dimension 8 (Ref.~\onlinecite{Read1995}).  Estimates of $\mu$ can be impacted by the effect of small inputs in the best-fit collapse; we therefore estimate $\mu$ using the three largest simulated sizes of each topology, as well as estimating using all sizes.  The results, along with related literature values, are summarized in Table~\ref{table:exponents}.

Precise estimates of $U$ require averaging over many spin-glass realizations (often called ``samples'' in the spin-glass community).  Therefore instead of focusing on 20-instance ensembles, which allow us to keep classical compute resources within reason, we simulate many realizations.  For each of 300 QPU calls, we generate a distinct random realization for each individual parallel embedding into the qubit lattice.  Thus the number of spin-glass instances simulated in QA for this analysis ranges from 300 (for the largest square and cubic instances) to 20,400 (for the smallest instances, where many random realizations are run in parallel in each QPU call).

Previous studies have provided independent estimates of $z$ and $1/\nu$, allowing us to validate the scaling of correlations in the system beyond the classically verifiable regime.  We compare our experimental estimates of $\mu$ against the results of classical Monte Carlo simulations (for square and cubic) and mean-field theory (for biclique).

For 2D and 3D, there has been some controversy over whether or not $z$ is finite.  For 2D, this was settled very recently\cite{Bernaschi2023} through the consideration of distinct parity sectors in the state space, using MC simulations of low-precision $L{\times} L$ square lattices (with fully periodic boundaries) with $L$ ranging from $8$ to $24$. The authors extracted distinct estimates for the finite-size measurement (fixing their fitting parameter $\eta_e^0=0$) and the large-system limit of, respectively,
\begin{equation}
    \mu = z + 1/\nu = 1.7(2) + 0.9(1) = 2.6(3)
\end{equation}
and 
\begin{equation}
    \mu = z + 1/\nu = 2.46(17) + 0.71(24) = 3.27(41),
\end{equation}
both of which are in remarkably good agreement with our all-size and largest-size estimates of $\mu$ from low-precision glasses ($2.67(30)$ and $2.99(19)$ respectively).  Future experiments in a larger QPU might clarify finite-size effects in these estimates.  The classical estimates of Ref.~\onlinecite{Bernaschi2023} updated the previous literature estimates\cite{Rieger1994} of $z=1.50(5)$ and $\nu=1.0(1)$---which are by now outdated in terms of computational reach---with extensive state-of-the-art GPU simulations.  When considering small-system effects, the three estimates of $\mu$ ($2.6(3)$, $2.5(2)$ and $2.67(30)$ from, respectively, Ref.~\onlinecite{Bernaschi2023}, Ref.~\onlinecite{Rieger1994}, and this work) concur extremely well.

For 3D, the finite-$z$ scenario is supported by considerable evidence, including the above result for 2D and large-scale experiments in quantum annealers (this work itself and Ref.~\onlinecite{King2023}), and we may reasonably take the position that $z$ is indeed finite.  The prevailing classical estimate of $z=1.3$ is, like the previous 2D estimate\cite{Rieger1994}, computationally outdated and could benefit from an extension of the methods and computational resources brought to bear in Ref.~\onlinecite{Bernaschi2023}.  The estimate of $1/\nu=1.55(11)$ has recently been updated for the purposes of evaluating QA\cite{King2023}.  Our estimates of $\mu$, summarized as $\mu = 2.72(16)$, are again in line with previous estimates.  The fact that this estimate is slightly lower than that reported by Ref.~\onlinecite{King2023} may be due to reduced noise in the ADV2 processor as compared to the ADV1 processor used previously, with possible influence of smaller system sizes used in the collapse in this work.

For biclique, we expect behavior in line with mean-field critical exponents $z=2$ and $1/\nu=4$, which would mean a collapse upon rescaling $t_a$ by $M^{-\mu/8} \approx M^{-3/4}$ (Ref.~\onlinecite{Read1995}).  There are large per-size fluctuations relative to the other topologies; this may be related to per-size configurations of inter-dimer coupling structure.  The estimates based on $M=8$ to $24$ are nonetheless very close to mean-field expectations of an infinite-dimensional topology.

\begin{table}[t!]
\centering\setlength{\tabcolsep}{12pt}
\begin{tabular}{r|cc}
& Cubic no-dimer & Diamond\\\hline
$\mu$, literature & {\bf 2.85}  &{\bf 2.85} \\
$\mu$, experiment, largest sizes, low precision  & $2.61(23)$ & $2.95(62)$  \\
$\mu$, experiment, largest sizes, high precision  & $2.98(12)$ & $2.96(15)$\\
$\mu$, experiment  & {\bf 2.80}$(18)$& {\bf 2.96}$(39)$\\
\end{tabular}
\caption{{\bf Critical exponents for cubic no-dimer and diamond glasses}. Estimates of $\mu$ are expected to be consistent with the 3D Ising value.}
\label{table:exponents_extra}
\end{table}

\begin{figure*}
    \includegraphics[scale=1]{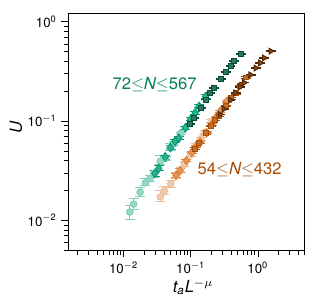}
  \caption{{\bf Dynamic finite-size scaling collapse for diamond and cubic no-dimer instances.}  Curves of collapsed Binder cumulant $U$ for diamond (green) and cubic no-dimer instances (orange).}\label{fig:collapse_sm}
\end{figure*}

In addition to the problem classes considered in the main text, we apply the same analysis to cubic no-dimer lattices, with collapses shown in Fig.~\ref{fig:collapse_sm} and compared with diamond lattices.  We expect both of these to exhibit behavior consistent with the 3D quantum Ising spin glass universality class.  For cubic no-dimer, we consider the same problem sizes as for cubic dimer.  Both cubic no-dimer and diamond inputs admit estimates of $\mu$ consistent with the 3D quantum Ising spin-glass universality class, as summarized in Table~\ref{table:exponents_extra}.  The slightly higher estimate compared to cubic dimer---while still within uncertainties---may be related to the larger effective system sizes.

\begin{figure*}
    \includegraphics[scale=1]{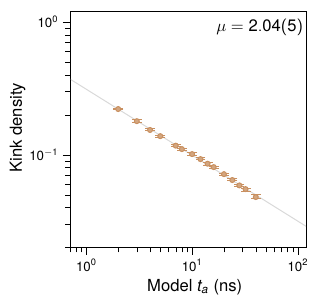}
  \caption{{\bf Kink density in pure 1D quantum Ising chain.}  Density of kinks (frustrated bonds) are shown as a function of $t_a$ for a pure antiferromagnetic 1D chain with length $L=512$.  Assuming a kink density proportional to $t_a^{-1/\mu}$ yields an estimate $\mu=2.04(5)$, in agreement with theory.}\label{fig:1dkink}
\end{figure*}

For 1D systems, we estimate $\mu$  by measuring kink density $\bar n$ as a function of $t_a$ for a large system ($L=512$, Fig.~\ref{fig:1dkink}).  As in previous works~\cite{King2022,Keesling2019}, the measured scaling exponent is consistent with the expected KZ defect scaling $\bar n \propto t_a^{-1/\mu}$ with $z=1$ and $\nu=1$.

Taken as a whole, the agreement in critical scaling, which relates time and correlation length via the Kibble-Zurek mechanism, provides indirect validation of QA results beyond the classically simulable regime.  This provides a complement to the precise quantitative agreement shown for small problems.

%% file: sm_monte_carlo_methods.tex

\section{Monte Carlo dynamics}
\label{app:MC}

\begin{figure*}
  \includegraphics[scale=1]{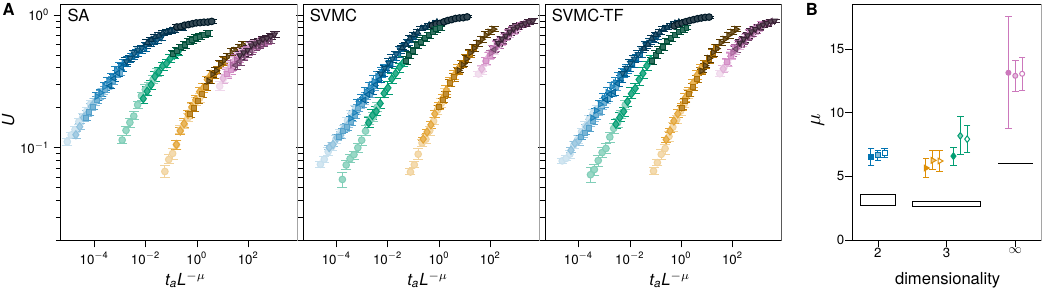}
  \caption{{\bf Dynamic finite-size scaling for Monte Carlo dynamics.}  ({\bf A}) Binder cumulant collapses for high-precision instances as in Fig.~\ref{fig:5}B for SA, SVMC, and SVMC-TF dynamics.  Blue, yellow, green, and purple data are from square, cubic, diamond, and biclique lattices respectively. ({\bf B}) Extracted KZ exponents $\mu$, with quantum estimates indicated in black.  For biclique problems we use the quantum upper-critical dimension of 8 rather than the classical value of 6, since we are testing consistency with quantum dynamics.  Dark, medium, and white shaded markers indicate SA, SVMC, and SVMC-TF values respectively.  Estimates are taken from high-precision instances only.}\label{fig:mc_collapse}
\end{figure*}

Tensor-network approaches provide faithful approximations to quantum dynamics, and generate accurate sample distributions in the appropriate parametric limits.  There are other methods, especially various Monte Carlo (MC) dynamics, which have been proposed to mimic the output of quantum annealers.  Here we discuss several such MC dynamics and show that they cannot emulate coherent out-of-equilibrium quantum dynamics.  The dynamics we discuss here are simulated annealing (SA)\cite{Kirkpatrick1983}, simulated quantum annealing (SQA) with path-integral Monte Carlo\cite{Boixo2013,Roennow2014}, spin-vector MC (SVMC)\cite{Smolin2013,Shin2014}, and SVMC with transverse-field freeze-out (SVMC-TF)\cite{Albash2021}.

On one hand, these methods are computationally inexpensive compared to tensor-network approaches, and therefore can be run on all system sizes used in our QA experiments.  On the other hand, they have already been shown to fail various emulation tests in toy problems\cite{Albash2015,Albash2015a}.  In particular, all of these dynamics fail in one way or another to qualitatively reproduce quantum dynamics even in a pure one-dimensional system~\cite{King2022} (see SM for Ref.~\onlinecite{King2022}, Figs.~S4 and S5).

Since these methods can be run at large scale, we can collapse the Binder cumulant for these dynamics, as in Fig.~\ref{fig:5}B, and compare against quantum mechanical values.  This was already done in Ref.~\cite{King2023}, which showed clearly distinct Kibble-Zurek exponents for QA, SQA, and SA in (dimerized) cubic problems.  As a consequence, QA was shown to have a dynamical advantage in optimization for these spin glasses.  We now repeat a similar test for SVMC and SVMC-TF to further solidify the deviation between these classical dynamics and the Schr\"odinger dynamics realized within QA.  For SA we use a geometric cooling schedule as in Ref.~\cite{King2023}.  For SVMC and SVMC-TF we use the model quench schedule with a low temperature equal to $\Gamma(s^*)/64$, where $s*$ is the crossing point with $\Gamma(s^*)=\mathcal J(s^*)$.  SA is run up to $t_a=2^{15}$ Monte Carlo sweeps, and SVMC and SVMC-TF are run up to $2^{17}$ Monte Carlo sweeps.  For such slow anneals we do not expect sensitivity to the shape of the annealing schedule.

Results are shown in Fig.~\ref{fig:mc_collapse} for SA, SVMC, and SVMC-TF for high-precision problems only.  SQA is far slower, so we disregard it since its consistency with quantum quench dynamics has already been refuted in 1D and 3D systems.  For the three dynamics, we find consistent Binder cumulant estimates that are much higher than quantum values.

%% file: sm_spatial_decomposition.tex
\section{Spatial decomposition techniques}\label{app:decomposition}
Experimental evidence and theory presented throughout the paper are consistent with the notion that in some of our experimental systems the correlation length, $\xi$\footnote{As discussed elsewhere, a correlation length scale might originate in the Kibble-Zurek mechanism, we call the Kibble Zurek correlation length $\hat \xi$. For purposes of this section we need not assume the Kibble-Zurek mechanism to be the source of correlations, or assume a scaling form.}, is not system spanning, particularly in higher precision (lower-energy scale) models at shorter time scales, and in lower dimensions. Intuition suggests that correlation length, more than system size, should dictate the complexity of approximate simulation.  In other words, we must consider the possibility that locality can be exploited through spatial decomposition of the simulation task.

This is natively in our PEPS methods, which, as a result, find success in square lattice models on short time scales. PEPS provides a globally consistent ansatz for the quantum state.  Given efficient contraction, this allows for estimation of both local and global observables. For calculation of local observables, some neighborhood contraction (proportional to $\xi$) is expected to be sufficient. 

For the purposes of estimating local observables, even simpler approximations than neighborhood-limited PEPS contraction may suffice. If correlation length $\xi$ is short throughout some dynamical process, we might truncate the system to a cluster of scale proportional to $\xi$ and only simulate dynamics on such a cluster, with statistics at its center well approximated. Unlike a PEPS neighborhood contraction, there needn't be an approximation of quantum phenomena beyond the cluster boundary, or at the boundary (in the following we assume that the clusters have open boundary conditions). Such subsystems might be simulated with an MPS or PEPS method, or by the QPU itself. In the context of MPS, the complexity of simulation can then be reduced in proportion to the area-law of the cluster with the error being controlled by a reduced bond dimension due to the smaller cluster size. In the context of the QPU, we might aim to estimate statistics associated with systems that are beyond a programmable scale.  Such a ``cluster approximation'' has the benefit of being much simpler than PEPS.

In both QPU and MPS cases, we might simulate many clusters centered at various points on a lattice, and then estimate correlations in the interior of such clusters. We study this for MPS sampling in \ref{app:travellingcluster} and for QPU subsampling in sections \ref{app:travellingcluster} and \ref{app:qpucluster}.  In Section \ref{app:boltzmannextrapolation} we consider a simple extrapolation method, based on such statistics, that allows accurate estimates of global statistics in some circumstances.

\subsection{Cluster approximations with MPS}
\label{app:travellingcluster}

\begin{figure*}
\includegraphics[width=1.0\columnwidth]{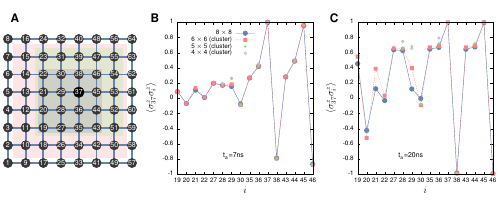}
\caption{{\bf Short-range correlations}. Panel {\bf A} shows the $6 {\times} 6$ (red), $5 {\times} 5$ (green), $4 {\times} 4$ (gray) clusters embedded in the $8 {\times} 8$ system, around site number 37. Correlations with respect to site=37, $c_{37,i}$, are shown for various cluster sizes in {\bf B} ($t_a=\SI{7}{ns}$) and {\bf C} ($\SI{20}{ns}$). Correlations are only shown with respect to the central sites compatible with all clusters.}\label{fig:tca1}
\end{figure*}

We already showed in the main text (Figs.~\ref{fig:2} and \ref{fig:4}) that the bond dimension required for MPS simulations to achieve the QPU accuracy follows area-law growth. Hence, naively it is expected that with increasing number of qubits the cost of MPS simulations attaining QPU accuracy should eventually go beyond the capability of classical computational resources. 
In this section we use MPS simulations on multiple open-boundary clusters around a chosen site inside an 
$L_x {\times} L_y$ 
square lattice cylinder, periodic in $y$, at a scale $L_x=L_y=8$; see Fig.\ref{fig:tca1}A. Panels B ($t_{a}=\SI{7}{ns}$) and C ($t_{a}=\SI{20}{ns}$) compare a subset of ground-truth spin-spin correlation estimates against those obtained with square clusters of sizes $L_c{\times} L_c$ with $L_c=4,5,6$. For $t_{a}=\SI{7}{ns}$, increasing accuracy is apparent as $L_c$ grows, with a good quantitative match in the majority of nearest and next-nearest neighbor correlations already at $L_c=5$.
Accuracy is reduced as $t_{a}$ increases towards $\SI{20}{ns}$ (Fig.~\ref{fig:tca1}C). This gives a qualitative understanding that for higher $t_{a}$ we need larger clusters to reproduce accurate short-range correlations, in line with increasing correlation lengths.

\begin{figure*}[!ht]
\includegraphics[width=1.0\columnwidth]{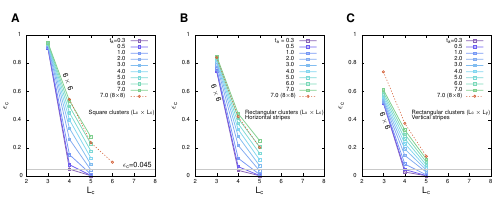}
\caption{{\bf Correlation matrix error}. Panels {\bf A}, {\bf B}, and {\bf C} show $\epsilon_c$ for various $t_{a}$ as a function of cluster size using the square cluster, open stripes, and periodic stripes, respectively. All results are for the full system of $6 {\times} 6$, except red dashed lines, which depict the $\epsilon_{c}$ of TCA on the $8 {\times} 8$ system. Legend values are provided in nanoseconds.}\label{fig:tca2}
\end{figure*}

The cluster position on the full lattice can be displaced to attain all short-range correlations. The quantum dynamics of all these small clusters can be trivially parallelized leading to only linear growth in the numerical cost with the system size ($N$ for square clusters, $L$ for boundary-spanning stripe clusters). 
A similar technique has been used extensively for performing Monte Carlo simulations on quantum-classical hybrid models~\cite{Kumar2006} under the name of traveling cluster approximation (TCA). We study the accuracy of TCA for quantum annealing by calculating $\epsilon_c$ attained by the TCA against MPS simulation of the full system (Fig.~\ref{fig:tca2}). We tried three different cluster shapes: square clusters ($L_{c} {\times} L_{c}$), open stripes ($L_{c}{\times} L_{y}$), and periodic stripes ($L_{x}{\times} L_{c}$). We noticed all three cases perform well for small $t_{a}$. For example, for $t_{a}=\SI{0.3}{ns}$, all three TCAs for the $6\times6$ system with cluster of linear length $4$ give QPU-level accuracy, and the clusters of length $5$ can produce trustworthy results up to $t_{a}=\SI{2}{ns}$. The correlation error grows as we increase $t_{a}$. This is expected because the correlation length increases with $t_{a}$ and the longer-range correlations, inaccessible to the small cluster, also becomes relevant. We also found periodic stripes perform best, followed by the open stripes and then square clusters. However, even at equivalent size, the cost of simulation of obtaining high accuracy results is a non-trivial function of the topology that must be considered. Periodic stripes require support for boundary spanning entanglement that must be accommodated with higher bond dimension. Similarly, the open stripes are costlier than square clusters of equivalent width not only in proportion to their additional length, but due to the higher bond dimension required to accommodate higher entanglement entropy (anticipated logarithmic corrections in the entropy as the system is passing through the QCP in the annealing process).

\subsection{Cluster approximations with the QPU}  
\label{app:qpucluster}
As shown in the previous section, MPS methods become computationally expensive for large systems because even though correlation lengths are short (for the annealing times of our annealing experiments), they can still encompass many qubits. 
When MPS simulations become unfeasible, one can then think of using the QPU as a \emph{solver} for each the traveling cluster.
It is therefore possible to use QPU subsampling to explore the relationship between large systems and the larger clusters required for their approximation.
Moreover, we can compare the QPU statistics with and without clustering approximations. For small systems, we have established that the cluster scale required for accurate estimation in this context is a good proxy for the cluster scale required for emulation of ground-truth statistics.

In this section we show that clustered approximations at these larger scales are indeed sufficient when the cluster volume is large compared to cluster-boundary, and when the correlation length (controlled primarily by annealing time $t_a$) is relatively small. In some scenarios we show that even clusters at a scale comparable to the full system are insufficient for estimation of local correlations, indicating the presence of significant system-spanning correlations.

\subsubsection{Two dimensions}
\label{app:two2stripes}

\begin{figure}
\includegraphics[scale=.8]{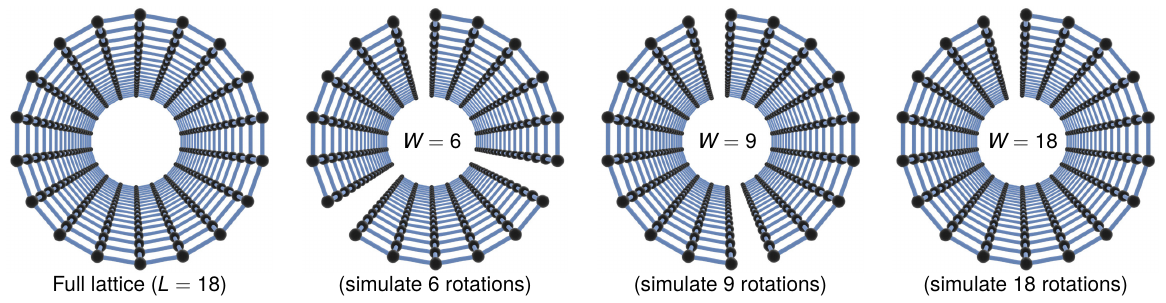}
  \caption{{\bf Rotating clusters in 2D.} For fast quenches, where final states have only short-ranged correlations, spoofing approaches can be based on simulating cheaper clusters---with one dimension having width $W$ smaller than the linear system size $L$---within the full lattice, then recombining output.  We evaluate this approach for square, cubic, and cubic no-dimer lattices.  Shown are examples of cluster shapes simulated in an effort to spoof $18\times 18$ square lattice results.}\label{fig:strips}
\end{figure}

For large 2D systems of size $L {\times} L$ with one periodic dimension, we use the following spatial decomposition to determine spin-spin correlations:\begin{itemize}
\item Choose a block width $W\leq L$.
\item In the periodic dimension, delete every $W$th row of couplers, leaving behind a set of $W {\times} L$ blocks that span from one open boundary to the other.  Do this for every possible offset ($0$ to $W$, ignoring divisibility issues which should be handled in the obvious way).
\item Simulate all the $W {\times} L$ blocks and determine spin-spin correlations within the blocks.
\item Combine the correlations.  For two spins that are in a block together, we take the weighted average of blocks that maximize the minimum distance of the two spins to the block cuts.  Correlations with spins near the block boundaries can be discarded, or not.  Long-distance correlations can be inferred from short-ranged correlations, or simply set to zero.
\end{itemize}
The cluster shapes are shown in \ref{fig:strips} for an $18{\times}18$ lattice; we might call this approach a rotating-cluster approximation (RCA). 

\begin{figure}
\includegraphics[scale=.99]{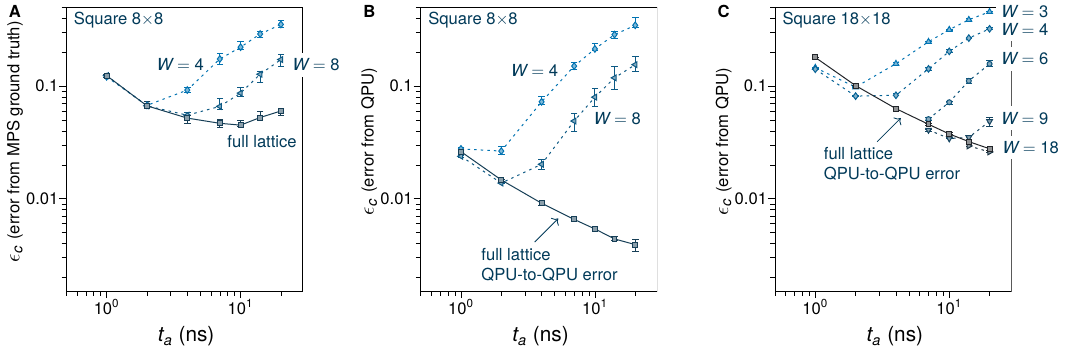}
  \caption{{\bf Error from rotating clusters in 2D.} ({\bf A}) QPU simulation is performed for $8\times 8$ square lattices, on the full lattice (solid line) and using a rotating cluster approximation (dashed lines).  We can compute MPS ground truths and measure $\epsilon_c$ directly for varying values of $W$.  For $W=8$, where each cluster is the same size as the full lattice, significant deviations from the full-lattice error level are seen for $t_a>\SI{4}{ns}$.  ({\bf B}) Here, error for the same experiments as in A is computed not from a ground truth, but from a full-lattice quantum simulation.  An independently-run full-lattice simulation gives a  noise floor marked by a solid line and square markers (which can be beaten due to averaging of clusters and smoothing of statistical noise).  Error against QA appears to lower-bound error against ground truth.  ({\bf C}) $18\times18$ system for which ground truth cannot be computed.  At $W=18$, $\epsilon_c$ is indistinguishable from the full lattice simulation even for $t_a=\SI{20}{ns}$.}\label{fig:strips2}
\end{figure}

In Fig.~\ref{fig:strips2}A--B we show experimental results for $L=8$ and $W=4,8$, performing cluster simulations and the full-lattice simulation with QPU; at this scale we can compute error against the MPS ground truth (Fig.~\ref{fig:strips2}A).  For fast quenches ($t_a < \SI{4}{ns}$) correlations are sufficiently short-ranged that RCA is indistinguishable from the full simulation, even for $W=4$.  As $t_a$ increases to $\SI{20}{ns}$, RCA error becomes significantly larger than full-lattice error.  Under the simplifying assumption that error from cluster size and QPU-to-model error are independent and add in quadrature for $\epsilon_c$, the fact that RCA error is larger than $\sqrt{2}$ times full-lattice error certifies that an RCA approach (at $W=8$) cannot produce simulation output with lower error than QA, even if the clusters were simulated exactly.

Fig.~\ref{fig:strips2}B represents the same experiments as Fig.~\ref{fig:strips2}A, but shows $\epsilon_c$ as calculated against a QPU ``ground truth'' rather than an MPS ground truth.  The full-lattice baseline becomes a measurement of $\epsilon_c$ between two independently-run but otherwise identical QPU experiments, essentially a noise floor that captures QPU device variation and noise as well as statistical noise (sampling error).  Here, for short $t_a$, errors are significantly smaller than in Fig.~\ref{fig:strips2}A, suggesting that QPU-to-model error is dominant.  However, for longer $t_a$ RCA is almost identical between Fig.~\ref{fig:strips2}A and B, indicating that cluster boundary effects become dominant.

We move to larger systems ($L=18$) in Fig.~\ref{fig:strips2}C, for which we were unable to compute a ground truth with MPS.  Note when $W=18$ we are performing 18 times as many simulations on the same number of spins, and combining the correlations determined from all rotations of the single cut.  Since we do not have a ground truth for $L>8$, we instead take our reference distribution (and correlations) from the full-lattice simulation as in Fig.~\ref{fig:strips2}B.  

As expected, this approach yields correlation errors that decrease as $W$ increases.  We also see that the signal deviates upwards from the noise floor as $t_a$ increases.  This is compatible with correlation length growth. The Kibble-Zurek mechanism anticipates a correlation length scaling as $\hat\xi \propto t_a^{1/\mu}$, where $\mu$ is the Kibble-Zurek exponent.  
This is loosely corroborated by the fact that we require $W\approx 6$ to approach the noise floor at $t_a=\SI{7}{ns}$, and $W\approx 9$ to approach it at $t_a=\SI{20}{ns}$, given that $\mu\approx 3$ in 2D systems.

These experiments are based on 20 inputs and 50,000 samples each, and the noise floor (solid line with square markers) is much higher than for the smaller system in Fig.~\ref{fig:strips2}B: Correlations at large distance (large compared to the correlation length) are noise-dominated, and the number of contributions scales as $N^2$. These are integrated with equal weight in the numerator and denominator of $\epsilon_c$, which is the origin of the sampling error. 

What conclusions can we draw from this data?  Error appears to be weakly dependent on system size (Figs.\ref{fig:4}A and \ref{fig:4_20ns}A), thus we anticipate that QPU-to-model error for $L=18$ should be close to that for $L=8$.  Observe that $W=6$ RCA error for $t_a=\SI{20}{ns}$ in Fig.~\ref{fig:strips2}C is more than twice that of the $L=8$ full-lattice simulation in Fig.~\ref{fig:strips2}A.  We take this as evidence that $W=6$ is insufficient for an RCA approach to match QPU error, even assuming exact simulation of the clusters.  Conversely, the fact that $L=18$, $W=9$ error in Fig.~\ref{fig:strips2}C is slightly less than $L=8$ QPU-to-model error in Fig.~\ref{fig:strips2}A suggests that an RCA approach might achieve comparable $\epsilon_c$ to QPU full-lattice simulation.  The scaling of $\chi_Q$ for strips of fixed width is anticipated to be at most logarithmic (in strip length), so only modest growth in MPS resources might be required to simulate local correlations accurately for $L \gtrsim 9$.

This analysis only applies to $\epsilon_c$, and does not imply that full-lattice states can be sampled routinely based on cluster output.  However, we find this to at least support the possibility that for square lattices with $L=18$ and $t_a=\SI{20}{ns}$, spin-spin correlations might be spoofed via spatial decomposition.

As already shown in the context of PEPS, square lattice problems are relatively easier than other lattices we simulate.  We now use the same RCA approach to establish that for three-dimensional lattices at the maximum experimental scale, such clustering approaches are not successful.

\begin{figure}
\includegraphics[scale=1]{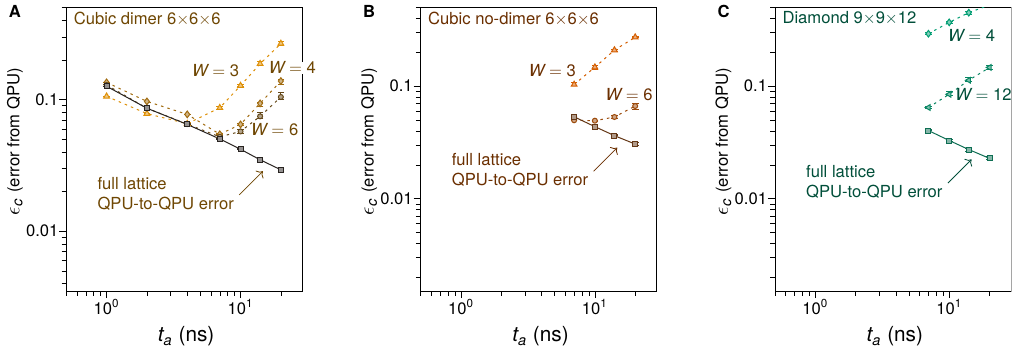}
  \caption{{\bf Error from rotating clusters in 3D.} ({\bf A}) Cubic dimer and ({\bf B}) cubic no-dimer at the largest simulated size (432 qubits), and ({\bf C}) diamond inputs at the largest defect-free simulated size (243 qubits).  As with 2D systems, correlations from full-lattice simulation are compared against composed correlations from multiple simulations with varying cluster size.  For $W=L$, the deviation of $\epsilon_c$ from the statistical floor (solid lines) indicates that a moving cluster containing all qubits, but with a single row of couplers deleted, is insufficient to capture correlations in the system for all three lattice types beyond short time scales.}\label{fig:strips3}
\end{figure}

\subsubsection{Three dimensions}
\label{app:rotatingcluster3d}
One can extend the square lattice approach to three dimensions in a straightforward way, decomposing a 3D system with planar cuts through the periodic dimension just as we decomposed a 2D system with linear cuts through the periodic dimension.  Here we do so, first showing results in Fig.~\ref{fig:strips3}A for an $L\times L\times L$ dimerized cubic system with $L=6$.  As in 2D, we see that deviation from the noise floor increases with $t_a$.  Indeed, it appears possible to spoof correlations fairly accurately with $W=4$ up to $t_a=\SI{7}{ns}$.  However, at $\SI{20}{ns}$ there is significant deviation even at $W=6$, where any two spins are inside the bulk of a cluster.

We repeat the approach for cubic no-dimer inputs, which---exactly like the diamond lattice---are four-regular and have girth (shortest cycle) six.  As in the cubic dimer case, we see in Fig.~\ref{fig:strips3}B a separation for $W=L=6$ at $t_a=\SI{20}{ns}$, although it is less pronounced.  For diamond lattices, we apply this approach to the largest defect-free lattice size, and find in Fig.~\ref{fig:strips3}C a very strong effect even at $W=L=12$---this is possibly in part due to the nature of the cluster boundary, which sees each spin lose two of four incident couplers, rather than one of four in cubic no-dimer lattices, and one of six neighboring dimers in the dimerized cubic lattice. Unlike in the square lattice case, errors for RCA at $t_a=\SI{20}{ns}$ are in all cases well above the threshold we anticipate to characterize the QPU. 

These results suggest that spoofing results for cubic, cubic no-dimer, and diamond systems by spatial decomposition is unpromising in this parametric regime. The simplest explanation for this is the presence of significant system-spanning correlations, the presence of which place limitations on a variety of related approximation methods.

\subsection{Classical extrapolation of statistics from local to global}\label{app:boltzmannextrapolation}

Spatial decomposition techniques typically allow estimation of local observables such as correlations and marginal probability distributions. However, inference of global properties is technically challenging on the basis of this information, particularly in the absence of a simple parametric model for the global distribution. In this section, we show that simple heuristics can, in some circumstances, extrapolate from (sufficiently accurate and numerous) local statistics of the projected distribution to macroscopic properties, with some tolerable error. In particular, we demonstrated examples in which fully visible Boltzmann distributions trained on accurate local pairwise correlations can be used to infer long-range correlations and approximately sample full states with an accuracy comparable to the QPU (in sampling a ground truth). 

A fully visible classical Boltzmann machine, also called an exponential model, takes the form of $P(x) = \frac{1}{Z}\exp(\theta_i f_i(x))$, where $Z$ is a normalization constant also called the partition function. The parameterization of this model can be inferred from expectations $\langle f_i(x)\rangle$. Supposing the statistics are geometrically local (e.g. $f_i(x) = x_i x_j$, for variables nearby on some lattice), then only local statistics are necessary. For smooth probability distribution described by essentially-classical correlation decay at larger geometric distances, a BM might provide good approximations (provided suitable choices are made for the set $f_i(x)$). Whilst it is technically NP-complete to train from sufficient statistics, training of a lattice model at the scales we present is practically straightforward, and methods such as maximum pseudolikelihood inference offer alternative efficient routes to model inference. 

Assuming long-range correlations are weak and well described classically, and that we have access to accurate local observables, then we can train our Boltzmann machine from local observables. A complex pattern of local correlations in the projected distribution, perhaps induced by entanglement, might be reproduced by a sufficiently complex set of $f_i(x)$. The full distribution, and statistics thereof, are then smooth extrapolations consistent with the local observables. The Boltzmann distribution can, for practical purposes, be sampled by Markov chain Monte Carlo (short correlation length implies short autocorrelation time) and so we can reproduce any statistic amenable to Monte-Carlo sampling with some systematic error reflecting the restriction in the choice of $f_{i}$ and quality of the training data.

Sections \ref{app:two2stripes}-\ref{app:travellingcluster} demonstrate MPS and QPU methods for obtaining local correlation estimates. PEPS methods based on the neighborhood contraction might also provide a basis for extrapolation in principle. 

\begin{figure}
\includegraphics[scale=1]{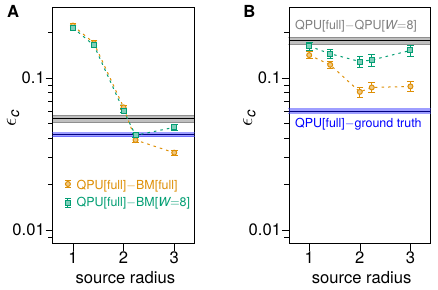}
  \caption{{\bf Error by extrapolation from local correlation estimates on $8{\times}8$ cylinders.} ({\bf A}) $t_a=\SI{7}{ns}$ and ({\bf B}) $t_a=\SI{20}{ns}$. 
  The divergence of the QPU from the ground-truth distribution is shown by the blue horizontal line. If the QPU distribution can be reconstructed from a rotating-cluster approximation with similar accuracy, we anticipate an equal-scale clustering approach can also be used for ground-truth estimation, as verified at smaller scale. 
  Correlations on $8 {\times} 8$ cylinder models can be estimated by rotating clusters with one cut (Fig. \ref{fig:strips} right); correlations are more accurate furthest from the cut. For each correlation, we can select a rotating cluster with the cut maximally separated from the site coordinates, yielding an error shown by the black horizontal line. We can select the best (spatially local) correlations from either the approximation (QPU[W=8]), or the target distribution (QPU), and train a Boltzmann distribution. As we incorporate correlations of increasing geometric distance, we can improve the quality of estimation relative to the raw correlation data. However, approximation data becomes increasingly errant, and so this only works for QPU[W=8] up to a limited scale. At $\SI{7}{ns}$ inclusion of ideal data up to geometric distance 3 is sufficient to reproduce the distribution with comparable accuracy to the ground truth, the effects of the boundary in the rotating cluster lead to a deviation upwards, so that $\epsilon_c$ remains just above the target threshold. At $\SI{20}{ns}$, longer distance is required for the ideal data [QPU] to achieve low error, compatible with a longer correlation length, and the approximating data QPU[W=8] is relatively poor leading to weak approximation.  }\label{fig:extrapolationLocalToGlobal}
\end{figure}

We found in our analysis of $L {\times} L$ square high and low-precision systems up to a scale $8 {\times} 8$ that achieving small $\epsilon_c$, for emulation of either the QPU or ground-truth distribution, was equally challenging (the error threshold obtained in training was similar for a fixed bound on the scale of local observables). Furthermore we were able to establish a strong correlation between $\epsilon_c$, classical fidelity and other distributional measures---measuring long range correlations as captured by $\epsilon_c$ was established as a good proxy for accurate reproduction of the full distribution. With this assumed, we present an analysis based on accurate estimation of $\epsilon_c$ only. Since observables approximated by large clusters are less expensive to create for QPU distributions, relative to MPS-based data, we present the approximation for the QPU generated approximations (\ref{app:two2stripes}), understanding this to hold qualitatively also for approximation of the ground truth. 

Figure \ref{fig:extrapolationLocalToGlobal} shows the result of training an Boltzmann machine to infer all correlations (and state probabilities) from local correlations within some geometrically constrained distance.\footnote{Geometric distance being the Euclidean distance in the standard square lattice representation. For geometric distance 3 we include only pairs separated by at most 2 couplers (graph distance 2); in other cases all correlations are included.}
For the $8 {\times} 8$ cylinder, the quality of the QPU distribution is captured by $\epsilon_c$, measuring the distance from the ground truth to the QPU. For our approximation, we measure the gap between the cluster-based approximation and the QPU (as the ground truth). When the latter error is smaller than the former, our cluster method can be considered to have succeeded in emulating the global distribution. At $\SI{7}{ns}$ we show that sufficiently accurate calculation of pairwise correlations up to Euclidean distance $\lesssim 3$ on square lattice at $8{\times}8$ scale is sufficient to train a Boltzmann machine with high accuracy. At $\SI{20}{ns}$, we do not see such a success, presumably due to longer correlation lengths. We present the training result based on both ideal data (from the full QPU distribution) and the estimates based on the rotating-cluster approximation. When incorporating data from sufficiently large geometric distance we expect the boundary effect to degrade performance of Boltzmann models trained with the latter set of correlations; this is found to be the case. At $\SI{7}{ns}$ the accuracy of local statistics is excellent, and we see little degradation in the performance out to distance 3. By contrast at $\SI{20}{ns}$ the correlation estimates are weaker, and we see an immediate deviation between the trained estimates. 

As shown in the main text, our distribution can become less entangled at larger $t_a$, particularly as we cross over to the adiabatic regime with system-spanning correlations. Less-entangled distributions favor this classical approximation method---assuming local observables may be accurately estimated. However, correlation lengths increase with $t_a$, degrading the quality of local estimates attainable by cluster approximations. The quality of local estimations is also impacted by dimensionality: for fixed correlation length the boundary effect (relative to the cluster bulk) has a more favorable ratio in low dimensions, in addition to the fact that correlation lengths can grow more quickly. In this section we have shown that at $\SI{7}{ns}$, 2D high-precision spin-glass problems at scale $8 {\times} 8$ and larger may be susceptible to clustering approximations, even where global statistics are concerned. For other experimental systems with larger correlation length relative to linear scale ($\SI{20}{ns}$ experiments, lower-precision, and/or higher-dimensional models) the method is understood to fail.